\documentclass[twocolumn, superscriptaddress]{aastex631}
\usepackage{CJK}
\usepackage{url}
\usepackage{amsmath}
\usepackage[super]{nth}
\usepackage{microtype}
\usepackage{graphicx}
\usepackage{rotating}
\usepackage{threeparttable}
\usepackage{hyperref}

\shorttitle{Detecting Storms and Spots with Extremely Large Telescopes}
\shortauthors{Plummer and Wang}

\graphicspath{{./}{figures/}}

\begin{document}
\begin{CJK*}{UTF8}{gbsn}

\title{Mapping the Skies of Ultracool Worlds: Detecting Storms and Spots with Extremely Large Telescopes}

\author[0000-0002-4831-0329]{Michael K. Plummer}
\affiliation{The Ohio State University \\
Department of Astronomy \\
140 W. 18th Ave. \\
Columbus, OH 43210, USA}

\correspondingauthor{Michael K. Plummer}
\email{plummer.323@osu.edu}

\author[0000-0002-4361-8885]{Ji Wang (王吉)}
\affiliation{The Ohio State University \\
Department of Astronomy \\
140 W. 18th Ave. \\
Columbus, OH 43210, USA}

\begin{abstract}

Extremely large telescopes (ELTs) present an unparalleled opportunity to study the magnetism, atmospheric dynamics, and chemistry of very low mass stars (VLMs), brown dwarfs, and exoplanets. Instruments such as the Giant Magellan Telescope - Consortium Large Earth Finder (GMT/GCLEF), the Thirty Meter Telescope's Multi-Objective Diffraction-limited High-Resolution Infrared Spectrograph (TMT/MODHIS), and the European Southern Observatory's Mid-Infrared ELT Imager and Spectrograph (ELT/METIS) provide the spectral resolution and signal-to-noise (S/N) necessary to Doppler image ultracool targets’ surfaces based on temporal spectral variations due to surface inhomogeneities. Using our publicly-available code, \texttt{Imber}, developed and validated in \citet{Plummer2022}, we evaluate these instruments’ abilities to discern magnetic star spots and cloud systems on a VLM star (TRAPPIST-1); two L/T transition ultracool dwarfs (VHS J1256$-$1257 b and SIMP J0136+0933); and three exoplanets (Beta Pic b and HR 8799 d and e). We find that TMT/MODHIS and ELT/METIS are suitable for Doppler imaging the ultracool dwarfs and Beta Pic b over a single rotation. Uncertainties for longitude and radius are typically $\lesssim 10^{\circ}$, and latitude uncertainties range from $\sim 10^{\circ} \ \rm{to} \ 30^{\circ}$. TRAPPIST-1's edge-on inclination and low $\upsilon \sin i$ provide a challenge for all three instruments while GMT/GCLEF and the HR 8799 planets may require observations over multiple rotations. We compare the spectroscopic technique, photometry-only inference, and the combination of the two. We find combining spectroscopic and photometric observations can lead to improved Bayesian inference of surface inhomogeneities and offers insight into whether ultracool atmospheres are dominated by spotted or banded features. 

\end{abstract}


\section{Introduction}\label{sec:Introduction}

\par Over the next decade, first light instruments are planned to come online for extremely large telescopes (ELTs) including the Giant Magellan Telescope - Consortium Large Earth Finder (GMT/GCLEF) \citep{Szentgyorgyi2018}, the Thirty Meter Telescope's Multi-Objective Diffraction-limited High-Resolution Infrared Spectrograph (TMT/MODHIS) \citep{Mawet2019}, and the European Southern Observatory's (ESO) Mid-Infrared ELT Imager and Spectrograph (ELT/METIS) \citep{Brandl2021}. Spectra collected with these instruments will have the requisite spectral resolution and signal-to-noise ratio (S/N) to map temperature variations and chemical inhomogeneities created by magnetic spots and clouds in ultracool atmospheres. 

\par Ultracool objects include very low mass stars (VLMs), brown dwarfs (BDs), and directly-imaged extrasolar giant planets (EGPs). With effective temperatures $\lesssim 2700$ K, these ultracool atmospheres allow the formation of metal and silicate condensates \citep{Tsuji1996a,Jones&Tsuji1997,Allard2001,Helling2008,Witte2011}. The hottest ultracool objects, such as late M dwarfs and early L dwarfs, may contain both magnetic star spots and mineral clouds, driving optical and near infrared (NIR) variability \citep{Lane2007,Heinze2013,Gizis2015,Dulaimi2023}. Cooler BDs and EGPs likely experience variability driven by cloud systems and planetary banding \citep{Reiners&Basri2008,Apai2013,Apai2017,apai21,Zhou2022}. 

\par Magnetic star spots and associated stellar activity can influence the atmospheres and habitability of planets with VLM stellar hosts. Although M dwarfs are more likely to host terrestrial planets than solar-type stars \citep{Mulders2015a,Tuomi2019,Zink2020,Sabotta2021}, late-type stars possess enhanced stellar activity and close-in habitable zones which could promote atmospheric escape and hinder the development of life \citep{kasting93,scalo07,zendejas10,kopparapu13,gunther20}. The orientation of high-energy events could affect the radiation experienced by short-period planets. Doppler imaging \citep{barnes15,Barnes2017} and TESS photometry \citep{Ilin2021,Martin2023} tentatively suggest a bias towards high-latitude star spot formation and flare eruption in M dwarfs, respectively. Follow-on observations and a statistical study can be accomplished by future ELT instruments.

 \par The condensation and subsequent precipitation of atmospheric mineral clouds has been proposed to explain both the reddening of L dwarf spectra as BDs cool across their lifetimes and the transition starting at $\sim 1300$ K to relatively cloud-free and blue mid-to-late T dwarf spectra \citep{Tsuji1996b,ackerman&marley01,Allard2001,Saumon2008}. This L/T transition is associated with both the emergence of methane absorption \citep{Oppenheimer1995,Noll2000} and an increase in spectral and photometric variability \citep{Radigan2014a,Radigan2014b,Eriksson2019} potentially due to  patchy clouds \citep{ackerman&marley01,Marley2002,Marley2010} or layers with varying temperatures and opacities \citep{Radigan2012, Apai2013,Buenzli2014}. A Doppler imaging map of Luhman 16B, an early T dwarf, supports these patchy and layered models \citep{crossfield14}.
 
\par Due to their similar temperature and spectral characteristics, field BDs have often been treated as high surface gravity analogs and used as stepping stones for EGPs. Studies of exoplanet-analog, planetary mass objects such as 2MASS J2139+02 \citep{Apai2013,Yang2016,Vos2022b}, PSO J318.5$-$22 \citep{Biller2015,Biller2018}, 2MASS 1207 b \citep{Zhou2016}, VHS J1256$-$1257 b (henceforth VHS 1256 b) \citep{Bowler2020,Zhou2020b,Zhou2022}, and SIMP J0136+0933 (henceforth SIMP 0136) \citep{Artigau2009,Apai2013,Yang2016,Vos2022b} have demonstrated variability to be common for these objects. However, upcoming telescopes will directly address EGPs' spectral and photometric variations with comparable S/N. 

\par Similar to BDs, EGPs demonstrate increasingly red spectra with evidence of extensive cloud coverage across the L class, but for planets such as HR 8799 bcde this track appears to continue to temperatures below $\sim1300$ K \citep{Bonnefoy2014,Zurlo2016,Allers&Liu2013,Zhang2020}. Cooler EGPs such as 51 Eri b \citep{Macintosh2015} and GJ 504 b \citep{Liu2016} manifest the color and $J$ band absolute magnitude commensurate with T dwarfs, perhaps indicating that the L/T transition occurs later for low-gravity planetary objects \citep{Biller2018}. 


\par Previously in \citet{Plummer2022}, we developed a unified spectroscopic and photometric analytical technique to infer surface inhomogeneities such as magnetic star spots, cloud systems, and atmospheric vortices in ultracool objects. In this paper, we use that technique to estimate the ability of ELTs to detect spots and storms in ultracool targets. We also introduce our publicly-available Python code, \texttt{Imber}\footnote{\url{https://github.com/mkplummer/Imber}} which uses the methods detailed in \citet{Plummer2022} to both numerically simulate spectroscopic and photometric surface inhomogeneities and to analytically infer such features on astrophysical targets. 

\par In \S \ref{sec:Instruments}, we describe GMT/GCLEF, TMT/MODHIS, and ELT/METIS and demonstrate \texttt{Imber}'s internal S/N calculator. In \S \ref{sec:Methods}, we describe our methods including how we produce simulated observations for each target, a review of our analytical technique, and how the method is used to generate and fit time-resolved spectroscopic and photometric observations. We then apply our method to six targets (TRAPPIST-1, VHS 1256 b, SIMP 0136, Beta Pic b, and HR 8799 d and e) in \S \ref{sec:Application2Targets} to determine the feasibility of Doppler imaging with each instrument. We explore combining spectroscopy and photometry to improve retrievals in \S \ref{sec:Spectral/Photo} and summarize our results in \S \ref{sec:summary}.

\section{Instruments}\label{sec:Instruments}

\par We will provide an overview of each telescopes' attributes and the instruments' scientific capabilities. We will then outline our process for adapting the S/N calculator originally created for the ESO ELT Spectroscopic Exposure Time Calculator (ETC) into a sub-module within \texttt{Imber}. To demonstrate our S/N computations validity, we compare \texttt{Imber}'s outputs to those obtained from the ELT Spectroscopic ETC. Instrument and telescope parameters are summarized in Table \ref{tab:InstrComp}.

\par To limit the scope of this work, we focus on first light, high spectral resolution instruments. The GMT/Near-Infrared (IR) Spectrograph (GMT/NIR) \citep{Jaffe2016} offers reasonably high spectral resolution ($R\sim 65,000 \ \rm{(JHK)}, \ 85,000 \ \rm{(LM)}$ and ideal wavelength coverage ($1.1 < \lambda < 5.4 \ \rm{\mu m}$), but as the instrument requires adaptive optics (AO) unavailable at first light, we do not include it. Planned (and conceptual) second generation, high spectral resolution instruments such as the TMT's Planet System Imager (TMT/PSI) \citep{Fitzgerald2019}; Mid-IR Camera, High-Disperser and IFU Spectrograph (TMT/b-MICHI) \citep{Packham2014}; High-Resolution Optical Spectrometer (TMT/HROS) \citep{Froning2006}; and the ELT's Armazones High Dispersion Echelle Spectrograph (ELT/ANDES) \citep{Marconi2022} are also left to future works. Our publicly-available and open source Python code \texttt{Imber} will conveniently allow second generation and beyond instrument performance to be evaluated using the methodology outlined in this paper.

\subsection{GMT/GCLEF}\label{ssec:GCLEF}

\par GMT is an integral component of the United States' ELT Program (US-ELTP) with a first light expected in the mid-2030s \citep{Fanson2020,Decadal2021}. Its primary mirror consists of seven 8.4 m diameter segments with a deformable secondary mirror comprised of seven, conjugate-paired 1.05 m diameter mirrors \citep{Johns2006}. GMT provides the IR diffraction-limited performance of a 24.5 m aperture and the collecting area ($368 \ \rm{m}^2$) equivalent to a 21.9 m telescope \citep{Johns2006}. Notably, unlike the two larger ELTs, AO/high-contrast imaging (HCI) are not scheduled to be implemented at GMT's first light \citep{McCarthy2016,Mawet2019,Brandl2021}.   

\par GMT/GCLEF is a visible light Echelle spectrograph and first light instrument for the telescope \citep{Szentgyorgyi2018}. Operating in its precision radial velocity (PRV) mode, GMT/GCLEF has a top spectral resolution of 105,000 (corresponding to an average velocity per resolution element of $\rm{\Delta v_{avg} = 2.86 \ km \ s^{-1}}$). The instrument has peak throughput of 11$\%$ and possesses both blue (350 to 540 nm) and red (540 to 900 nm) spectral channels \citep{Szentgyorgyi2018}. Due to the relatively cool effective temperatures of the targets considered in this paper, we will only be using the GMT/GCLEF red channel for our analysis.

\par GMT will be built at the Las Campanas Observatory (LCO) \citep{Thomas-Osip2008}. Due to lack of publicly-available sky emission and transmission models for GMT site, for this paper we will approximate the LCO sky model with ESO's SkyCalc tool evaluated using the Cerro Paranal Advanced Sky Model \citep{Noll2012,Jones2013}. We assume an airmass of 1.5, precipitable water vapor (PWV) of 2.5 mm, and include scattered starlight, zodiacal light, upper and lower atmospheric emission, and airglow in the radiance model.

\subsection{TMT/MODHIS}\label{ssec:MODHIS}

\par TMT is also part of the US-ELTP, and similar to GMT, has an expected first light in the mid-2030s \citep{Decadal2021}. TMT will be the only ELT located in the Northern Hemisphere \citep{Decadal2021}. TMT has a diameter of 30 m with a total collecting area of 664.2 m$^2$ \citep{Skidmore2015}. The telescope is capable of observing in wavelengths ranging from 0.31 to 28 $\mu$m \citep{Skidmore2015}.

\par TMT/MODHIS is a diffraction-limited, high-resolution spectrograph (R $\sim 100,000$, $\rm{\Delta v_{avg} = 3.00 \ km \ s^{-1}}$) \citep{Mawet2019}. With simultaneous NIR wavelength coverage ranging from 0.95 to 2.4 $\mu$m \citep{Mawet2019}, the instrument has an estimated throughput of $\sim 10 \%$\footnote{\url{www.tmt.org/page/modhis}}. 

\par The building site for TMT has yet to be determined, but the two final contenders are Mauna Kea, Hawaii and the Observatorio del Roque de Los Muchachos (ORM) in La Palma, Canary Islands (Spain)\footnote{\url{www.tmt.org/page/site}}. For the purposes of this paper, we will utilize the publicly-available sky emission and transmission tables for Mauna Kea \citep{Lord1992,Maihara1993}, but our code will allow a simple change to accommodate the potential ORM site. These models use an atmospheric temperature of 273 K and we assume an airmass of 1.5 and PWV of 1.0 mm for both the emission and transmission models.

\subsection{ELT/METIS}\label{ssec:METIS}

The European ELT (E-ELT), designed and operated by ESO, expects first light in 2028\footnote{\url{https://elt.eso.org/about/timeline/}}. E-ELT will have a diameter of 39 m and is comprised of 798 segments\footnote{\url{https://elt.eso.org/mirror}}. With a 978 $\rm{m}^2$ collecting area, at the time of its completion, it will be the largest optical and IR telescope in the world\footnote{\url{www.eso.org/public/archives/annualreports/pdf/ar_2021.pdf}}. 

ELT/METIS is capable of high-resolution spectroscopy (R $\sim 100,000$, $\rm{\Delta v_{avg} = 3.00 \ km \ s^{-1}}$) and HCI with coronagraphy in the NIR and mid-IR (MIR) regions (3 to 13 $\mu$m) \citep{Brandl2021}. For Doppler imaging, we are interested in its integral field unit (IFU) spectroscopy modes in the $L$ and $M$ bands which can also be combined with coronagraphy. We will use the extended wavelength option which covers 300 nm \citep{Brandl2021} to increase the effectiveness of spectral deconvolution (discussed with more detail in \S \ref{sec:Methods}). 

\par ELT/METIS also contains an imager with both continuum and spectral feature modes \citep{Brandl2021}. Notably, ELT/METIS is capable of switching between spectroscopic and imaging modes within seconds, allowing for near-simultaneous, multi-modal observations (Bernhard Brandl, Private Communication). 

E-ELT will be built at Cerro Armazones in the Chilean Desert. For our sky emission and transmission models, we again use the ESO SkyCalc Tool \citep{Noll2012,Jones2013} and select the high altitude location to best approximate Cerro Armazones. We make the same airmass, PWV, and radiance model assumptions as described for GMT/GCLEF.

\begin{table*}[t]
\caption{\label{tab:InstrComp} Instrument Comparison Summary}
\centering
    \begin{tabular}{l l c c c c c l}
    \hline
    \hline
    Instrument & Mode & $\lambda_{range}$  & R & $\rm{\Delta v_{avg}}$ & Throughput & Area & Sources\\
    & & [$\mu$m] & & [$\rm{km \ s^{-1}}$] & & [$\rm{m^{2}}$] \\
    \hline
    GMT/GCLEF & PRV$^a$ & 0.35-0.95$^a$ & 105,000$^a$ & 2.86 & 0.11$^a$ & 368$^b$ & $^a$\citet{Szentgyorgyi2018} \\
    & & & & & & &  $^b$\citet{Johns2006} \\
    \\
    \hline
    TMT/MODHIS & N/A & 0.95-2.4$^a$ & 100,000$^a$ & 3.00 & 0.1$^b$ & 664.2$^c$ & $^a$\cite{Mawet2019} \\
    & & & & & & & $^b$\url{www.tmt.org/page/modhis}  \\
    & & & & & & & $^c$\citet{Skidmore2015}  \\
    \\
    \hline
    ELT/METIS & IFU$^a$ & 3.65-3.95$^a$ & 100,000$^a$ & 3.00 & 0.25$^b$ & 978$^c$ & $^a$\citet{Brandl2021} \\
    & L' Imaging$^a$ & 3.475-4.105$^a$ & N/A & N/A  & 0.5$^b$ &  & 
    $^b$\url{www.eso.org/observing/etc/}  \\
    & & & & & &  & $^c$\url{https://elt.eso.org/about/facts/}  \\
    \\
    \hline
    \hline
    \end{tabular}
\label{tab:datasets}
\end{table*}

\subsection{Instrument S/N Computation}\label{ssec:SNRcomparison}

\par Computing the three instruments' S/N for each target described in \S \ref{sec:Application2Targets} is key to generating accurate simulated spectra, line profiles (LPs), and light curves. In this paper, we adopt the methodology used in the ESO ELT Spectroscopic and Imaging ETCs which use the following formula (modified here to account for host starlight suppression)\footnote{\url{www.eso.org/observing/etc/}},

\begin{equation}
    S/N = \frac{\sqrt{n_{exp}}N_{obj}}{\sqrt{N_{obj}+CN_{host}+N_{sky}+N_{pix}r^2+N_{pix}dT}}
\end{equation}

where $n_{exp}$ is the number of exposures and $N_{obj}$ is the number of electrons measured from the target. For targets requiring HCI, $C$ is the starlight suppression level and $N_{host}$ is the number of electrons from the host star. $N_{sky}$ is the number of electrons detected from the background sky, $N_{pix}$ is the number of detector pixels for which the observed light is distributed, $r$ is readout noise, $d$ is detector dark current, and $T$ is exposure time. The exposure time and quantity is set by the observer while readout noise and dark current depend on the detector used in the instrument. $N_{pix}$ is also instrument dependent while $N_{sky}$ is computed using the sky emission and transmission tables in tandem with the instrument's S/N reference area, the angular extent of the sky for which photons are collected\footnote{\label{SpectralETC}\url{www.eso.org/observing/etc/doc/elt/etc_spec_model.pdf}}.

\par Calculating $N_{obj}$ requires the target's apparent magnitude and template (synthetic) spectrum as well as the observing instrument's wavelength band, collecting area, throughput/efficiency, and effective diameter. The apparent magnitude determines how much flux from the target is available in a particular photometric band. The template spectrum is then scaled to the photometric flux to produce the mocked observation flux. This flux, along with the collecting area, throughput, and efficiency determine the number of photons available to the instrument detector which can be converted to electrons. 


\par
\begin{figure*}
\centering
\includegraphics[width=1.0\textwidth]{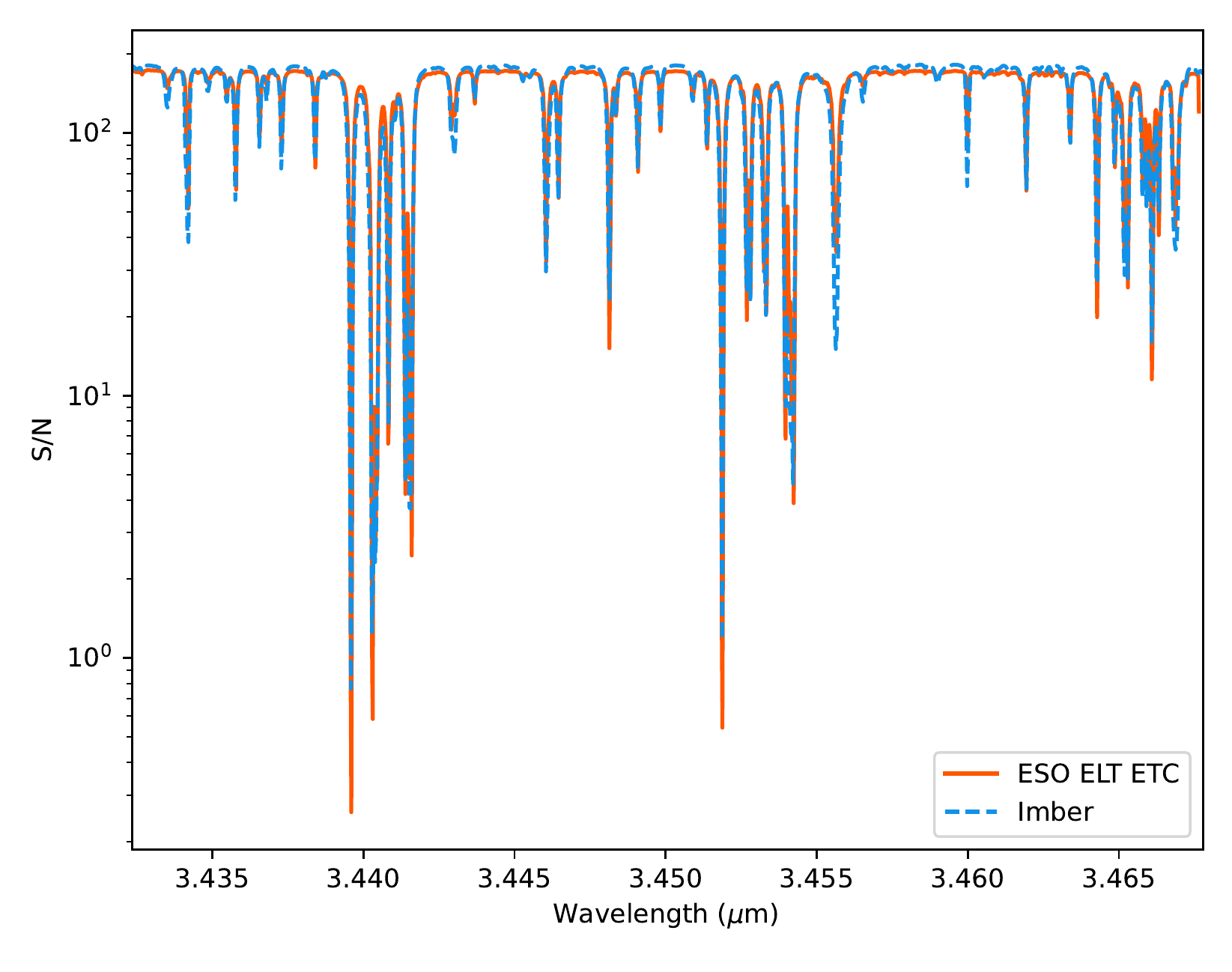}
\caption{\label{fig:SNRcomp} Spectral S/N comparison between ESO's ELT Spectroscopic Exposure Time Calculator (ETC) and \texttt{Imber} using same methodology. For comparison, a 1500 K object with apparent $J$ magnitude of 15 was used with a 300 s exposure. The spectral S/N were output in the $L$ band to demonstrate \texttt{Imber}'s ability to scale apparent magnitudes based on input spectra and to demonstrate performance in a similar band as ELT/METIS. Vertical axis is logarithmic.}
\end{figure*}

 To confirm \texttt{Imber}'s internal S/N calculator matches the ELT Spectroscopic ETC, we directly compare their results for the same input in Figure \ref{fig:SNRcomp}. We input a 1500 K blackbody spectra with a $J$ band apparent magnitude of 15 along with an exposure time of 300 seconds. We output the spectral S/N in the $L$ band to demonstrate \texttt{Imber}'s ability to scale instrument flux based on input apparent magnitude's band and also to confirm this ability near the spectral range of an instrument such as ELT/METIS. \texttt{Imber}'s computed S/N closely matches the values calculated by ESO's ELT Spectroscopic ETC with the small deviations primarily due to differences in sky transmission and emission models.

\section{Methods}\label{sec:Methods}

\par Using \texttt{Imber}, we predict the selected ELT instrument's ability to Doppler image ultracool targets. We begin by simulating observed LPs for each astronomical target; these observations are then fit via the analytical model described in \citet{Plummer2022}. As discussed in \citet{Plummer2022} in greater detail, \texttt{Imber} can also be used to create light curves and infer surface inhomogeneities based solely on photometric data or in combination with spectroscopic data to provide a unified solution.

\subsection{Simulating Observations}\label{ssec:SimulatingObservations}

\par To provide simulated spectroscopic observations, we implement a numerical simulation for each astrophysical target. The target's gridded (250 latitude by 500 longitude) photosphere is first simulated using an appropriate limb darkening effect to create a baseline flux map which is the basis for the rotational broadening kernel (BK). For this paper, we select a linear limb darkening law to lower computational costs based on \citet{Plummer2022}'s findings that rotational BKs created with linear and Claret coefficients agree within 2$\%$.

\par
\begin{figure*}
\centering
\includegraphics[width=0.5\textwidth]{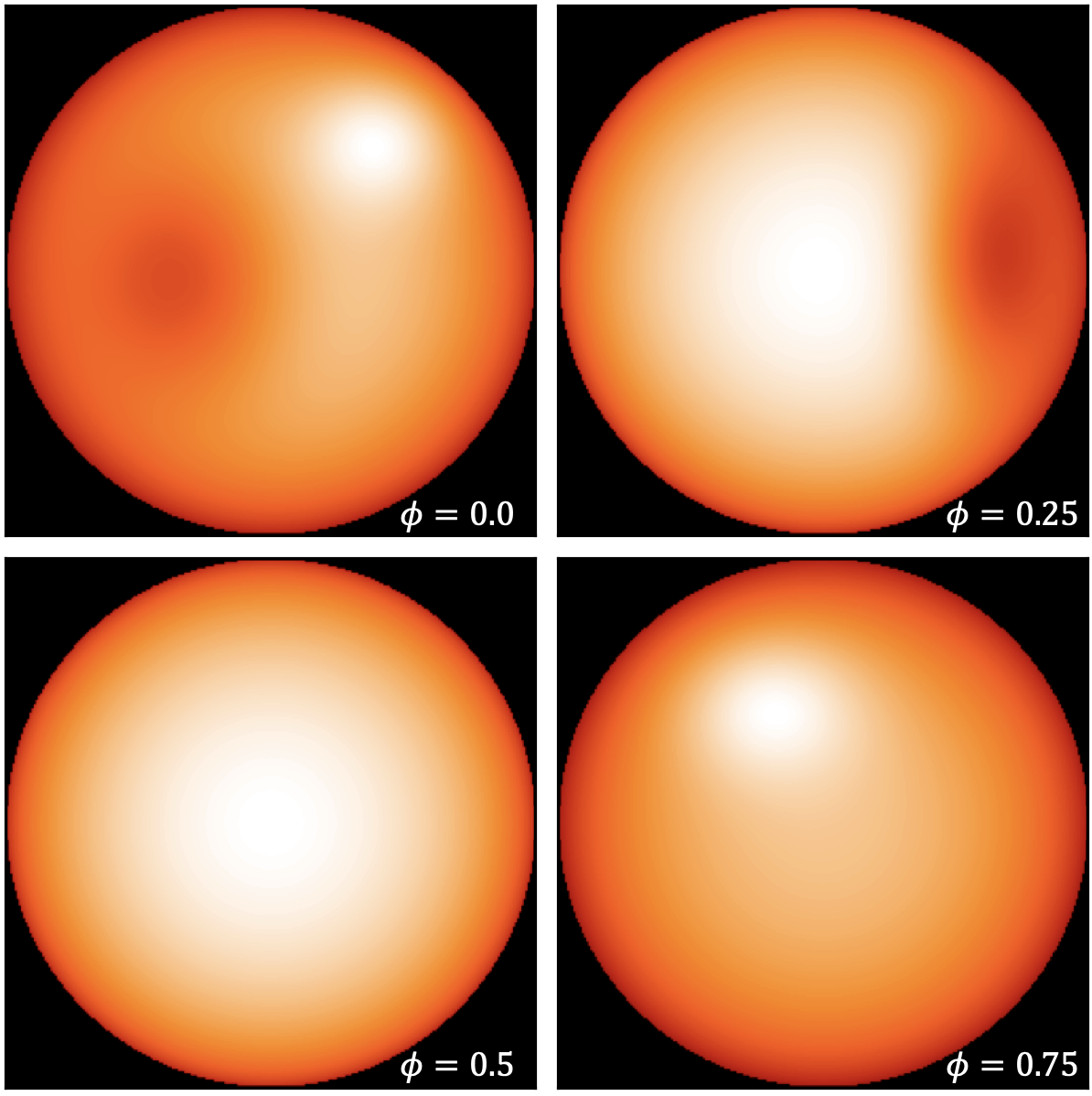}
\caption{\label{fig:StarMap} Orthographic view of example target ($i = 50^{\circ}$) over a full rotation with phases of 0.0, 0.25, 0.5, and 0.75. Both a dark spot (latitude = $15^{\circ}$, radius = $30^{\circ}$, and contrast = 0.3) and bright spot (latitude = $60^{\circ}$, radius = $25^{\circ}$, and contrast = -0.2) are added to the surface. Flux field shown as implemented by \texttt{Imber} to numerically generate simulated spectra, LPs, and photometric light curves.}
\end{figure*}

 Surface inhomogeneities are included via 2D Gaussian spots with positive (bright spots) or negative (dark spots) fluxes to create a time-varying spot map. In both numerical and analytical models, contrast can vary from twice the background brightness (contrast = -1) to a perfectly dark surface (contrast = +1). These 2D Gaussian spots are projected onto the orthographic view of the target (see Figure \ref{fig:StarMap}) from the observer's perspective via the Euler-Rodrigues formula \citep{Shuster1993}, discussed in finer detail in \citet{Plummer2022}. After this coordinate transformation, summing the flux from each longitudinal column and interpolating onto the desired number of LP elements creates a BK incorporating both rotational broadening and spotted features.

\par With rapidly rotating objects, rotation is the dominant broadening mechanism. But for slow rotators such as TRAPPIST-1, instrument, thermal, pressure, and natural broadening as well as microturbulence and macroturbulence can induce non-negligible line broadening. To address this, for our template, we adopt BT-Settl \citep{Allard2012} spectral models for VLMs and BD (further discussed in \S \ref{ssec:TestMethodology}) which use the PHOENIX stellar atmosphere code \citep{Hauschildt1992,Hauschildt1993,Allard&Hauschildt1995,Hauschildt1995,Hauschildt1996,Hauschildt1997,Hauschildt&Baron1995,Baron1996}. This lineage of stellar models accounts for thermal broadening, pressure broadening, natural broadening, and microturbulence \citep{Schweitzer1996}. For instrument broadening, we assume a normalized Gaussian profile with full-width-at-half-maximum (FWHM) equal to the central wavelength divided by the resolving power of the spectrograph.

\par
\begin{figure*}
\centering
\includegraphics[width=1.0\textwidth]{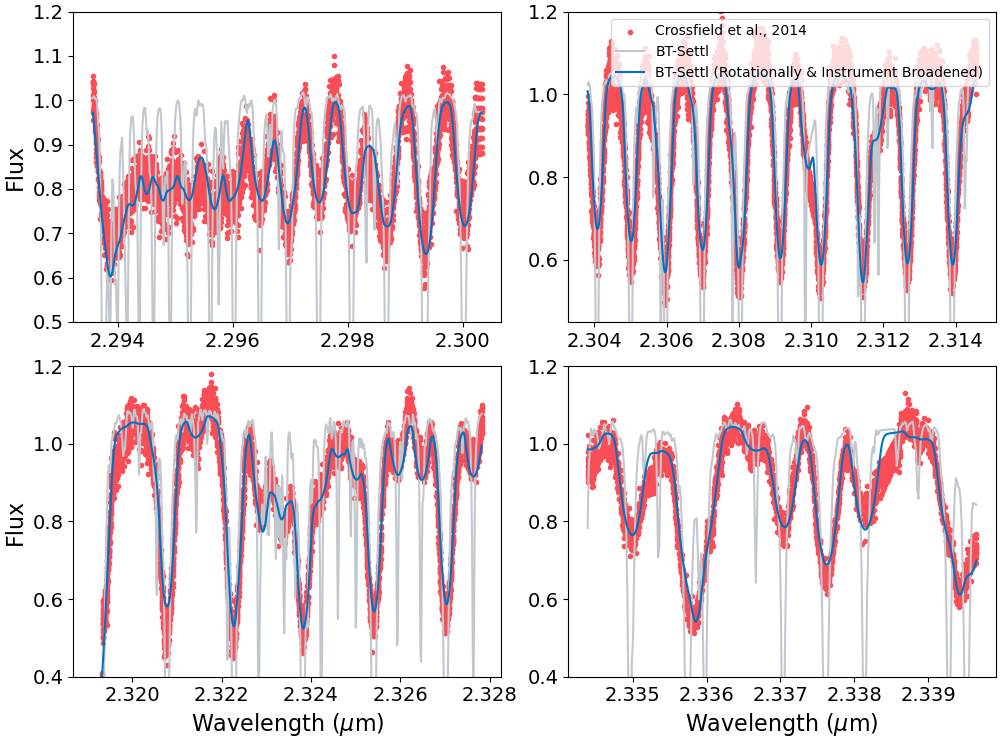}
\caption{\label{fig:L16Spectra} Demonstration of the rotational and instrument broadening (cyan) of a BT-Settl \citep{Allard2012} model (gray) to fit real-world, Luhman 16B spectra (red). Luhman 16B observations originally published by \citet{crossfield14} and reprocessed by \citet{luger21a}. Rotational broadening kernel assumes $\upsilon \sin i = 26.1 \ \rm{km \ s^{-1}}$ per \citet{crossfield14}.}
\end{figure*}

 The rotational and instrument BKs are convolved with the astronomical target's BT-Settl model to create a noiseless synthetic spectrum. Figure \ref{fig:L16Spectra} demonstrates the process of broadening a template, synthetic spectra (BT-Settl) to model real-world spectroscopic observations. Fourteen NIR spectroscopic observations of Luhman 16B (originally published by \citet{crossfield14} and re-processed in \citet{luger21a}) are over-plotted by an appropriate BT-Settl model (T = 1450 K, $\rm{\log(g) = 5.0}$) both before and after rotational and instrument broadening. Similar results to those two works are achieved and shown.

It is at this stage that Gaussian noise is added to the spectrum to account for the simulated observation's S/N. As described in \S \ref{ssec:SNRcomparison}, the S/N depends on the selected target, instrument, and atmospheric conditions. In \texttt{Imber}, S/N is computed at each wavelength, and noise level is sampled at each epoch of observation to provide spectrally and temporally varying noise. At this point, we have a complete set of simulated spectra for mocked observations.

\par Adopting a realistic procedure, we use Least Square Deconvolution (LSD), initially introduced by \citet{Donati1997} and refined by \citet{Kochukhov2010}, to compute LPs for the simulated observed spectra.  We use the following implementation of the technique as demonstrated in prior works \citep{Wang17,Wang18,PaiAsnodkar2021,Plummer2022},

\begin{equation}\label{eqn:LSDeqn}
\mathbf{Z(\upsilon_i) = \left (M^{T} \cdot M + \Lambda R \right )^{-1} \cdot M^{T} \cdot Y^{0}}
\end{equation}

\par where $Z(\upsilon)$ is the LP and $\upsilon_i$ is the RV corresponding to each element. $M$ is a line mask Toeplitz matrix computed using a template spectrum as described in \citet{Donati1997}. $\Lambda R$ are the scalar regularization parameter and regularization Toeplitz matrix which dampen noise amplification associated with LSD \citep{donatelli&reichel14,Kochukhov2010}.

\subsection{Analytical Model}\label{ssec:AnalyticalMethod}

\par \texttt{Imber} analytically models the astronomical target as an instrumentally and rotationally broadened LP with scaled 1D Gaussian perturbations corresponding to surface inhomogeneities either adding or subtracting flux from the background profile. The residuals of this analytical model and the LSD-computed observed LP can be used to implement Bayesian inference on spot parameters: latitude(s), longitude(s), radius(i), and contrast. 

\par Bayesian retrieval with this technique will identify the most prominent surface features for the set number of spots. Typically, during analysis, the number of spots is increased until the logarithmic evidence plateaus. Depending on the target's geometry and the data quality, degeneracies will begin to manifest as the number of spots retrieved increases. Gaussian priors can be implemented to break degeneracies but should be used with caution. For this reason, the computationally economical method summarized here and described in detail within \citet{Plummer2022} should be thought to complement numerical cell-by-cell Doppler imaging with smoothing via max entropy methods \citep{vogt87} or Tikhonov regularization \citep{Piskunov1990}.

\par When applying this technique to real-world data, it may be necessary to account for additional broadening processes such as microturbulence or Doppler broadening \citep{Gray2006}. For these mechanisms, a convolution with an additional Gaussian BK whose width is determined either through Bayesian analysis or least squares methods has been demonstrated to produce a reasonable fit \citep{Plummer2022}.

\subsection{ Photometric Light Curves and Analytical Fit}\label{ssec:PhotometricMethod}

\par As presented in \citet{Plummer2022}, \texttt{Imber} can also be used to both generate and fit photometric light curves. For the numerical simulation, the sum of the background photosphere flux is used to normalize to unity. That is, an unspotted surface will have a flux of 1. Bright spots add flux while dark spots subtract flux from the photometric curve. The analytical model works very similarly. The instrumentally and rotationally broadening kernel is used to normalize to unity such that the 1D Gaussian spots will either add or subtract from the light curve flux. In this manner, \texttt{Imber} can both simulate observed light curves and fit those light curves with an analytical model with similar caveats as the spectroscopic method with regards to degeneracies. 

\section{Application to Astronomical Targets}\label{sec:Application2Targets}

We will now apply our technique to six representative astronomical targets including TRAPPIST-1, VHS 1256 b, SIMP 0136, Beta Pic b, and HR 8799 d and e. A summary of each target's parameters pertinent to Doppler imaging are included in Table \ref{tab:TargetParams}. Each target's apparent magnitudes, organized by photometric bands corresponding to instrument wavelength range, are summarized in Table \ref{tab:TargetMagnitudes}. Because $I$ band magnitudes were not available for VHS 1256 b and SIMP 0136, we display the scaled $I$ band magnitude computed by \texttt{Imber} based on observed $J$ band magnitudes, their effective temperatures, and BT-Settl spectral models. The average computed S/N across each instruments' wavelength band are summarized numerically in Table \ref{tab:TargetMagnitudes} and graphically in Figure \ref{fig:TargetErrorPlots}. Retrieved spot parameter uncertainties are shown in Figure \ref{fig:TargetErrorPlots}.

\begin{table*}[t]
\caption{\label{tab:TargetParams} Astronomical Target Parameters}
\centering
    \begin{tabular}{l c c c c c c l}
    \hline
    \hline
    Name &  T$\rm{_{eff}}$ & $\log$(g) & Inclination & P & $i$ & $\upsilon \sin i$ & Sources \\
    (SpT)&  [K] & [$\log$(cm s$^{-2}$)] & Posterior$^*$ & [h] & [deg] & [km s$^{-1}$] \\
    \\
    \hline
    TRAPPIST-1 &  2566$\pm$26$^b$ & 5.2396$^{+0.0056}_{-0.0073}$$^b$& & $\sim79$$^{c,d}$ & $\sim90$$^{e}$ & $2.1\pm0.3$$^e$  & $^a$\citet{Costa2006} \\
    (M8V$^a$) & & & & & & & $^b$\citet{Agol2021} \\
    & & & & & & &  $^c$\citet{Luger2017} \\ 
    & & & & & & &  $^d$\citet{Vida2017} \\ 
    & & & & & & &  $^e$\citet{Brady2022} \\ 
    \\
    \hline
    VHS 1256 b  & $\sim1200^b$ & $\sim 4.5^b$ & & $22.04 \pm 0.05$$^c$ & $54^{+33}_{-39}$  & 13.5$^{+3.6}_{-4.1}$$^d$ & $^a$\citet{Gauza2015}\\
    (L7$^a$) & & & & & & & $^b$\citet{Miles2022}\\
    & & & & & & & $^e$\citet{Zhou2020b}\\
    & & & & & & & $^d$\citet{Bryan2018}\\
    \\
    \hline
    SIMP 0136 & $1150 \pm 70^{b}$ & $4.5 \pm 0.4^{b}$ & & $2.425 \pm 0.003$$^c$ & $80^{+10}_{-12}$$^{d}$& $52.8^{+1.0}_{-1.1}$$^{d}$  & $^a$\cite{Artigau2006} \\
    (T2.5$^{a}$) & & & & & &  & $^b$\cite{Vos2022b} \\
    & & & & & & & $^c$\cite{Artigau2009} \\
    & & & & & & & $^d$\cite{Vos2017} \\
    \\
    \hline
    Beta Pic b & $1724\pm15^a$ & $4.18\pm0.01^a$ & & $8.1 \pm 1.0$$^b$ & $57^{+18}_{-24}$ & $25.0\pm0.3^b$ &  $^a$\citet{Chilcote2017}\\
    & & & & & & &  $^b$\citet{Snellen2014} \\
    & & & & & & & $^c$\citet{Bryan2020b}\\
    \\
    \hline
    HR 8799 d & $1558.8^{+50.9}_{-81.4}$$^a$  & $5.1^{+0.3}_{-0.4}$$^a$ & Aligned & $6.0$$^{+2.5}_{-1.5}$$^a$ & $24^{+14}_{-13}$& $10.1^{+2.8}_{-2.7}$$^a$ &   $^a$\cite{JJWang2021} \\
    & & & Random & $12.0^{+5.3}_{-4.4}$ & $51.0^{+42}_{-29}$ \\
    \\
    \hline
    HR 8799 e & $1345.6^{+57.0}_{-53.3}$$^a$  & $3.7^{+0.3}_{-0.1}$$^a$ & Aligned & $4.1$$^{+1.1}_{-0.8}$$^a$ & $24\pm9$ & $15.0^{+2.3}_{-2.6}$$^a$  &   $^a$\cite{JJWang2021} \\
    & & & Random & $8.4^{+2.3}_{-3.0}$ & $56^{+40}_{-22}$ \\
    \\
    \hline
    \hline
    \end{tabular}
    \begin{tablenotes}
        \item{$^*$Two inclination prior scenarios are explored in \citet{JJWang2021}: (1) `Aligned' in which inclination priors are drawn based on the assumption that planetary spin inclination is aligned with orbital inclination (2) `Random' in which inclination priors are drawn from a uniform distribution.}
    \end{tablenotes}
\end{table*}

\begin{table*}[t]
\caption{\label{tab:TargetMagnitudes} Apparent Magnitudes, Integration Times, and S/N}
\centering
    \begin{tabular}{l c c c c l}
    \hline
    \hline
    Name & GMT/GCLEF & TMT/MODHIS & ELT/METIS & Integration Time & Sources \\
    & ($I$) & ($H$) & ($W1/L'$) & (Aligned/Random)\\
    \hline
    TRAPPIST-1 & $14.024 \pm 0.115^a$ & $10.718 \pm 0.021^b$  & $10.067 \pm 0.024^c$ & 5.28 h & $^a$\citet{Costa2006} \\
    & & & (W1) &  & $^b$\citet{Cutri2003} \\
    & & & & & $^c$\citet{WISE2010}\\
    S/N$_{\rm{(avg)}}$ & 229 & 1910 & 1640 \\
    \\
    \hline
    VHS 1256 b & $\sim 20.0$ & 15.595$\pm0.209^a$ & 13.6$\pm0.5^a$ & 1.46 h &  $^a$\citet{Gauza2015} \\
    & (computed) & & (W1) & \\
    \\
    S/N$_{\rm{(avg)}}$ & 7.39  & 97.7 & 53.5\\
    \\
    \hline
    SIMP 0136 & $\sim 16.6$ & $12.809 \pm 0.002^{a}$ & $11.94\pm 0.02^b$ & 9.65 m & $^a$\citet{Lawrence2012} \\
    & (computed) & & (W1) & & $^b$\citet{Cutri2014} \\
    \\
    S/N$_{\rm{(avg)}}$ & 13.1  & 118 & 79.0\\
    \\
    \hline
    Beta Pic b &  $N/A$ & $13.32\pm0.14^a$ & $11.24\pm0.08^a$ & 32.4 m&$^a$\citet{Currie2013} \\
    & & & (L') \\
    \\
    S/N$_{\rm{(avg)}}$ & N/A  & 119 & 107\\
    \\
    \hline
    HR 8799 d & N/A & $17.29\pm0.28^a$ & $14.59\pm0.17^b$ & 24/48 m &  $^a$\citet{Skemer2012}   \\
    & & & (L') &  & $^b$\citet{Currie2014}\\
    \\
    S/N$_{\rm{(avg)}}$ & N/A  & 8.13/11.6 & 4.33/6.14\\
    (Aligned/Random) \\
    \\
    \hline
    HR 8799 e & N/A & $16.94\pm0.28^a$ & $14.57\pm0.23^b$ & 16.4/33.6 m &  $^a$\citet{Skemer2012}   \\
    & & & (L') & & $^b$\citet{Currie2014}\\
    \\
    S/N$_{\rm{(avg)}}$ & N/A  & 9.48/13.7 & 5.77/8.30\\
    (Aligned/Random) \\
    \\
    \hline
    \hline
    \end{tabular}
\end{table*}

\subsection{Test Setup}\label{ssec:TestMethodology}

\par We will test each instrument's (GMT/GCLEF, TMT/MODHIS, and ELT/METIS) ability to infer spot location and radius for a 1-Spot scenario. TMT/MODHIS and ELT/METIS will be applied to all six targets. Due to GMT/GCLEF's lack of first light AO/HCI, it will not be applied to the companion EGPs (Beta Pic b and HR 8799 d and e). For both TMT/MODHIS and ELT/METIS, we conservatively assume $10^{-4}$ host star flux suppression in agreement with estimates from the instrument literature \citep{Mawet2019,Brandl2021}. 

Integrated exposures times (see Table \ref{tab:TargetMagnitudes}) for each target are computed by dividing the object's period (see Table \ref{tab:TargetParams}) by the number of samples. For our initial test, we adopt 15 samples, similar to \citet{crossfield14}. Higher number of samples will improve temporal resolution at the cost of lower S/N while the reverse is also true. These integrated exposure times are notional for objects with longer periods, as continuous viewing would not be feasible for periods $\gtrsim 10$ hours. 

\par Given the observed multi-rotational light curve evolution seen in ultracool dwarfs \citep{Buenzli15b,Buenzli15a,Zhou2020b,Zhou2022}, the temporal resolution may fail to capture the complete dynamic nature of spot parameters for slower rotators such as TRAPPIST-1 and VHS 1256 b. For this reason, dynamic spots with rotationally evolving size and contrast haven been implemented in the \texttt{Imber} code and are planned to be the focus of a future work but are not included here to limit our scope.



As mentioned in \S \ref{ssec:SimulatingObservations}, with all targets, we adopt BT-Settl model spectra described in \citet{Allard2012}. These high-resolution template spectra include H$_2$O, CH$_4$, NH$_3$, and CO$_2$ opacity line lists; revised solar oxygen abundances; and include cloud modeling which allows the code to be applied to astrophysical objects with effective temperatures ranging from 400 K to 70,000 K and $\log(g)$ of -0.5 to 5.5 \citep{Allard2012}. These ranges match well with our intended ultracool targets.

For a fair comparison, we enforce the same input truth parameters (latitude = $30^{\circ}$, longitude = $30^{\circ}$, radius = $30^{\circ}$, and contrast = +0.25) across all targets. While the assumed values may not be representative for TRAPPIST-1’s surface features (which are currently unknown), the values are typical for other ultracool dwarfs in the literature. A radius of $30^{\circ}$ agrees with inferred values for an equatorial dark spot on Luhman 16B from both \citet{crossfield14} and \citet{Plummer2022}. A contrast of $+0.25$ is consistent with spotted models fitting photometry for Luhman 16B \citep{karalidi16,Plummer2022} and VHS 1256 b \citep{Zhou2022}.

\par For Bayesian inference, we employ dynamic nested sampling \citep{Skilling2004,skilling06,higson19} via the publicly-available Python module \texttt{Dynesty}  \citep{speagle20}. For both spectroscopic and photometric Bayesian retrievals, we adopt uniform priors. Spot latitude and longitude distributions are $0^\circ \pm 90^\circ$ and $0^\circ \pm 180^\circ$ respectively. Radii distributions are uniform with lower and upper bounds of $5^{\circ}$ and $55^{\circ}$. For our instrument comparison, contrast will be fixed at +0.25 (indicating a spot 25\% darker than the background surface), but contrast as a free parameter is further explored in \S \ref{sec:Spectral/Photo}. 

\subsection{Application to TRAPPIST-1}\label{ssec:Trappist1}

\subsubsection{TRAPPIST-1 Background}\label{sssec:Trap1Bkgd}

TRAPPIST-1 is a late M dwarf, hosting at least seven rocky exoplanets \citep{Gillon2016,Gillon2017}, located 12.47$\pm0.01 \ \rm{pc}$ \citep{Gaia2021} from our solar system. With a spectral type of M8V \citep{Costa2006}, TRAPPIST-1 has an effective temperature of $\rm{T_{eff}}=2566\pm26 \ \rm{K}$ and a surface gravity of $\log\rm{(g)} = 5.2396^{+0.0056}_{-0.0073}$ \citep{Agol2021}. Life would likely have had sufficient time to develop on the star's planets due to the system's estimated age of over 7 Gyr \citep{Burgasser&Mamajek2017} although this value is not without controversy as the star possesses conflicting spectral features and kinematics associated with both young stars and older, field main-sequence dwarfs respectively \citep{Gonzales2019}.

Our interest in TRAPPIST-1 lies in its status as both a planetary host and an ultracool dwarf. Potentially three or four of the TRAPPIST-1 exoplanets lie within the star system's habitable zone \citep{O'Malley-James2017,Wilson2021}, making it an exciting target for the search of extraterrestrial life. However, like many late M dwarfs, TRAPPIST-1 appears to be active, exhibiting solar flares \citep{Vida2017,Paudel2018} as well as H$_\alpha$ \citep{Reiners&Basri2010} (correlated with chromospheric activity) and XUV emissions \citep{Wheatley2017}. Star spots have been inferred in TRAPPIST-1's photosphere by several studies using \textit{K2} data \citep{Luger2017,Morris2018}. Bright spots appear to be correlated with flaring events \citep{Morris2018}, likely impacting the system's habitability and planetary atmospheric escape.

\subsubsection{TRAPPIST-1 Setup and Results}\label{sssec:Trap1Results}

As with TRAPPIST-1's age, until recently the star's rotational period also appeared to be unsettled in the literature. \citet{Reiners&Basri2010} measured a $\upsilon \sin i$ of $6\pm2 \ \rm{km \ s^{-1}}$ which aligned with the rotational period measurement of $\sim1.4$ days from \citet{Gillon2016}. However, photometric measurements using \textit{K2} and \textit{Spitzer} data supported a rotational period of $\sim3.3$ days, corresponding to a $\upsilon \sin i$ of $\sim1.8 \ \rm{km \ s^{-1}}$ \citep{Luger2017,Vida2017}. The discrepancy was thought to be due to photometric observations recording the characteristic timescale of stellar activity versus the true rotational period \citep{Roettenbacher2017,Morris2018}. Recent measurements from the extreme-precision radial velocity spectrograph MAROON-X suggest $\upsilon \sin i$ and period values of $2.1\pm0.3 \ \rm{km \ s^{-1}}$ and 3.3 days \citep{Brady2022}.

Due to TRAPPIST-1's assumed edge-on inclination ($i \sim 90^{\circ}$), spot degeneracies are naturally created in the Northern and Southern Hemispheres with respect to inferred latitude. Because the star's exact inclination is unknown, we will assume an inclination of $85^\circ$ to provide a notional value to compute theoretical results. 

Despite more than sufficient S/N, with TRAPPIST-1's $\upsilon \sin i$ of 2.1 $\rm{km \ s^{-1}}$ and the three instruments' spectral resolutions of $R \sim 100,000$, \texttt{Imber} is unable to sufficiently resolve deviations to the rotationally-broadened LP to the degree required to successfully infer all spot parameters within 1$\sigma$. Although the retrieved longitude and radius were reasonably accurate, latitude retrievals possessed biases $\gtrsim 30^{\circ}$. Increasing the temporal resolution from 15 to 100 samples, per \citet{Kochukhov2016}, resulted in minor to negligible improvements to the retrieved solution.

To determine the source of the retrieved latitude inaccuracies, we explored varying TRAPPIST-1's stellar parameters. Decreasing the inclination to $45^{\circ}$ and $60^{\circ}$ improves the solution, but not enough to be within 1$\sigma$ with retrieved biases of $\sim 5^{\circ} \ \rm{and} \ 10^{\circ}$ respectively. This result matches the Doppler imaging technique introduced in \citet{Hebrard2016} in which the injection-retrieval appears to have a retrieved bias of $\sim 10^{\circ}$ for spots on a simulated target with $i = 60^{\circ}$ and $\upsilon \sin i = 1 \ \rm{km \ s^{-1}}$. However, by increasing $\upsilon \sin i$, it was determined that a $\upsilon \sin i \sim 7 \ \rm{km \ s^{-1}}$ resulted in a successful retrieval with further improvement seen at higher $\upsilon \sin i$ values. Surveying NASA's Exoplanet Archive\footnote{\url{https://exoplanetarchive.ipac.caltech.edu/index.html}}, there are not currently any ultracool dwarfs with measured $\upsilon \sin i \gtrsim \ 3 \ \rm{km \ s^{-1}}$ hosting confirmed exoplanets.

\par Provided polarimetric observations are available, Zeeman Doppler Imaging (ZDI) offers the ability to map stellar magnetic structure, including star spots, for slower rotators. Such maps have been created for low $\upsilon \sin i$ ($\sim \rm{1 \ to \ 4} \ \rm{km \ s^{-1}})$ targets including Sun-like stars \citep{Petit2008} and early M dwarfs \citep{Hebrard2016}. ZDI might only be effective for the hottest ultracool objects as the enhanced magnetic activity associated with star spots have not yet been detected in objects cooler than L5 dwarfs \citep{Paudel2018,Paudel2020}. In the future, we intend to extend our method to include deviations to polarized spectral line profiles.

\subsection{Application to VHS 1256 b}\label{ssec:VHS1256b}

\subsubsection{VHS 1256 b Background}\label{sssec:VHS1256bBkgd}

VHS 1256 b is a planetary mass object (PMO) with mass $< 20 \ M_J$ near the L/T transition (L7$\pm1.5$) discovered and initially characterized by \citet{Gauza2015}. The ultracool dwarf orbits a binary M dwarf system at a distance of $179 \pm 9 \ \rm{AU}$ \citep{Dupuy2020}. Notably, VHS 1256 b is a James Webb Space Telescope (JWST) Early Release Science (ERS) target. As part of ERS, \citet{Miles2022} presented the highest fidelity spectrum of the object to date, with wavelength coverage ranging from 1 to 20 $\mu$m. In addition to identifying volatiles (H$_2$O, CH$_4$, CO, and CO$_2$) and alkali metals (Na and K) in VHS 1256 b's atmosphere, \citet{Miles2022} found strong spectral evidence for chemical disequilibrium, likely caused by vertical mixing, and achieved the first detection of silicate clouds in a planetary mass companion.

Beyond chemical disequilibrium, atmospheric circulation is also suspected in VHS 1256 b due to its status as the most variable substellar object observed to date \citep{Miles2022}. \citet{Bowler2010,Bowler2020} observed $\sim 20\%$ NIR spectroscopic variability using the \textit{Hubble Space Telescope}/Wide Field Camera 3 (HST/WFC3). Follow-on studies using \textit{Spitzer}'s Infrared Array Camera (IRAC) Channel 2 centered at $4.5 \ \mu$m \citep{Zhou2020b} and the HST $J$ band \citep{Zhou2022} detected variability of $ 5.76\pm0.04\%$ and up to $38\%$ respectively. As discussed in \citet{Zhou2020b}, substellar models \citep{Marley2010,Morley2014} predict variability to be larger at shorter wavelengths and deeper atmospheric levels than at longer wavelengths corresponding to molecular bands. 

\begin{figure*}
\centering
\includegraphics[width=1.0\textwidth]{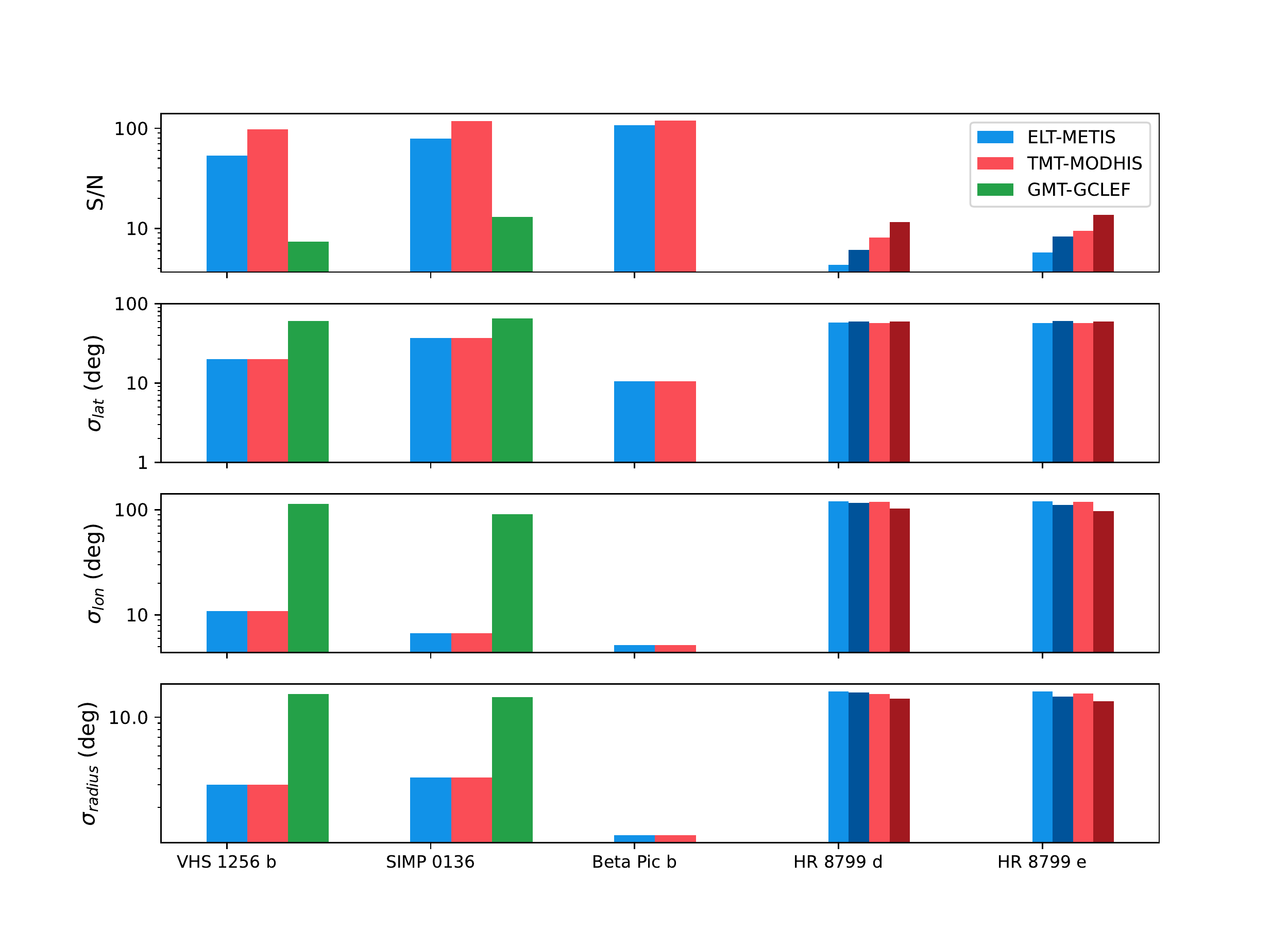}
\caption{\label{fig:TargetErrorPlots} Computed S/N and retrieved latitude, longitude, and radius uncertainties (1$\sigma$) for each target. Each case assumes a 1-Spot model with latitude = 30$^\circ$, longitude = 30$^\circ$, radius = 30$^\circ$, and a fixed contrast of +0.25, corresponding to a dark spot. For HR 8799 d and e, the lighter color bars denote retrievals run assuming inclinations aligned with the planets' orbital axes; darker bars denote inclinations derived from random uniform priors. All vertical axes are logarithmic. Retrievals conducted using \texttt{Imber} for numerical simulated observations and analytical model. \texttt{Dynesty} is used for dynamic nested sampling.}
\end{figure*}

\subsubsection{VHS 1256 b Setup and Results}\label{sssec:VHS1256bResults}

\par Studies have inferred a range of temperatures for VHS 1256 b from 1000 K to 1380 K \citep{Zhou2020,Dupuy2023,Hoch2022,Miles2022,Petrus2022}. For the BT-Settl template, we use an effective temperature of 1200 K as a compromise and $\log(\rm{g})=4.5$ from \citet{Miles2022}. In terms of rotation, we adopt a period of $22.04 \pm 0.05$ hours \citep{Zhou2020b} and $\upsilon \sin i = 13.5^{+3.6}_{-4.1} \ \rm{km \ s^{-1}}$ from \citet{Bryan2018}. Using this value, along with the inferred equatorial speed, $v_{eq} = 16.6^{+5.8}_{-7.0} \ \rm{km \ s^{-1}}$ from \cite{Bryan2020b}, we compute an inclination of $54^{\circ}$$^{+33^{\circ}}_{-39^{\circ}}$ by assuming the following expression,

\begin{equation}
    i = \sin^{-1} \bigg( \frac{\upsilon \sin i}{\upsilon_{r}} \bigg).
\end{equation}

We apply the same method for targets with unpublished inclinations such as Beta Pic b and the HR 8799 planets.

\begin{figure*}
\centering
\includegraphics[width=1\textwidth]{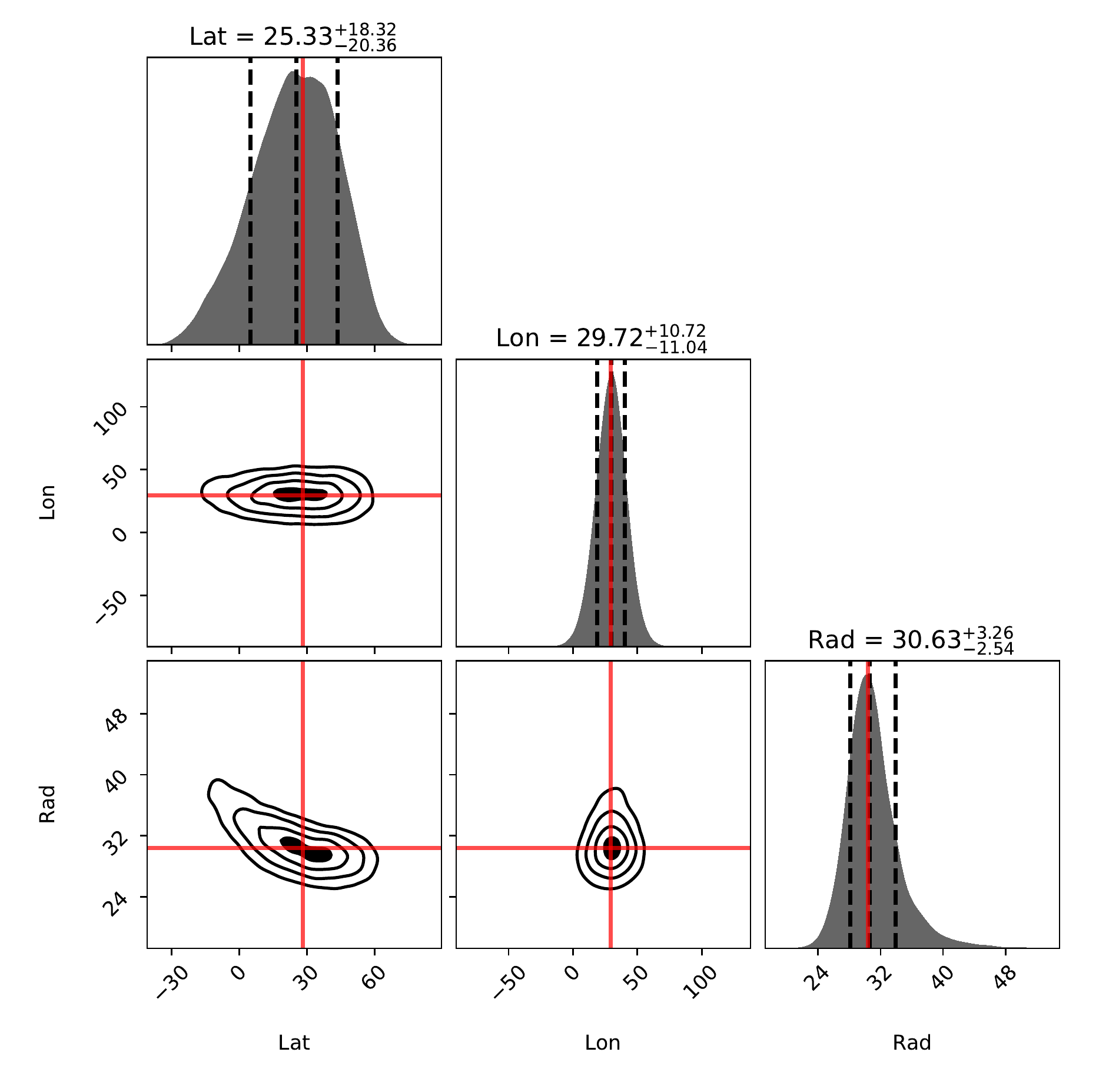}
\caption{\label{fig:Corner1Spot} VHS 1256 b example Bayesian retrieval of 1-Spot scenario with ELT/METIS via dynamic nested sampling. The mean inferred values along with $1\sigma$ quantiles displayed on the top of each column. Contours denote $0.5,1,1.5,$ and $2\sigma$ regions. Truth values are shown by the red line. Plot created with \texttt{Dynesty} \citep{speagle20}.
}
\end{figure*}

\begin{figure*}
\centering
\includegraphics[width=1\textwidth]{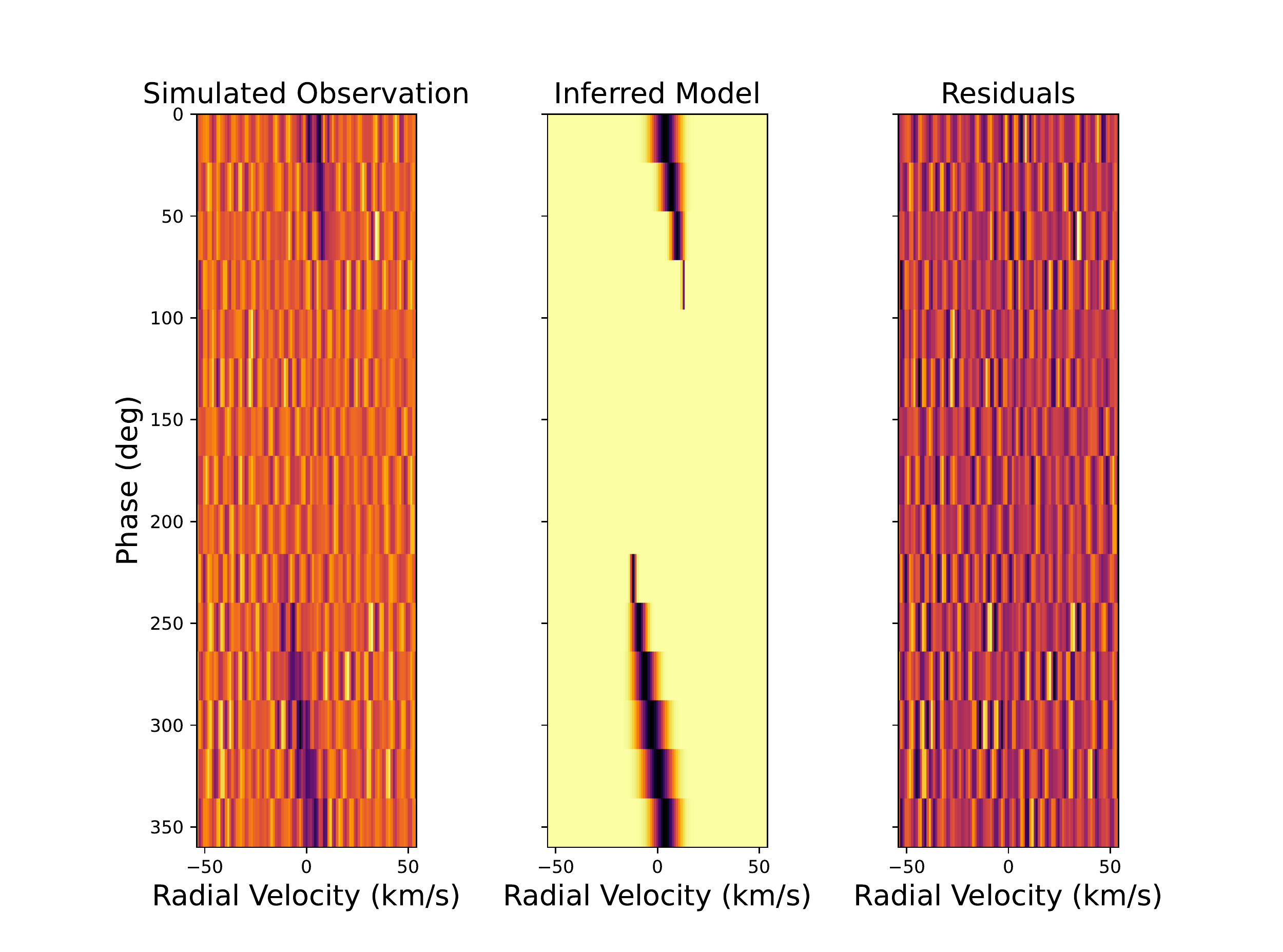}
\caption{\label{fig:DevPlotVHS} VHS 1256 b LP deviations. (Left) Deviations from simulated ELT/METIS observations. (Middle) Modeled deviations (reduced $\chi^{2} = 0.844$) for inferred dark spot. (Right) Residuals computed by subtracting modeled LP deviations from observed LP deviations.
}
\end{figure*}

To demonstrate the process used for each target, VHS 1256 b will be used as an example. Figure \ref{fig:Corner1Spot} is a corner plot showing the posterior distributions for the spot latitude, longitude, and radius for mocked observations using ELT/METIS. For each parameter, the input value is within the 68$\%$ credible range. Using the solution with the greatest likelihood, we generate LPs to compare the inferred analytical model to the the synthetic LP deviations created with our numerical simulation. The results can be seen in Figure \ref{fig:DevPlotVHS} with our simulated observation, inferred model (reduced $\chi^{2} = 0.844$), and residuals. 

\begin{figure*}
\centering
\includegraphics[width=1.0\textwidth]{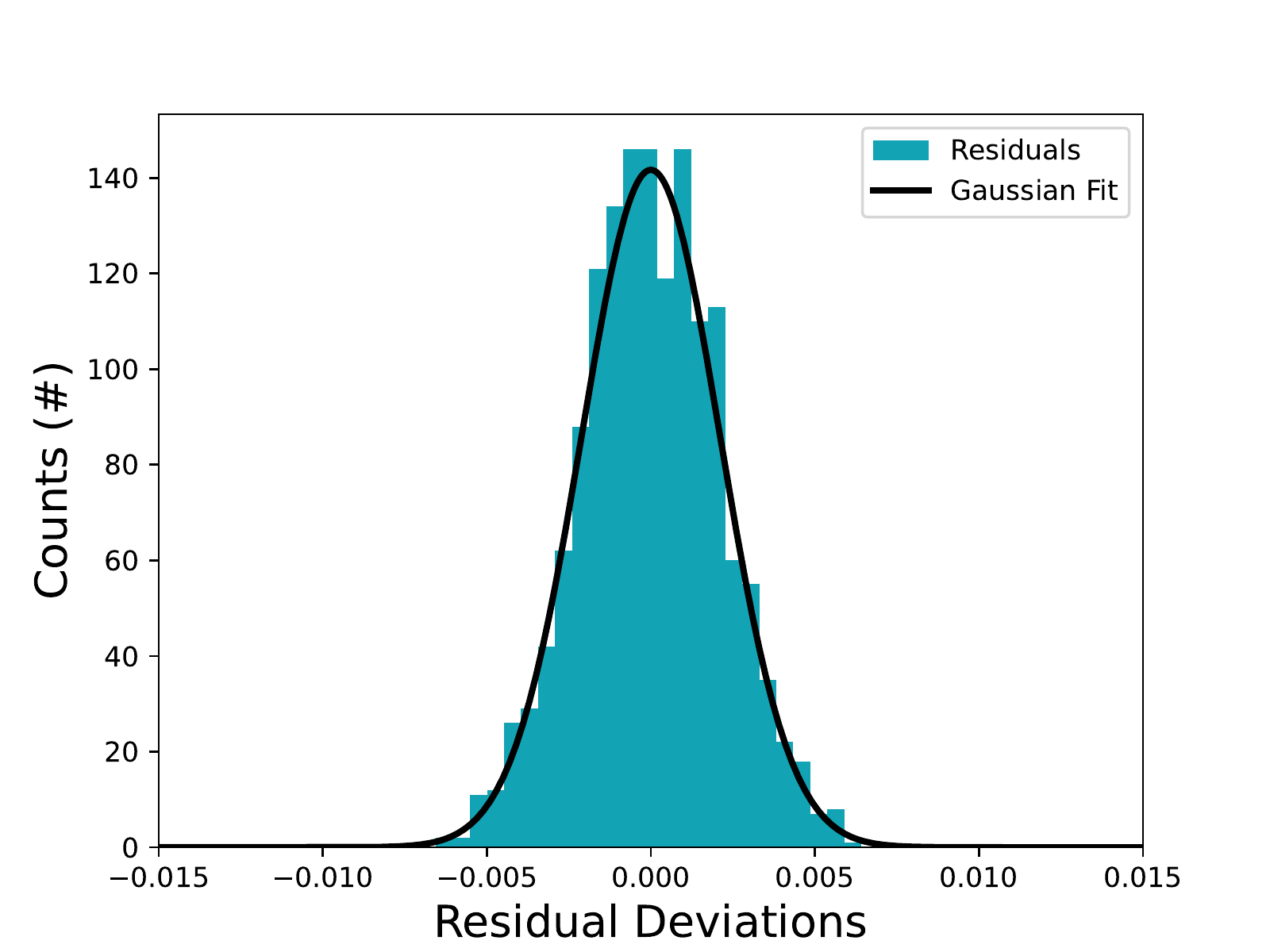}
\caption{\label{fig:ResidualDistributions} Residuals distribution of simulated VHS 1256 b LP deviations (for ELT/METIS) and inferred analytical model. Distribution is fit with Gaussian profile resulting in $\mu = -4.85 \times 10^{-7}$ and $\sigma = 0.00224$. The LP noise level corresponds to a S/N $\sim 450$, a significant increase over the mean spectral S/N ($\sim 55$). With a wider spectral band, TMT/MODHIS shows even higher gain with S/N improvement of $\sim100$ to $\sim1600$. As the S/N gain scales with the square root of the number of deep absorption lines (with wavelength coverage as a proxy), expectations are met as TMT/MODHIS (gain = 16, $\Delta \lambda = 1450$ nm) results in twice the S/N gain as ELT/METIS (gain $\sim8$, $\Delta \lambda = 300$ nm).} Although the distribution is only approximately Gaussian in the figure, tests indicate increasing time samples leads to convergence on the Gaussian.
\end{figure*}

The residuals in Figure \ref{fig:DevPlotVHS} appear to approximate Gaussian white noise. To confirm, we plot the residual distribution in Figure \ref{fig:ResidualDistributions}. The residuals approach a Gaussian distribution with a fit producing an approximately zero mean and a standard deviation ($\sigma = 0.00224$) corresponding to a LP S/N $\sim 450$, a factor of 8 gain over the mean spectra S/N ($\sim 55$). Due to its broader wavelength coverage, TMT/MODHIS shows an even more significant gain (factor of 16) between mean spectral and LP S/N ($\sim100$ to $\sim1600$ respectively). The S/N gain scales with the square root of the number of deep absorption lines (which can be estimated by the wavelength coverage). Expectations are met as the gain for TMT/MODHIS ($\Delta \lambda = 1450$ nm) is approximately twice that of ELT/METIS ($\Delta \lambda = 300$ nm).

TMT/MODHIS and ELT/METIS provide an order of magnitude greater mean spectra S/N than GMT/GCLEF as is expected based on each telescopes' relative aperture size and wavelength coverage with regard to predominantly infrared (IR) targets. The corresponding retrieved latitude, longitude, and radius uncertainties can be seen in Figure \ref{fig:TargetErrorPlots}. TMT/MODHIS provides improved S/N over ELT/METIS ($\sim 100$ vs. $\sim 55$), likely due to the enhanced thermal background in ELT/METIS's $L$ band, but this only results in a small difference in resulting uncertainties. 

For TMT/MODHIS and ELT/METIS, latitude uncertainties are approximately two to ten times higher than longitude and radius uncertainties. Difficulties with inferring latitude (vs. longitude, size, or contrast) via Doppler imaging has been documented \citep{Khokhlova1985,Khokhlova1986,vogt87}, and stems from the more subtle LP deviations that need to be resolved to precisely infer spot latitude.

\subsection{Application to SIMP 0136}\label{ssec:SIMP0136}

\subsubsection{SIMP 0136 Background}\label{sssec:SIMP0136Bkgd}

\par Similar to VHS 1256 b, SIMP 0136 (T2$\pm0.5$) \citep{Artigau2006} is an ultracool object at the L/T transition with a mass (12.7$\pm1.0 \ \rm{M_J}$) \citep{Gagne2017} near the deuterium-burning mass limit. First discovered by \citet{Artigau2006}, SIMP 0136 was previously considered to be a higher-mass BD, but due to its likely ($>99.9\%$) membership in the Carina-Near moving group \citep{Gagne2017} and corresponding age (200$\pm50$ \ \rm{Myr}) \citep{Zuckerman2006}, its mass appears to be that of a PMO (sometimes denoted as a rogue planet).

\par Significant $J$ band photometric variability has been detected for SIMP 0136. \citet{Artigau2009} first detected peak-to-peak amplitude variability of $8\%$ with a rotational period of $\sim 2.4 \ \rm{hr}$. Follow-up observations by \citet{Radigan2014a} saw smaller amplitudes of 2.9$\%$ but did not capture a full period. \citet{Croll2016} performed observations over 15 nights and found variability to evolve from $<1\%$ to $>6\%$ on timescales of 1 to 10 rotational periods. More recent observations by \cite{Eriksson2019} confirmed strong variability. 
\par A two-layer model comprised of a cold, upper layer and hot, deeper level had been proposed \citep{Apai2013} and demonstrated \citep{Yang2016} to explain SIMP 0136's variability. Recently, \citet{Vos2022b} found that observations best match overlying, patchy forsterite clouds and a deeper, thick iron cloud deck.

\subsubsection{SIMP 0136 Setup and Results}\label{sssec:SIMP0136Results}

Astronomical parameters (see Table \ref{tab:TargetParams}) for SIMP 0136 necessary for selecting a BT-Settl spectral model (T$\rm{_{eff}}$, $\log \rm{(g)}$) and relevant to Doppler imaging ($\upsilon \sin i$, $i$) are explored in 
\citet{Gagne2017} and \citet{Vos2017,Vos2022b}. For effective temperature and surface gravity we adopt the latest estimates from \citet{Vos2022b}, $T_{\rm{eff}} = 1150 \pm 70$ K and $\log \rm{(g)} = 4.5 \pm 0.4$. \citet{Gagne2017} constrains the $\upsilon \sin i = 50.9\pm0.8 \ \rm{km \ s^{-1}}$ and inclination, $i = 55.9^{\circ}\pm0.8^{\circ}$. \citet{Vos2017} finds a similar $\upsilon \sin i$ ($52.8$$^{+1.0}_{-1.1} \ \rm{km \ s^{-1}}$ ), but a higher inclination ($80^{\circ}$$^{+10^{\circ}}_{-12^{\circ}}$). We adopt the \citet{Vos2017} values due to the more conservative (for Doppler imaging) inclination. SIMP 0136 has a measured rotational period of $2.425 \pm 0.05$ hours \citep{Artigau2009}.

The S/N and retrieved parameter uncertainties for SIMP 0136 are qualitatively alike to VHS 1256 b with the exception of latitude uncertainty. The retrieved latitude uncertainty for SIMP 0136 is approximately twice that of VHS 1256; a result likely due to SIMP 0136's near edge-on inclination and the degeneracies created by this geometry. A similar but lesser effect is seen in SIMP 0136's retrieved radius when compared to the VHS 1256 b results. Small improvements to SIMP 0136's longitude (not affected by high inclination) are due to its higher S/N.

\subsection{Application to Beta Pic b}\label{ssec:BetaPicb}

\subsubsection{Beta Pic b Results}\label{sssec:BetaPicBkgd}

Beta Pic is a $24\pm3 \ \rm{Myr}$ \citep{Bell2015} star system located relatively close to our own solar system at $19.44\pm0.05 \ \rm{pc}$ in the aptly-named Beta Pic Moving Group \citep{vanLeeuwen2007}. It is host to Beta Pic b, a directly-imaged EGP with favorable contrast to its host star due to its youth and large, though disputed, mass ($\sim10 \ \rm{to} \ 15 \ \rm{M_J}$) \citep{Lagrange2010,Lagrange2012,Lagrange2020,Bonnefoy2013,Bonnefoy2014,Morzinski2015,Chilcote2017,2018Snellen&Brown,Dupuy2019,Kervella2019,GravityCollaboration2020,Vandal2020,Brandl2021}. Notionally identified in \citet{Lagrange2009} and then confirmed by \citet{Lagrange2010}, Beta pic b orbits its host star with a semi-major axis of $9.8\pm0.4 \ \rm{AU}$ \citep{Lagrange2020}.

\subsubsection{Beta Pic b Setup and Results}\label{sssec:BetaPicResults}

For our spectral template, we reference the effective temperature and surface gravity inferred by \citet{Chilcote2017} ($\rm{T_{eff} = 1724\pm15}$ K, $\log\rm{(g)=4.18\pm0.01}$) and \citet{GravityCollaboration2020} ($\rm{T_{eff} = 1742\pm10}$ K, $\log\rm{(g)=4.34^{+0.08}_{-0.09}}$) to select a BT-Settl model with $\rm{T_{eff} = 1700}$ K and $\log\rm{(g)=4.0}$. Beta Pic b has an observed period of $8.1 \pm 1.0$ hrs and $\upsilon \sin i$ of $25.0\pm3.0 \ \rm{km \ s^{-1}}$ \citep{Snellen2014}. Similar to how we proceeded for VHS 1256 b, we can use Beta Pic b's rotational velocity, $29.7^{+6.1}_{-8.8} \ \rm{km \ s^{-1}}$ \citep{Bryan2020b} and $\upsilon \sin i$ to compute its inclination, $57^{\circ}$$^{+18^{\circ}}_{-24^{\circ}}$.

For a directly-imaged EGP, Beta Pic b is relatively bright which is reflected in its larger S/N and smaller retrieved parameter uncertainties than the more massive L/T transition ultracool objects (VHS 1256 b and SIMP 0136). As mentioned above, we conservatively assume both TMT/MODHIS and ELT/METIS suppress starlight with a contrast of $10^{-4}$. Actual contrast values are likely to be improved in the final instrument design; however, starlight contamination above these levels would degrade the predicted performance.

\subsection{Application to HR 8799 d and e}\label{ssec:HR8799e}

\subsubsection{HR 8799 d and e Background}\label{sssec:HR8799Bkgd}

HR 8799 is a young star system with an estimated age of $30^{+20}_{-10} \ \rm{Myr}$ \citep{Doyon2010,Zuckerman2011} derived from its likely membership within the Columba Association moving group. HR 8799 hosts at least four EGPs \citep{Marois2008,Marois2010}, inner and outer debris disks (interior to HR 8799 e and exterior to HR 8799 b), and a halo consisting of small dust grains \citep{Su2009}. The two inner-most detected planets in the system to date, at $\sim 24$ and $15$ AU respectively \citep{Marois2008,Marois2010,GravityCollab2019}, HR 8799 d and e likely have masses between 7 to 10 $M_{J}$ \citep{Marois2008,Marois2010,JJWang2018,GravityCollab2019}. 

Clouds and chemical disequilibrium appear to be prevalent among the gas giant planets of HR 8799. Within the outer three planets (HR 8799 bcd), observations indicate the likely presence of clouds and chemical disequilibrium \citep{Bowler2010,Currie2014,Hinz2010,Janson2010,Barman2011,Madhusudhan2011,Barman2015,Marley2012,Skemer2014,Currie2014,Bonnefoy2016,Lavie2017,Wang2020,Ruffio2021,Wang2022,JJWang2022}. Furthermore, \citet{Marley2012} presents evidence that patchy cloud models fit HR 8799 bcd observations, and \citet{Currie2014} argues that all four gas giants' photometry fit thick, patchy cloud models better than uniform cloud models. Chemical disequilibrium and thick clouds have also been inferred in HR 8799 e in relatively recent literature \citep{Bonnefoy2016,GravityCollab2019,Molliere2020}.

\subsubsection{HR 8799 d and e Setup and Results}\label{sssec:HR8799Results}

For HR 8799 d's and e's effective temperature ($1558.8^{+50.9}_{-81.4}$ K/$1345.6^{+57.0}_{-53.3}$ K) and logarithmic surface gravity ($5.1^{+0.3}_{-0.4}$ /$3.7^{+0.3}_{-0.1}$), we adopt values retrieved by \citet{JJWang2021} using BT-Settl models. As discussed in \citet{JJWang2021}, these temperatures and HR 8799 d's surface gravity are higher than the values cited in the preponderance of studies \citep{Marois2008,Marois2010,Bonnefoy2016,Greenbaum2018,JJWang2022}. However, because we are also using BT-Settl models for our templates due to their high-resolution, we cautiously adopt \citet{JJWang2021}'s values.

\par Spin parameters for both HR 8799 d and e derive from \citet{JJWang2021} in which two inclination  scenarios are considered: the first in which inclination priors are based on the orbital inclination ($\sim 26^{\circ}$) \citep{JJWang2018,Gozdziewski2020} and the second in which the inclination priors derive from a random uniform distribution of $\upsilon \sin i$ as in \citet{Bryan2020b}. These assumptions lead to varying periods, computed inclinations (see Table \ref{tab:TargetParams}), and ultimately, integration times (see Table \ref{tab:TargetMagnitudes}). \citet{JJWang2021} infers $\upsilon \sin i$ values of $10.1^{+2.8}_{-2.7} \ \rm{km \ s^{-1}} / 15.0^{+2.3}_{-2.6} \ \rm{km \ s^{-1}}$ for HR 8799 d and e.

\par Due to the faint apparent magnitudes of HR 8799 d and e, the planets have the lowest S/N and highest average uncertainties out of the targets evaluated with TMT/MODHIS and ELT/METIS. The `random' inclination scenarios result in longer periods (with fixed temporal sampling this means longer integration times) and more edge-on viewing angles. As a result, these scenarios yield comparatively higher S/N and overall more constrained parameters than when their inclinations are aligned with their orbits. 

\par
\begin{figure*}
\centering
\includegraphics[width=1\textwidth]{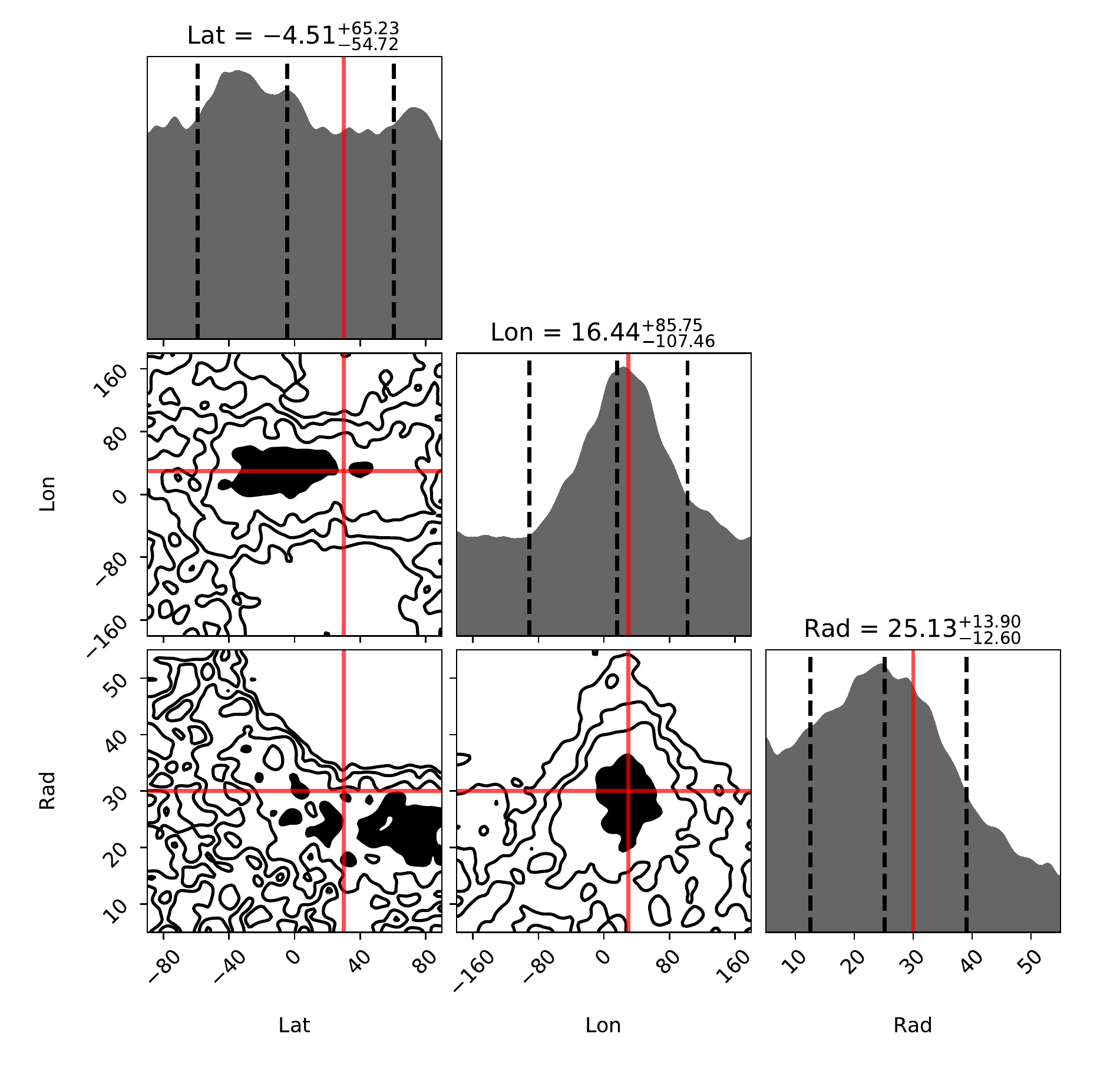}
\caption{\label{fig:HR8799e(1Rot)} HR 8799 e (Random Prior Inclination, 1-Rotation) Bayesian retrieval of 1-Spot scenario with TMT/MODHIS via dynamic nested sampling. The mean inferred values along with $1\sigma$ quantiles displayed on the top of each column. Contours denote $0.5,1,1.5,$ and $2\sigma$ regions. Truth values are shown by the red line. Plot created with \texttt{Dynesty} \citep{speagle20}.
}
\end{figure*}

\begin{figure*}
\centering
\includegraphics[width=1\textwidth]{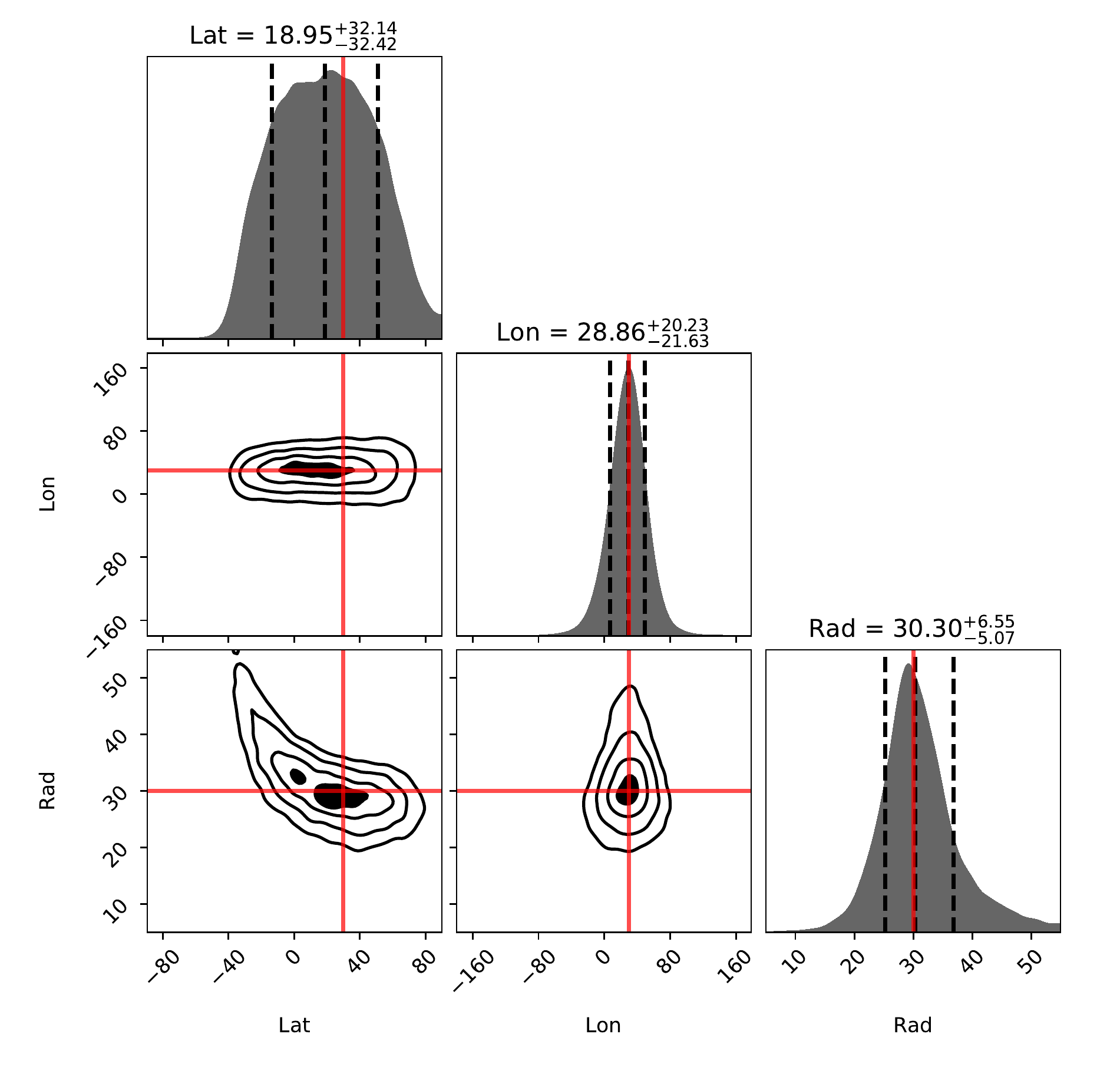}
\caption{\label{fig:HR8799e(5Rots)} HR 8799 e (Random Prior Inclination, 5-Rotations) Bayesian retrieval of 1-Spot scenario with TMT/MODHIS via dynamic nested sampling. The mean inferred values along with $1\sigma$ quantiles displayed on the top of each column. Contours denote $0.5,1,1.5,$ and $2\sigma$ regions. Truth values are shown by the red line. Plot created with \texttt{Dynesty} \citep{speagle20}.
}
\end{figure*}

 Figure \ref{fig:HR8799e(1Rot)} demonstrates that although the uncertainties are high, longitude and radius are beginning to be constrained for 1 rotation. Simulating 5 rotations and stacking the integrated exposures (S/N $\sim 31$ for TMT/MODHIS) leads to all three parameters being constrained (see Figure \ref{fig:HR8799e(5Rots)}). Although this technique would not identify atmospheric evolution on the timescales observed for the most variable ultracool dwarfs (VHS 1256 b, SIMP 0136, etc.), it would likely identify enduring features such as Jupiter's Great Red Spot. It should be noted observations over 5 rotations would require a significant investment in observation time.

\section{Improved Doppler Imaging Performance by Combined Spectroscopy/Photometry}\label{sec:Spectral/Photo}

\par Near-simultaneous, multi-modal observations have proven to be beneficial in astronomical observations \citep{Albrecht2022}. As demonstrated in \citet{Plummer2022}, the framework used in \texttt{Imber} can be used to generate both synthetic spectroscopic and photometric data via numerical simulation and also infer spot location, size, and contrast with computationally inexpensive analytical modeling using both spectral LPs and photometric light curves. 

\par As discussed in \S \ref{ssec:METIS}, ELT/METIS has both IFU spectrograph and imaging modes. Here both modes are applied to VHS 1256 b (with properties as described in \S \ref{ssec:VHS1256b}) to compare each mode's ability to retrieve spot parameters. We also consider a near-simultaneous, combined spectroscopic and photometric observation.

\subsection{Test Setup}\label{ssec:Photo/SpectralSetup}

\par Simulating ELT/METIS's imaging mode is conducted very similarly to the spectroscopic mode described in \S \ref{ssec:SNRcomparison}. An airmass of 1.5 is assumed at Cerro Armazones. To improve inference using light curves, the number of time samples is increased from 15 to 30. For photometry, an exposure time of 60 seconds is used, resulting in a S/N $\sim 1830$, considerably higher than those achieved with the spectroscopic instruments.

\par Increasing the number of time samples reduces each spectroscopic exposure (from 1.46 hours to 44 minutes) and corresponding S/N (from $\sim 53$ to $\sim 38$), but enhances temporal resolution. When spectroscopy is combined with photometry, integrated time for spectroscopy is 43 minutes per exposure (as 60 seconds is used for photometry) with minimal impact on the resulting spectroscopic S/N ($\sim 37$).  

\par
\begin{figure*}
\centering
\includegraphics[width=0.75\textwidth]{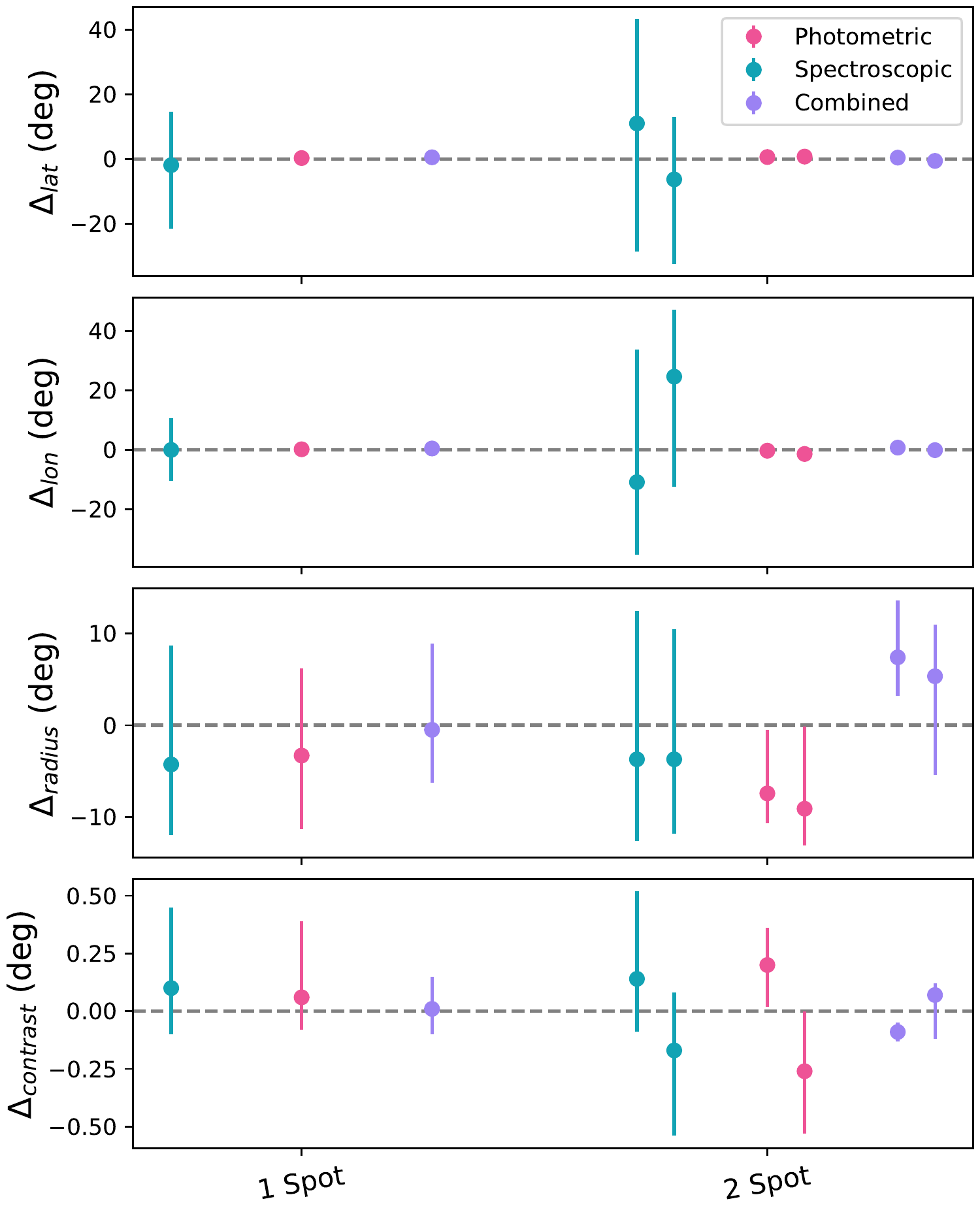}
\caption{\label{fig:PhotoComp} VHS 1256 b retrieval results for spectroscopic (aqua), photometric (magenta), and combined spectroscopic/photometric (indigo) modes. Spectroscopy and photometry are conducted with ELT/METIS's IFU and $L$ band imager respectively. Central point depicts maximum likelihood value and error bars represent 1$\sigma$ uncertainty. For the 2-Spot model, for each mode, the left error bar corresponds to the input dark spot and the right values corresponds to the input bright spot. Vertical axes show retrieved deviations from input truth values. 
}
\end{figure*}

 For this test, we consider two cases: Case 1 with one dark spot ($lat = 30^{\circ}$, $lon = 30^{\circ}$, $radius = 30^{\circ}$, $contrast = +0.25$) and Case 2 with two spots, bright and dark, ($lat = 30^{\circ}, \ 60^{\circ}$, $lon = -60^{\circ}, \ 60^{\circ}$, $radius = 30^{\circ}, \ 30^{\circ}$, $contrast = +0.25, \ -0.25$). Latitude, longitude, radius, and contrast are free parameters in all scenarios. Each retrievals' deviations from the input truth parameter is shown in Figure \ref{fig:PhotoComp}.

\par For 1-Spot models, the contrast priors vary uniformly from $+1.0$ to $-1.0$, but for 2-Spot model priors, we use both a bright spot ($-0.5 \pm 0.5$) and dark spot ($+0.5 \pm 0.5$). All other priors are as described in \S \ref{ssec:TestMethodology}. 

\subsection{Combined Spectroscopy/Photometry Results}\label{ssec:Spectral/PhotoResults}

\par Incorporating photometry into the retrieval results in improved inferred values in nearly all scenarios. For 1-Spot models, although photometry-alone improves both the retrieval's accuracy and uncertainty, a combined spectroscopic/photometric solution is even better. The 2-Spot model introduces a more complicated picture. Photometry-alone observations introduce a bias in retrieved the radius and contrast which is not fully resolved in the combined solution. These biases could perhaps be addressed by weighting the spectroscopic observations higher in the combined solution.

\subsection{Opportunities from Combined Spectroscopy and Photometry}

Ongoing tension exists on whether the spectral and photometric variability seen near the L/T transition is primarily due to planetary waves within banded structures or spotted-features such as vortices, cloud-systems, and rain-out patches. \citet{Apai2017} and \cite{apai21} argue the former case based on long term photometric observations of ultracool dwarfs. Polarimetric observations of the binary BD system, Luhman 16, also support banding \citep{Millar-Blanchaer2020}. There is furthermore a growing consensus in the BD general circulation model (GCM) community that atmospheric zonal bands form in nearly all radiative and convective forcing scenarios, even if only in the form of a single equatorial jet \citep{zhang14,showman19}.

On the other hand, there may also be a strong argument in favor of spotted features driving the observed variability and color and magnitude behavior at the L/T transition \citep{ackerman&marley01}. Recent analysis by \citet{Zhou2022} of VHS 1256 b multi-rotational photometry identified a degeneracy where both wave and spot-based models could dictate rotational modulation in ultracool objects. Furthermore, \citet{crossfield14} inferred and mapped spotted features on the visible surface of Luhman 16B using spectroscopy-based Doppler imaging, a technique insensitive to planetary banding. These results were replicated in \citet{luger21a} and \citet{Plummer2022}.

Using a combined spectroscopic and photometric approach, we could break this degeneracy and gain greater insight into the atmospheric structure of ultracool dwarfs and EGPs. Spectroscopic observations can confirm the existence and number of spotted features deforming a spectral LP for a specific astrophysical target. Furthermore, based on our results in \S \ref{ssec:Spectral/PhotoResults} and shown in Figure \ref{fig:PhotoComp}, it can be seen that supplementing spectroscopic observations with near-simultaneous photometry has the potential to improve spot inference. This proposed technique would be ideal for either coordinated efforts between observatories or instruments such as ELT/METIS that contain both spectroscopic and imaging modes.

\section{Summary}\label{sec:summary}

We have applied the unified spectroscopic and photometric numerical and analytical model developed in \citet{Plummer2022} to estimate Doppler imaging performance for ELT instruments using our publicly-available Python code \texttt{Imber}. We simulated spectral LPs for six targets including a VLM star (TRAPPIST-1), two ultracool dwarfs (VHS 1256 b and SIMP 0136), and three directly-imaged exoplanets (Beta Pic b and HR 8799 d and e). With dynamic nested sampling, we performed \textit{injection-retrieval} for surface inhomogeneities, inferring spot parameters via our analytical method. Here are our primary findings:
\begin{enumerate}

    \item TMT/MODHIS and ELT/METIS are suitable instruments for Doppler imaging VHS 1256 b, SIMP 0136, and Beta Pic b over 1 rotation. HR 8799 d and e may require multiple rotations with stacked spectra.

    \item GMT/GCLEF results in uncertainties significantly greater than TMT/MODHIS and ELT/METIS for the selected targets due to its wavelength coverage and aperture size. Stacking multiple rotation observations could allow for the identification of surface features using GMT/GCLEF.

    \item TRAPPIST-1 appears to be a less suitable target for all three instruments due to its relatively low $\upsilon \sin i$ ($2.1 \ \rm{km \ s^{-1}}$) and near edge-on inclination ($i \sim 90^{\circ}$). With a notional $85^{\circ}$ inclination, tests indicate $\upsilon \sin i$ values of $\gtrsim 7 \ \rm{km \ s^{-1}}$ are required to successfully retrieve spot parameters.

    \item Instruments with both spectroscopic and imaging modes, such as ELT/METIS, may be able to improve their solutions by augmenting spectroscopic data with short exposure time photometry.

\end{enumerate}

Moving forward, \texttt{Imber} can provide an estimation of spectroscopic and photometric instruments' abilities to detect magnetic spots, cloud structures, and storm systems in stars and ultracool objects to address questions such as habitability and the exact nature of the spectral L/T transition. The code can also be applied to real-world observations to detect surface inhomogeneities as demonstrated in \citet{Plummer2022}. In terms of future astrophysical applications, the unified spectroscopic and photometric approach of \texttt{Imber} makes it particularly well-suited to solve problems such as the degeneracy between spotted features and planetary banding seen in long-term ultracool object photometry. 

\section*{Acknowledgements}

\par The authors would like to thank the United States Air Force Academy, Department of Physics for sponsoring the graduate work of the first author. Additionally, we want to acknowledge the instrument expertise of Bernhard Brandl, Jochen Liske, Dimitri Mawet, and Andrew Szentgyorgyi. We want to thank the anonymous referee for constructive feedback and suggestions during the review process. We would like to thank the Group for Studies of Exoplanets (GFORSE) at The Ohio State University, Department of Astronomy for continuous feedback throughout the development of this research. J.W. acknowledges the support by the National Science Foundation under Grant No. 2143400.

\par This publication makes use of data products from the Wide-field Infrared Survey Explorer, which is a joint project of the University of California, Los Angeles, and the Jet Propulsion Laboratory/California Institute of Technology, funded by the National Aeronautics and Space Administration.

\par The views expressed in this article are those of the author and do not necessarily reflect the official policy or position of the Air Force, the Department of Defense, or the U.S. Government.

\software{Astropy \citep{astropy:2013, astropy:2018}, Dynesty \citep{speagle20}, Matplotlib \citep{Matplotlib}, Pandas \citep{Pandas}, Scipy \citep{scipy2020}}

\clearpage
\bibliography{references}{}

\begin{thebibliography}{}
\expandafter\ifx\csname natexlab\endcsname\relax\def\natexlab#1{#1}\fi
\providecommand{\url}[1]{\href{#1}{#1}}
\providecommand{\dodoi}[1]{doi:~\href{http://doi.org/#1}{\nolinkurl{#1}}}
\providecommand{\doeprint}[1]{\href{http://ascl.net/#1}{\nolinkurl{http://ascl.net/#1}}}
\providecommand{\doarXiv}[1]{\href{https://arxiv.org/abs/#1}{\nolinkurl{https://arxiv.org/abs/#1}}}

\bibitem[{{Ackerman} \& {Marley}(2001)}]{ackerman&marley01}
{Ackerman}, A.~S., \& {Marley}, M.~S. 2001, \apj, 556, 872,
  \dodoi{10.1086/321540}

\bibitem[{{Agol} {et~al.}(2021){Agol}, {Dorn}, {Grimm}, {Turbet}, {Ducrot},
  {Delrez}, {Gillon}, {Demory}, {Burdanov}, {Barkaoui}, {Benkhaldoun},
  {Bolmont}, {Burgasser}, {Carey}, {de Wit}, {Fabrycky}, {Foreman-Mackey},
  {Haldemann}, {Hernandez}, {Ingalls}, {Jehin}, {Langford}, {Leconte},
  {Lederer}, {Luger}, {Malhotra}, {Meadows}, {Morris}, {Pozuelos}, {Queloz},
  {Raymond}, {Selsis}, {Sestovic}, {Triaud}, \& {Van Grootel}}]{Agol2021}
{Agol}, E., {Dorn}, C., {Grimm}, S.~L., {et~al.} 2021, \psj, 2, 1,
  \dodoi{10.3847/PSJ/abd022}

\bibitem[{{Albrecht} {et~al.}(2022){Albrecht}, {Jensen}, {Jensen}, {Wilson},
  {Plummer}, {Key}, {O'Keefe}, {Chun}, {Strong}, \&
  {Schuetz-Christy}}]{Albrecht2022}
{Albrecht}, E.~M., {Jensen}, A.~M., {Jensen}, E.~G., {et~al.} 2022, Journal of
  the Astronautical Sciences, 69, 120, \dodoi{10.1007/s40295-021-00292-x}

\bibitem[{{Allard} \& {Hauschildt}(1995)}]{Allard&Hauschildt1995}
{Allard}, F., \& {Hauschildt}, P.~H. 1995, \apj, 445, 433,
  \dodoi{10.1086/175708}

\bibitem[{{Allard} {et~al.}(2001){Allard}, {Hauschildt}, {Alexander},
  {Tamanai}, \& {Schweitzer}}]{Allard2001}
{Allard}, F., {Hauschildt}, P.~H., {Alexander}, D.~R., {Tamanai}, A., \&
  {Schweitzer}, A. 2001, \apj, 556, 357, \dodoi{10.1086/321547}

\bibitem[{{Allard} {et~al.}(2012){Allard}, {Homeier}, \&
  {Freytag}}]{Allard2012}
{Allard}, F., {Homeier}, D., \& {Freytag}, B. 2012, Philosophical Transactions
  of the Royal Society of London Series A, 370, 2765,
  \dodoi{10.1098/rsta.2011.0269}

\bibitem[{{Allers} \& {Liu}(2013)}]{Allers&Liu2013}
{Allers}, K.~N., \& {Liu}, M.~C. 2013, \apj, 772, 79,
  \dodoi{10.1088/0004-637X/772/2/79}

\bibitem[{{Apai} {et~al.}(2021){Apai}, {Nardiello}, \& {Bedin}}]{apai21}
{Apai}, D., {Nardiello}, D., \& {Bedin}, L.~R. 2021, \apj, 906, 64,
  \dodoi{10.3847/1538-4357/abcb97}

\bibitem[{{Apai} {et~al.}(2013){Apai}, {Radigan}, {Buenzli}, {Burrows}, {Reid},
  \& {Jayawardhana}}]{Apai2013}
{Apai}, D., {Radigan}, J., {Buenzli}, E., {et~al.} 2013, \apj, 768, 121,
  \dodoi{10.1088/0004-637X/768/2/121}

\bibitem[{{Apai} {et~al.}(2017){Apai}, {Karalidi}, {Marley}, {Yang}, {Flateau},
  {Metchev}, {Cowan}, {Buenzli}, {Burgasser}, {Radigan}, {Artigau}, \&
  {Lowrance}}]{Apai2017}
{Apai}, D., {Karalidi}, T., {Marley}, M.~S., {et~al.} 2017, Science, 357, 683,
  \dodoi{10.1126/science.aam9848}

\bibitem[{{Artigau} {et~al.}(2009){Artigau}, {Bouchard}, {Doyon}, \&
  {Lafreni{\`e}re}}]{Artigau2009}
{Artigau}, {\'E}., {Bouchard}, S., {Doyon}, R., \& {Lafreni{\`e}re}, D. 2009,
  \apj, 701, 1534, \dodoi{10.1088/0004-637X/701/2/1534}

\bibitem[{{Artigau} {et~al.}(2006){Artigau}, {Doyon}, {Lafreni{\`e}re},
  {Nadeau}, {Robert}, \& {Albert}}]{Artigau2006}
{Artigau}, {\'E}., {Doyon}, R., {Lafreni{\`e}re}, D., {et~al.} 2006, \apjl,
  651, L57, \dodoi{10.1086/509146}

\bibitem[{{Astropy Collaboration} {et~al.}(2013){Astropy Collaboration},
  {Robitaille}, {Tollerud}, {Greenfield}, {Droettboom}, {Bray}, {Aldcroft},
  {Davis}, {Ginsburg}, {Price-Whelan}, {Kerzendorf}, {Conley}, {Crighton},
  {Barbary}, {Muna}, {Ferguson}, {Grollier}, {Parikh}, {Nair}, {Unther},
  {Deil}, {Woillez}, {Conseil}, {Kramer}, {Turner}, {Singer}, {Fox}, {Weaver},
  {Zabalza}, {Edwards}, {Azalee Bostroem}, {Burke}, {Casey}, {Crawford},
  {Dencheva}, {Ely}, {Jenness}, {Labrie}, {Lim}, {Pierfederici}, {Pontzen},
  {Ptak}, {Refsdal}, {Servillat}, \& {Streicher}}]{astropy:2013}
{Astropy Collaboration}, {Robitaille}, T.~P., {Tollerud}, E.~J., {et~al.} 2013,
  \aap, 558, A33, \dodoi{10.1051/0004-6361/201322068}

\bibitem[{{Astropy Collaboration} {et~al.}(2018){Astropy Collaboration},
  {Price-Whelan}, {Sip{\H{o}}cz}, {G{\"u}nther}, {Lim}, {Crawford}, {Conseil},
  {Shupe}, {Craig}, {Dencheva}, {Ginsburg}, {Vand erPlas}, {Bradley},
  {P{\'e}rez-Su{\'a}rez}, {de Val-Borro}, {Aldcroft}, {Cruz}, {Robitaille},
  {Tollerud}, {Ardelean}, {Babej}, {Bach}, {Bachetti}, {Bakanov}, {Bamford},
  {Barentsen}, {Barmby}, {Baumbach}, {Berry}, {Biscani}, {Boquien}, {Bostroem},
  {Bouma}, {Brammer}, {Bray}, {Breytenbach}, {Buddelmeijer}, {Burke},
  {Calderone}, {Cano Rodr{\'\i}guez}, {Cara}, {Cardoso}, {Cheedella}, {Copin},
  {Corrales}, {Crichton}, {D'Avella}, {Deil}, {Depagne}, {Dietrich}, {Donath},
  {Droettboom}, {Earl}, {Erben}, {Fabbro}, {Ferreira}, {Finethy}, {Fox},
  {Garrison}, {Gibbons}, {Goldstein}, {Gommers}, {Greco}, {Greenfield},
  {Groener}, {Grollier}, {Hagen}, {Hirst}, {Homeier}, {Horton}, {Hosseinzadeh},
  {Hu}, {Hunkeler}, {Ivezi{\'c}}, {Jain}, {Jenness}, {Kanarek}, {Kendrew},
  {Kern}, {Kerzendorf}, {Khvalko}, {King}, {Kirkby}, {Kulkarni}, {Kumar},
  {Lee}, {Lenz}, {Littlefair}, {Ma}, {Macleod}, {Mastropietro}, {McCully},
  {Montagnac}, {Morris}, {Mueller}, {Mumford}, {Muna}, {Murphy}, {Nelson},
  {Nguyen}, {Ninan}, {N{\"o}the}, {Ogaz}, {Oh}, {Parejko}, {Parley}, {Pascual},
  {Patil}, {Patil}, {Plunkett}, {Prochaska}, {Rastogi}, {Reddy Janga},
  {Sabater}, {Sakurikar}, {Seifert}, {Sherbert}, {Sherwood-Taylor}, {Shih},
  {Sick}, {Silbiger}, {Singanamalla}, {Singer}, {Sladen}, {Sooley},
  {Sornarajah}, {Streicher}, {Teuben}, {Thomas}, {Tremblay}, {Turner},
  {Terr{\'o}n}, {van Kerkwijk}, {de la Vega}, {Watkins}, {Weaver}, {Whitmore},
  {Woillez}, {Zabalza}, \& {Astropy Contributors}}]{astropy:2018}
{Astropy Collaboration}, {Price-Whelan}, A.~M., {Sip{\H{o}}cz}, B.~M., {et~al.}
  2018, \aj, 156, 123, \dodoi{10.3847/1538-3881/aabc4f}

\bibitem[{{Barman} {et~al.}(2015){Barman}, {Konopacky}, {Macintosh}, \&
  {Marois}}]{Barman2015}
{Barman}, T.~S., {Konopacky}, Q.~M., {Macintosh}, B., \& {Marois}, C. 2015,
  \apj, 804, 61, \dodoi{10.1088/0004-637X/804/1/61}

\bibitem[{{Barman} {et~al.}(2011){Barman}, {Macintosh}, {Konopacky}, \&
  {Marois}}]{Barman2011}
{Barman}, T.~S., {Macintosh}, B., {Konopacky}, Q.~M., \& {Marois}, C. 2011,
  \apj, 733, 65, \dodoi{10.1088/0004-637X/733/1/65}

\bibitem[{{Barnes} {et~al.}(2017){Barnes}, {Jeffers}, {Haswell}, {Jones},
  {Shulyak}, {Pavlenko}, \& {Jenkins}}]{Barnes2017}
{Barnes}, J.~R., {Jeffers}, S.~V., {Haswell}, C.~A., {et~al.} 2017, \mnras,
  471, 811, \dodoi{10.1093/mnras/stx1482}

\bibitem[{{Barnes} {et~al.}(2015){Barnes}, {Jeffers}, {Jones}, {Pavlenko},
  {Jenkins}, {Haswell}, \& {Lohr}}]{barnes15}
{Barnes}, J.~R., {Jeffers}, S.~V., {Jones}, H.~R.~A., {et~al.} 2015, \apj, 812,
  42, \dodoi{10.1088/0004-637X/812/1/42}

\bibitem[{{Baron} {et~al.}(1996){Baron}, {Hauschildt}, {Branch}, {Kirshner}, \&
  {Filippenko}}]{Baron1996}
{Baron}, E., {Hauschildt}, P.~H., {Branch}, D., {Kirshner}, R.~P., \&
  {Filippenko}, A.~V. 1996, \mnras, 279, 799, \dodoi{10.1093/mnras/279.3.799}

\bibitem[{{Bell} {et~al.}(2015){Bell}, {Mamajek}, \& {Naylor}}]{Bell2015}
{Bell}, C. P.~M., {Mamajek}, E.~E., \& {Naylor}, T. 2015, \mnras, 454, 593,
  \dodoi{10.1093/mnras/stv1981}

\bibitem[{{Biller} {et~al.}(2015){Biller}, {Vos}, {Bonavita}, {Buenzli},
  {Baxter}, {Crossfield}, {Allers}, {Liu}, {Bonnefoy}, {Deacon}, {Brandner},
  {Schlieder}, {Dupuy}, {Kopytova}, {Manjavacas}, {Allard}, {Homeier}, \&
  {Henning}}]{Biller2015}
{Biller}, B.~A., {Vos}, J., {Bonavita}, M., {et~al.} 2015, \apjl, 813, L23,
  \dodoi{10.1088/2041-8205/813/2/L23}

\bibitem[{{Biller} {et~al.}(2018){Biller}, {Vos}, {Buenzli}, {Allers},
  {Bonnefoy}, {Charnay}, {B{\'e}zard}, {Allard}, {Homeier}, {Bonavita},
  {Brandner}, {Crossfield}, {Dupuy}, {Henning}, {Kopytova}, {Liu},
  {Manjavacas}, \& {Schlieder}}]{Biller2018}
{Biller}, B.~A., {Vos}, J., {Buenzli}, E., {et~al.} 2018, \aj, 155, 95,
  \dodoi{10.3847/1538-3881/aaa5a6}

\bibitem[{{Bonnefoy} {et~al.}(2013){Bonnefoy}, {Boccaletti}, {Lagrange},
  {Allard}, {Mordasini}, {Beust}, {Chauvin}, {Girard}, {Homeier}, {Apai},
  {Lacour}, \& {Rouan}}]{Bonnefoy2013}
{Bonnefoy}, M., {Boccaletti}, A., {Lagrange}, A.~M., {et~al.} 2013, \aap, 555,
  A107, \dodoi{10.1051/0004-6361/201220838}

\bibitem[{{Bonnefoy} {et~al.}(2014){Bonnefoy}, {Marleau}, {Galicher}, {Beust},
  {Lagrange}, {Baudino}, {Chauvin}, {Borgniet}, {Meunier}, {Rameau},
  {Boccaletti}, {Cumming}, {Helling}, {Homeier}, {Allard}, \&
  {Delorme}}]{Bonnefoy2014}
{Bonnefoy}, M., {Marleau}, G.~D., {Galicher}, R., {et~al.} 2014, \aap, 567, L9,
  \dodoi{10.1051/0004-6361/201424041}

\bibitem[{{Bonnefoy} {et~al.}(2016){Bonnefoy}, {Zurlo}, {Baudino}, {Lucas},
  {Mesa}, {Maire}, {Vigan}, {Galicher}, {Homeier}, {Marocco}, {Gratton},
  {Chauvin}, {Allard}, {Desidera}, {Kasper}, {Moutou}, {Lagrange}, {Antichi},
  {Baruffolo}, {Baudrand}, {Beuzit}, {Boccaletti}, {Cantalloube}, {Carbillet},
  {Charton}, {Claudi}, {Costille}, {Dohlen}, {Dominik}, {Fantinel},
  {Feautrier}, {Feldt}, {Fusco}, {Gigan}, {Girard}, {Gluck}, {Gry}, {Henning},
  {Janson}, {Langlois}, {Madec}, {Magnard}, {Maurel}, {Mawet}, {Meyer},
  {Milli}, {Moeller-Nilsson}, {Mouillet}, {Pavlov}, {Perret}, {Pujet}, {Quanz},
  {Rochat}, {Rousset}, {Roux}, {Salasnich}, {Salter}, {Sauvage}, {Schmid},
  {Sevin}, {Soenke}, {Stadler}, {Turatto}, {Udry}, {Vakili}, {Wahhaj}, \&
  {Wildi}}]{Bonnefoy2016}
{Bonnefoy}, M., {Zurlo}, A., {Baudino}, J.~L., {et~al.} 2016, \aap, 587, A58,
  \dodoi{10.1051/0004-6361/201526906}

\bibitem[{{Bowler} {et~al.}(2010){Bowler}, {Liu}, {Dupuy}, \&
  {Cushing}}]{Bowler2010}
{Bowler}, B.~P., {Liu}, M.~C., {Dupuy}, T.~J., \& {Cushing}, M.~C. 2010, \apj,
  723, 850, \dodoi{10.1088/0004-637X/723/1/850}

\bibitem[{{Bowler} {et~al.}(2020){Bowler}, {Zhou}, {Morley}, {Kataria},
  {Bryan}, {Benneke}, \& {Batygin}}]{Bowler2020}
{Bowler}, B.~P., {Zhou}, Y., {Morley}, C.~V., {et~al.} 2020, \apjl, 893, L30,
  \dodoi{10.3847/2041-8213/ab8197}

\bibitem[{{Brady} {et~al.}(2022){Brady}, {Bean}, {Seifahrt}, {Kasper}, {Luque},
  {Reiners}, {Benneke}, {Stef{\'a}nsson}, \& {St{\"u}rmer}}]{Brady2022}
{Brady}, M., {Bean}, J., {Seifahrt}, A., {et~al.} 2022, arXiv e-prints,
  arXiv:2211.11841.
\newblock \doarXiv{2211.11841}

\bibitem[{{Brandl} {et~al.}(2021){Brandl}, {Bettonvil}, {van Boekel},
  {Glauser}, {Quanz}, {Absil}, {Amorim}, {Feldt}, {Glasse}, {G{\"u}del}, {Ho},
  {Labadie}, {Meyer}, {Pantin}, {van Winckel}, \& {METIS
  Consortium}}]{Brandl2021}
{Brandl}, B., {Bettonvil}, F., {van Boekel}, R., {et~al.} 2021, The Messenger,
  182, 22, \dodoi{10.18727/0722-6691/5218}

\bibitem[{{Bryan} {et~al.}(2018){Bryan}, {Benneke}, {Knutson}, {Batygin}, \&
  {Bowler}}]{Bryan2018}
{Bryan}, M.~L., {Benneke}, B., {Knutson}, H.~A., {Batygin}, K., \& {Bowler},
  B.~P. 2018, Nature Astronomy, 2, 138, \dodoi{10.1038/s41550-017-0325-8}

\bibitem[{{Bryan} {et~al.}(2020){Bryan}, {Ginzburg}, {Chiang}, {Morley},
  {Bowler}, {Xuan}, \& {Knutson}}]{Bryan2020b}
{Bryan}, M.~L., {Ginzburg}, S., {Chiang}, E., {et~al.} 2020, \apj, 905, 37,
  \dodoi{10.3847/1538-4357/abc0ef}

\bibitem[{{Buenzli} {et~al.}(2014){Buenzli}, {Apai}, {Radigan}, {Reid}, \&
  {Flateau}}]{Buenzli2014}
{Buenzli}, E., {Apai}, D., {Radigan}, J., {Reid}, I.~N., \& {Flateau}, D. 2014,
  \apj, 782, 77, \dodoi{10.1088/0004-637X/782/2/77}

\bibitem[{{Buenzli} {et~al.}(2015{\natexlab{a}}){Buenzli}, {Marley}, {Apai},
  {Saumon}, {Biller}, {Crossfield}, \& {Radigan}}]{Buenzli15b}
{Buenzli}, E., {Marley}, M.~S., {Apai}, D., {et~al.} 2015{\natexlab{a}}, \apj,
  812, 163, \dodoi{10.1088/0004-637X/812/2/163}

\bibitem[{{Buenzli} {et~al.}(2015{\natexlab{b}}){Buenzli}, {Saumon}, {Marley},
  {Apai}, {Radigan}, {Bedin}, {Reid}, \& {Morley}}]{Buenzli15a}
{Buenzli}, E., {Saumon}, D., {Marley}, M.~S., {et~al.} 2015{\natexlab{b}},
  \apj, 798, 127, \dodoi{10.1088/0004-637X/798/2/127}

\bibitem[{{Burgasser} \& {Mamajek}(2017)}]{Burgasser&Mamajek2017}
{Burgasser}, A.~J., \& {Mamajek}, E.~E. 2017, \apj, 845, 110,
  \dodoi{10.3847/1538-4357/aa7fea}

\bibitem[{{Chilcote} {et~al.}(2017){Chilcote}, {Pueyo}, {De Rosa}, {Vargas},
  {Macintosh}, {Bailey}, {Barman}, {Bauman}, {Bruzzone}, {Bulger}, {Burrows},
  {Cardwell}, {Chen}, {Cotten}, {Dillon}, {Doyon}, {Draper}, {Duch{\^e}ne},
  {Dunn}, {Erikson}, {Fitzgerald}, {Follette}, {Gavel}, {Goodsell}, {Graham},
  {Greenbaum}, {Hartung}, {Hibon}, {Hung}, {Ingraham}, {Kalas}, {Konopacky},
  {Larkin}, {Maire}, {Marchis}, {Marley}, {Marois}, {Metchev},
  {Millar-Blanchaer}, {Morzinski}, {Nielsen}, {Norton}, {Oppenheimer},
  {Palmer}, {Patience}, {Perrin}, {Poyneer}, {Rajan}, {Rameau},
  {Rantakyr{\"o}}, {Sadakuni}, {Saddlemyer}, {Savransky}, {Schneider}, {Serio},
  {Sivaramakrishnan}, {Song}, {Soummer}, {Thomas}, {Wallace}, {Wang},
  {Ward-Duong}, {Wiktorowicz}, \& {Wolff}}]{Chilcote2017}
{Chilcote}, J., {Pueyo}, L., {De Rosa}, R.~J., {et~al.} 2017, \aj, 153, 182,
  \dodoi{10.3847/1538-3881/aa63e9}

\bibitem[{{Costa} {et~al.}(2006){Costa}, {M{\'e}ndez}, {Jao}, {Henry},
  {Subasavage}, \& {Ianna}}]{Costa2006}
{Costa}, E., {M{\'e}ndez}, R.~A., {Jao}, W.~C., {et~al.} 2006, \aj, 132, 1234,
  \dodoi{10.1086/505706}

\bibitem[{{Croll} {et~al.}(2016){Croll}, {Muirhead}, {Lichtman}, {Han},
  {Dalba}, \& {Radigan}}]{Croll2016}
{Croll}, B., {Muirhead}, P.~S., {Lichtman}, J., {et~al.} 2016, arXiv e-prints,
  arXiv:1609.03587.
\newblock \doarXiv{1609.03587}

\bibitem[{{Crossfield} {et~al.}(2014){Crossfield}, {Biller}, {Schlieder},
  {Deacon}, {Bonnefoy}, {Homeier}, {Allard}, {Buenzli}, {Henning}, {Brandner},
  {Goldman}, \& {Kopytova}}]{crossfield14}
{Crossfield}, I.~J.~M., {Biller}, B., {Schlieder}, J.~E., {et~al.} 2014, \nat,
  505, 654, \dodoi{10.1038/nature12955}

\bibitem[{{Currie} {et~al.}(2013){Currie}, {Burrows}, {Madhusudhan},
  {Fukagawa}, {Girard}, {Dawson}, {Murray-Clay}, {Kenyon}, {Kuchner},
  {Matsumura}, {Jayawardhana}, {Chambers}, \& {Bromley}}]{Currie2013}
{Currie}, T., {Burrows}, A., {Madhusudhan}, N., {et~al.} 2013, \apj, 776, 15,
  \dodoi{10.1088/0004-637X/776/1/15}

\bibitem[{{Currie} {et~al.}(2014){Currie}, {Burrows}, {Girard}, {Cloutier},
  {Fukagawa}, {Sorahana}, {Kuchner}, {Kenyon}, {Madhusudhan}, {Itoh},
  {Jayawardhana}, {Matsumura}, \& {Pyo}}]{Currie2014}
{Currie}, T., {Burrows}, A., {Girard}, J.~H., {et~al.} 2014, \apj, 795, 133,
  \dodoi{10.1088/0004-637X/795/2/133}

\bibitem[{{Cutri}(2014)}]{Cutri2014}
{Cutri}, R.~M. 2014, VizieR Online Data Catalog, 2328

\bibitem[{{Cutri} {et~al.}(2003){Cutri}, {Skrutskie}, {van Dyk}, {Beichman},
  {Carpenter}, {Chester}, {Cambresy}, {Evans}, {Fowler}, {Gizis}, {Howard},
  {Huchra}, {Jarrett}, {Kopan}, {Kirkpatrick}, {Light}, {Marsh}, {McCallon},
  {Schneider}, {Stiening}, {Sykes}, {Weinberg}, {Wheaton}, {Wheelock}, \&
  {Zacarias}}]{Cutri2003}
{Cutri}, R.~M., {Skrutskie}, M.~F., {van Dyk}, S., {et~al.} 2003, VizieR Online
  Data Catalog, II/246

\bibitem[{Donatelli \& Reichel(2014)}]{donatelli&reichel14}
Donatelli, M., \& Reichel, L. 2014, Journal of Computational and Applied
  Mathematics, 272, 334

\bibitem[{{Donati} {et~al.}(1997){Donati}, {Semel}, {Carter}, {Rees}, \&
  {Collier Cameron}}]{Donati1997}
{Donati}, J.~F., {Semel}, M., {Carter}, B.~D., {Rees}, D.~E., \& {Collier
  Cameron}, A. 1997, \mnras, 291, 658, \dodoi{10.1093/mnras/291.4.658}

\bibitem[{{Doyon} {et~al.}(2010){Doyon}, {Lafreni{\`e}re}, {Artigau}, {Malo},
  \& {Marois}}]{Doyon2010}
{Doyon}, R., {Lafreni{\`e}re}, D., {Artigau}, E., {Malo}, L., \& {Marois}, C.
  2010, in In the Spirit of Lyot 2010, ed. A.~{Boccaletti}, E42

\bibitem[{{Dulaimi} {et~al.}(2023){Dulaimi}, {Golden}, {Boyle}, \&
  {Butler}}]{Dulaimi2023}
{Dulaimi}, S., {Golden}, A., {Boyle}, R.~P., \& {Butler}, R.~F. 2023, \mnras,
  518, 4428, \dodoi{10.1093/mnras/stac2894}

\bibitem[{{Dupuy} {et~al.}(2019){Dupuy}, {Brandt}, {Kratter}, \&
  {Bowler}}]{Dupuy2019}
{Dupuy}, T.~J., {Brandt}, T.~D., {Kratter}, K.~M., \& {Bowler}, B.~P. 2019,
  \apjl, 871, L4, \dodoi{10.3847/2041-8213/aafb31}

\bibitem[{{Dupuy} {et~al.}(2023){Dupuy}, {Liu}, {Evans}, {Best}, {Pearce},
  {Sanghi}, {Phillips}, \& {Bardalez Gagliuffi}}]{Dupuy2023}
{Dupuy}, T.~J., {Liu}, M.~C., {Evans}, E.~L., {et~al.} 2023, \mnras, 519, 1688,
  \dodoi{10.1093/mnras/stac3557}

\bibitem[{{Dupuy} {et~al.}(2020){Dupuy}, {Liu}, {Magnier}, {Best}, {Baraffe},
  {Chabrier}, {Forveille}, {Metchev}, \& {Tremblin}}]{Dupuy2020}
{Dupuy}, T.~J., {Liu}, M.~C., {Magnier}, E.~A., {et~al.} 2020, Research Notes
  of the American Astronomical Society, 4, 54, \dodoi{10.3847/2515-5172/ab8942}

\bibitem[{{Eriksson} {et~al.}(2019){Eriksson}, {Janson}, \&
  {Calissendorff}}]{Eriksson2019}
{Eriksson}, S.~C., {Janson}, M., \& {Calissendorff}, P. 2019, \aap, 629, A145,
  \dodoi{10.1051/0004-6361/201935671}

\bibitem[{{Fanson} {et~al.}(2020){Fanson}, {Bernstein}, {Angeli}, {Ashby},
  {Bigelow}, {Brossus}, {Bouchez}, {Burgett}, {Contos}, {Demers}, {Figueroa},
  {Fischer}, {Groark}, {Laskin}, {Millan-Gabet}, {Pi}, \&
  {Wheeler}}]{Fanson2020}
{Fanson}, J., {Bernstein}, R., {Angeli}, G., {et~al.} 2020, in Society of
  Photo-Optical Instrumentation Engineers (SPIE) Conference Series, Vol. 11445,
  Society of Photo-Optical Instrumentation Engineers (SPIE) Conference Series,
  114451F, \dodoi{10.1117/12.2561852}

\bibitem[{{Fitzgerald} {et~al.}(2019){Fitzgerald}, {Bailey}, {Baranec},
  {Batalha}, {Benneke}, {Beichman}, {Brandt}, {Chilcote}, {Chun}, {Crossfield},
  {Currie}, {Davis}, {Dekany}, {Delorme}, {Dong}, {Doyon}, {Dressing},
  {Echeverri}, {Fortney}, {Frazin}, {Guyon}, {Hashimoto}, {Hillenbrand},
  {Hinz}, {Howard}, {Jensen-Clem}, {Jovanovic}, {Kawahara}, {Knutson},
  {Konopacky}, {Kotani}, {Lafreni{\`e}re}, {Liu}, {Lozi}, {Lu}, {Males},
  {Marley}, {Marois}, {Mawet}, {Mazin}, {Millar-Blanchaer}, {Mondal},
  {Murakami}, {Murray-Clay}, {Narita}, {Pezzato}, {Pyo}, {Roberts}, {Ruane},
  {Sallum}, {Serabyn}, {Shields}, {Simard}, {Skemer}, {Stelter}, {Tamura},
  {Troy}, {Vasisht}, {Wallace}, {Wang}, {Wang}, \& {Wright}}]{Fitzgerald2019}
{Fitzgerald}, M., {Bailey}, V., {Baranec}, C., {et~al.} 2019, in Bulletin of
  the American Astronomical Society, Vol.~51, 251

\bibitem[{{Froning} {et~al.}(2006){Froning}, {Osterman}, {Beasley}, {Green}, \&
  {Beland}}]{Froning2006}
{Froning}, C., {Osterman}, S., {Beasley}, M., {Green}, J., \& {Beland}, S.
  2006, in Society of Photo-Optical Instrumentation Engineers (SPIE) Conference
  Series, Vol. 6269, Society of Photo-Optical Instrumentation Engineers (SPIE)
  Conference Series, ed. I.~S. {McLean} \& M.~{Iye}, 62691V,
  \dodoi{10.1117/12.669358}

\bibitem[{{Gagn{\'e}} {et~al.}(2017){Gagn{\'e}}, {Faherty}, {Burgasser},
  {Artigau}, {Bouchard}, {Albert}, {Lafreni{\`e}re}, {Doyon}, \& {Bardalez
  Gagliuffi}}]{Gagne2017}
{Gagn{\'e}}, J., {Faherty}, J.~K., {Burgasser}, A.~J., {et~al.} 2017, \apjl,
  841, L1, \dodoi{10.3847/2041-8213/aa70e2}

\bibitem[{{Gaia Collaboration} {et~al.}(2021){Gaia Collaboration}, {Brown},
  {Vallenari}, {Prusti}, {de Bruijne}, {Babusiaux}, {Biermann}, {Creevey},
  {Evans}, {Eyer}, {Hutton}, {Jansen}, {Jordi}, {Klioner}, {Lammers},
  {Lindegren}, {Luri}, {Mignard}, {Panem}, {Pourbaix}, {Randich}, {Sartoretti},
  {Soubiran}, {Walton}, {Arenou}, {Bailer-Jones}, {Bastian}, {Cropper},
  {Drimmel}, {Katz}, {Lattanzi}, {van Leeuwen}, {Bakker}, {Cacciari},
  {Casta{\~n}eda}, {De Angeli}, {Ducourant}, {Fabricius}, {Fouesneau},
  {Fr{\'e}mat}, {Guerra}, {Guerrier}, {Guiraud}, {Jean-Antoine Piccolo},
  {Masana}, {Messineo}, {Mowlavi}, {Nicolas}, {Nienartowicz}, {Pailler},
  {Panuzzo}, {Riclet}, {Roux}, {Seabroke}, {Sordo}, {Tanga}, {Th{\'e}venin},
  {Gracia-Abril}, {Portell}, {Teyssier}, {Altmann}, {Andrae}, {Bellas-Velidis},
  {Benson}, {Berthier}, {Blomme}, {Brugaletta}, {Burgess}, {Busso}, {Carry},
  {Cellino}, {Cheek}, {Clementini}, {Damerdji}, {Davidson}, {Delchambre},
  {Dell'Oro}, {Fern{\'a}ndez-Hern{\'a}ndez}, {Galluccio}, {Garc{\'\i}a-Lario},
  {Garcia-Reinaldos}, {Gonz{\'a}lez-N{\'u}{\~n}ez}, {Gosset}, {Haigron},
  {Halbwachs}, {Hambly}, {Harrison}, {Hatzidimitriou}, {Heiter},
  {Hern{\'a}ndez}, {Hestroffer}, {Hodgkin}, {Holl}, {Jan{\ss}en}, {Jevardat de
  Fombelle}, {Jordan}, {Krone-Martins}, {Lanzafame}, {L{\"o}ffler}, {Lorca},
  {Manteiga}, {Marchal}, {Marrese}, {Moitinho}, {Mora}, {Muinonen}, {Osborne},
  {Pancino}, {Pauwels}, {Petit}, {Recio-Blanco}, {Richards}, {Riello},
  {Rimoldini}, {Robin}, {Roegiers}, {Rybizki}, {Sarro}, {Siopis}, {Smith},
  {Sozzetti}, {Ulla}, {Utrilla}, {van Leeuwen}, {van Reeven}, {Abbas}, {Abreu
  Aramburu}, {Accart}, {Aerts}, {Aguado}, {Ajaj}, {Altavilla}, {{\'A}lvarez},
  {{\'A}lvarez Cid-Fuentes}, {Alves}, {Anderson}, {Anglada Varela}, {Antoja},
  {Audard}, {Baines}, {Baker}, {Balaguer-N{\'u}{\~n}ez}, {Balbinot}, {Balog},
  {Barache}, {Barbato}, {Barros}, {Barstow}, {Bartolom{\'e}}, {Bassilana},
  {Bauchet}, {Baudesson-Stella}, {Becciani}, {Bellazzini}, {Bernet}, {Bertone},
  {Bianchi}, {Blanco-Cuaresma}, {Boch}, {Bombrun}, {Bossini}, {Bouquillon},
  {Bragaglia}, {Bramante}, {Breedt}, {Bressan}, {Brouillet}, {Bucciarelli},
  {Burlacu}, {Busonero}, {Butkevich}, {Buzzi}, {Caffau}, {Cancelliere},
  {C{\'a}novas}, {Cantat-Gaudin}, {Carballo}, {Carlucci}, {Carnerero},
  {Carrasco}, {Casamiquela}, {Castellani}, {Castro-Ginard}, {Castro Sampol},
  {Chaoul}, {Charlot}, {Chemin}, {Chiavassa}, {Cioni}, {Comoretto}, {Cooper},
  {Cornez}, {Cowell}, {Crifo}, {Crosta}, {Crowley}, {Dafonte}, {Dapergolas},
  {David}, {David}, {de Laverny}, {De Luise}, {De March}, {De Ridder}, {de
  Souza}, {de Teodoro}, {de Torres}, {del Peloso}, {del Pozo}, {Delbo},
  {Delgado}, {Delgado}, {Delisle}, {Di Matteo}, {Diakite}, {Diener},
  {Distefano}, {Dolding}, {Eappachen}, {Edvardsson}, {Enke}, {Esquej}, {Fabre},
  {Fabrizio}, {Faigler}, {Fedorets}, {Fernique}, {Fienga}, {Figueras},
  {Fouron}, {Fragkoudi}, {Fraile}, {Franke}, {Gai}, {Garabato},
  {Garcia-Gutierrez}, {Garc{\'\i}a-Torres}, {Garofalo}, {Gavras}, {Gerlach},
  {Geyer}, {Giacobbe}, {Gilmore}, {Girona}, {Giuffrida}, {Gomel}, {Gomez},
  {Gonzalez-Santamaria}, {Gonz{\'a}lez-Vidal}, {Granvik},
  {Guti{\'e}rrez-S{\'a}nchez}, {Guy}, {Hauser}, {Haywood}, {Helmi}, {Hidalgo},
  {Hilger}, {H{\l}adczuk}, {Hobbs}, {Holland}, {Huckle}, {Jasniewicz},
  {Jonker}, {Juaristi Campillo}, {Julbe}, {Karbevska}, {Kervella}, {Khanna},
  {Kochoska}, {Kontizas}, {Kordopatis}, {Korn}, {Kostrzewa-Rutkowska},
  {Kruszy{\'n}ska}, {Lambert}, {Lanza}, {Lasne}, {Le Campion}, {Le Fustec},
  {Lebreton}, {Lebzelter}, {Leccia}, {Leclerc}, {Lecoeur-Taibi}, {Liao},
  {Licata}, {Lindstr{\o}m}, {Lister}, {Livanou}, {Lobel}, {Madrero Pardo},
  {Managau}, {Mann}, {Marchant}, {Marconi}, {Marcos Santos}, {Marinoni},
  {Marocco}, {Marshall}, {Martin Polo}, {Mart{\'\i}n-Fleitas}, {Masip},
  {Massari}, {Mastrobuono-Battisti}, {Mazeh}, {McMillan}, {Messina},
  {Michalik}, {Millar}, {Mints}, {Molina}, {Molinaro}, {Moln{\'a}r},
  {Montegriffo}, {Mor}, {Morbidelli}, {Morel}, {Morris}, {Mulone}, {Munoz},
  {Muraveva}, {Murphy}, {Musella}, {Noval}, {Ord{\'e}novic}, {Orr{\`u}},
  {Osinde}, {Pagani}, {Pagano}, {Palaversa}, {Palicio}, {Panahi}, {Pawlak},
  {Pe{\~n}alosa Esteller}, {Penttil{\"a}}, {Piersimoni}, {Pineau}, {Plachy},
  {Plum}, {Poggio}, {Poretti}, {Poujoulet}, {Pr{\v{s}}a}, {Pulone}, {Racero},
  {Ragaini}, {Rainer}, {Raiteri}, {Rambaux}, {Ramos}, {Ramos-Lerate}, {Re
  Fiorentin}, {Regibo}, {Reyl{\'e}}, {Ripepi}, {Riva}, {Rixon}, {Robichon},
  {Robin}, {Roelens}, {Rohrbasser}, {Romero-G{\'o}mez}, {Rowell}, {Royer},
  {Rybicki}, {Sadowski}, {Sagrist{\`a} Sell{\'e}s}, {Sahlmann}, {Salgado},
  {Salguero}, {Samaras}, {Sanchez Gimenez}, {Sanna}, {Santove{\~n}a},
  {Sarasso}, {Schultheis}, {Sciacca}, {Segol}, {Segovia}, {S{\'e}gransan},
  {Semeux}, {Shahaf}, {Siddiqui}, {Siebert}, {Siltala}, {Slezak}, {Smart},
  {Solano}, {Solitro}, {Souami}, {Souchay}, {Spagna}, {Spoto}, {Steele},
  {Steidelm{\"u}ller}, {Stephenson}, {S{\"u}veges}, {Szabados}, {Szegedi-Elek},
  {Taris}, {Tauran}, {Taylor}, {Teixeira}, {Thuillot}, {Tonello}, {Torra},
  {Torra}, {Turon}, {Unger}, {Vaillant}, {van Dillen}, {Vanel}, {Vecchiato},
  {Viala}, {Vicente}, {Voutsinas}, {Weiler}, {Wevers}, {Wyrzykowski}, {Yoldas},
  {Yvard}, {Zhao}, {Zorec}, {Zucker}, {Zurbach}, \& {Zwitter}}]{Gaia2021}
{Gaia Collaboration}, {Brown}, A.~G.~A., {Vallenari}, A., {et~al.} 2021, \aap,
  650, C3, \dodoi{10.1051/0004-6361/202039657e}

\bibitem[{{Gauza} {et~al.}(2015){Gauza}, {B{\'e}jar}, {P{\'e}rez-Garrido},
  {Zapatero Osorio}, {Lodieu}, {Rebolo}, {Pall{\'e}}, \& {Nowak}}]{Gauza2015}
{Gauza}, B., {B{\'e}jar}, V. J.~S., {P{\'e}rez-Garrido}, A., {et~al.} 2015,
  \apj, 804, 96, \dodoi{10.1088/0004-637X/804/2/96}

\bibitem[{{Gillon} {et~al.}(2016){Gillon}, {Jehin}, {Lederer}, {Delrez}, {de
  Wit}, {Burdanov}, {Van Grootel}, {Burgasser}, {Triaud}, {Opitom}, {Demory},
  {Sahu}, {Bardalez Gagliuffi}, {Magain}, \& {Queloz}}]{Gillon2016}
{Gillon}, M., {Jehin}, E., {Lederer}, S.~M., {et~al.} 2016, \nat, 533, 221,
  \dodoi{10.1038/nature17448}

\bibitem[{{Gillon} {et~al.}(2017){Gillon}, {Triaud}, {Demory}, {Jehin}, {Agol},
  {Deck}, {Lederer}, {de Wit}, {Burdanov}, {Ingalls}, {Bolmont}, {Leconte},
  {Raymond}, {Selsis}, {Turbet}, {Barkaoui}, {Burgasser}, {Burleigh}, {Carey},
  {Chaushev}, {Copperwheat}, {Delrez}, {Fernandes}, {Holdsworth}, {Kotze}, {Van
  Grootel}, {Almleaky}, {Benkhaldoun}, {Magain}, \& {Queloz}}]{Gillon2017}
{Gillon}, M., {Triaud}, A. H.~M.~J., {Demory}, B.-O., {et~al.} 2017, \nat, 542,
  456, \dodoi{10.1038/nature21360}

\bibitem[{{Gizis} {et~al.}(2015){Gizis}, {Dettman}, {Burgasser}, {Camnasio},
  {Alam}, {Filippazzo}, {Cruz}, {Metchev}, {Berger}, \& {Williams}}]{Gizis2015}
{Gizis}, J.~E., {Dettman}, K.~G., {Burgasser}, A.~J., {et~al.} 2015, \apj, 813,
  104, \dodoi{10.1088/0004-637X/813/2/104}

\bibitem[{{Gonzales} {et~al.}(2019){Gonzales}, {Faherty}, {Gagn{\'e}}, {Teske},
  {McWilliam}, \& {Cruz}}]{Gonzales2019}
{Gonzales}, E.~C., {Faherty}, J.~K., {Gagn{\'e}}, J., {et~al.} 2019, \apj, 886,
  131, \dodoi{10.3847/1538-4357/ab48fc}

\bibitem[{{Go{\'z}dziewski} \& {Migaszewski}(2020)}]{Gozdziewski2020}
{Go{\'z}dziewski}, K., \& {Migaszewski}, C. 2020, \apjl, 902, L40,
  \dodoi{10.3847/2041-8213/abb881}

\bibitem[{{GRAVITY Collaboration} {et~al.}(2019){GRAVITY Collaboration},
  {Lacour}, {Nowak}, {Wang}, {Pfuhl}, {Eisenhauer}, {Abuter}, {Amorim},
  {Anugu}, {Benisty}, {Berger}, {Beust}, {Blind}, {Bonnefoy}, {Bonnet},
  {Bourget}, {Brandner}, {Buron}, {Collin}, {Charnay}, {Chapron}, {Cl{\'e}net},
  {Coud{\'e} Du Foresto}, {de Zeeuw}, {Deen}, {Dembet}, {Dexter}, {Duvert},
  {Eckart}, {F{\"o}rster Schreiber}, {F{\'e}dou}, {Garcia}, {Garcia Lopez},
  {Gao}, {Gendron}, {Genzel}, {Gillessen}, {Gordo}, {Greenbaum}, {Habibi},
  {Haubois}, {Hau{\ss}mann}, {Henning}, {Hippler}, {Horrobin}, {Hubert},
  {Jimenez Rosales}, {Jocou}, {Kendrew}, {Kervella}, {Kolb}, {Lagrange},
  {Lapeyr{\`e}re}, {Le Bouquin}, {L{\'e}na}, {Lippa}, {Lenzen}, {Maire},
  {Molli{\`e}re}, {Ott}, {Paumard}, {Perraut}, {Perrin}, {Pueyo}, {Rabien},
  {Ram{\'\i}rez}, {Rau}, {Rodr{\'\i}guez-Coira}, {Rousset}, {Sanchez-Bermudez},
  {Scheithauer}, {Schuhler}, {Straub}, {Straubmeier}, {Sturm}, {Tacconi},
  {Vincent}, {van Dishoeck}, {von Fellenberg}, {Wank}, {Waisberg}, {Widmann},
  {Wieprecht}, {Wiest}, {Wiezorrek}, {Woillez}, {Yazici}, {Ziegler}, \&
  {Zins}}]{GravityCollab2019}
{GRAVITY Collaboration}, {Lacour}, S., {Nowak}, M., {et~al.} 2019, \aap, 623,
  L11, \dodoi{10.1051/0004-6361/201935253}

\bibitem[{{GRAVITY Collaboration} {et~al.}(2020){GRAVITY Collaboration},
  {Nowak}, {Lacour}, {Molli{\`e}re}, {Wang}, {Charnay}, {van Dishoeck},
  {Abuter}, {Amorim}, {Berger}, {Beust}, {Bonnefoy}, {Bonnet}, {Brandner},
  {Buron}, {Cantalloube}, {Collin}, {Chapron}, {Cl{\'e}net}, {Coud{\'e} Du
  Foresto}, {de Zeeuw}, {Dembet}, {Dexter}, {Duvert}, {Eckart}, {Eisenhauer},
  {F{\"o}rster Schreiber}, {F{\'e}dou}, {Garcia Lopez}, {Gao}, {Gendron},
  {Genzel}, {Gillessen}, {Hau{\ss}mann}, {Henning}, {Hippler}, {Hubert},
  {Jocou}, {Kervella}, {Lagrange}, {Lapeyr{\`e}re}, {Le Bouquin}, {L{\'e}na},
  {Maire}, {Ott}, {Paumard}, {Paladini}, {Perraut}, {Perrin}, {Pueyo}, {Pfuhl},
  {Rabien}, {Rau}, {Rodr{\'\i}guez-Coira}, {Rousset}, {Scheithauer},
  {Shangguan}, {Straub}, {Straubmeier}, {Sturm}, {Tacconi}, {Vincent},
  {Widmann}, {Wieprecht}, {Wiezorrek}, {Woillez}, {Yazici}, \&
  {Ziegler}}]{GravityCollaboration2020}
{GRAVITY Collaboration}, {Nowak}, M., {Lacour}, S., {et~al.} 2020, \aap, 633,
  A110, \dodoi{10.1051/0004-6361/201936898}

\bibitem[{{Gray} {et~al.}(2006){Gray}, {Corbally}, {Garrison}, {McFadden},
  {Bubar}, {McGahee}, {O'Donoghue}, \& {Knox}}]{Gray2006}
{Gray}, R.~O., {Corbally}, C.~J., {Garrison}, R.~F., {et~al.} 2006, \aj, 132,
  161, \dodoi{10.1086/504637}

\bibitem[{{Greenbaum} {et~al.}(2018){Greenbaum}, {Pueyo}, {Ruffio}, {Wang}, {De
  Rosa}, {Aguilar}, {Rameau}, {Barman}, {Marois}, {Marley}, {Konopacky},
  {Rajan}, {Macintosh}, {Ansdell}, {Arriaga}, {Bailey}, {Bulger}, {Burrows},
  {Chilcote}, {Cotten}, {Doyon}, {Duch{\^e}ne}, {Fitzgerald}, {Follette},
  {Gerard}, {Goodsell}, {Graham}, {Hibon}, {Hung}, {Ingraham}, {Kalas},
  {Larkin}, {Maire}, {Marchis}, {Metchev}, {Millar-Blanchaer}, {Nielsen},
  {Norton}, {Oppenheimer}, {Palmer}, {Patience}, {Perrin}, {Poyneer},
  {Rantakyr{\"o}}, {Savransky}, {Schneider}, {Sivaramakrishnan}, {Song},
  {Soummer}, {Thomas}, {Wallace}, {Ward-Duong}, {Wiktorowicz}, \&
  {Wolff}}]{Greenbaum2018}
{Greenbaum}, A.~Z., {Pueyo}, L., {Ruffio}, J.-B., {et~al.} 2018, \aj, 155, 226,
  \dodoi{10.3847/1538-3881/aabcb8}

\bibitem[{{G{\"u}nther} {et~al.}(2020){G{\"u}nther}, {Zhan}, {Seager},
  {Rimmer}, {Ranjan}, {Stassun}, {Oelkers}, {Daylan}, {Newton}, {Kristiansen},
  {Olah}, {Gillen}, {Rappaport}, {Ricker}, {Vanderspek}, {Latham}, {Winn},
  {Jenkins}, {Glidden}, {Fausnaugh}, {Levine}, {Dittmann}, {Quinn},
  {Krishnamurthy}, \& {Ting}}]{gunther20}
{G{\"u}nther}, M.~N., {Zhan}, Z., {Seager}, S., {et~al.} 2020, \aj, 159, 60,
  \dodoi{10.3847/1538-3881/ab5d3a}

\bibitem[{{Hauschildt}(1992)}]{Hauschildt1992}
{Hauschildt}, P.~H. 1992, \jqsrt, 47, 433, \dodoi{10.1016/0022-4073(92)90105-D}

\bibitem[{{Hauschildt}(1993)}]{Hauschildt1993}
---. 1993, \jqsrt, 50, 301, \dodoi{10.1016/0022-4073(93)90080-2}

\bibitem[{{Hauschildt} \& {Baron}(1995)}]{Hauschildt&Baron1995}
{Hauschildt}, P.~H., \& {Baron}, E. 1995, \jqsrt, 54, 987,
  \dodoi{10.1016/0022-4073(95)00118-5}

\bibitem[{{Hauschildt} {et~al.}(1997){Hauschildt}, {Baron}, \&
  {Allard}}]{Hauschildt1997}
{Hauschildt}, P.~H., {Baron}, E., \& {Allard}, F. 1997, \apj, 483, 390,
  \dodoi{10.1086/304233}

\bibitem[{{Hauschildt} {et~al.}(1996){Hauschildt}, {Baron}, {Starrfield}, \&
  {Allard}}]{Hauschildt1996}
{Hauschildt}, P.~H., {Baron}, E., {Starrfield}, S., \& {Allard}, F. 1996, \apj,
  462, 386, \dodoi{10.1086/177160}

\bibitem[{{Hauschildt} {et~al.}(1995){Hauschildt}, {Starrfield}, {Shore},
  {Allard}, \& {Baron}}]{Hauschildt1995}
{Hauschildt}, P.~H., {Starrfield}, S., {Shore}, S.~N., {Allard}, F., \&
  {Baron}, E. 1995, \apj, 447, 829, \dodoi{10.1086/175921}

\bibitem[{{H{\'e}brard} {et~al.}(2016){H{\'e}brard}, {Donati}, {Delfosse},
  {Morin}, {Moutou}, \& {Boisse}}]{Hebrard2016}
{H{\'e}brard}, {\'E}.~M., {Donati}, J.~F., {Delfosse}, X., {et~al.} 2016,
  \mnras, 461, 1465, \dodoi{10.1093/mnras/stw1346}

\bibitem[{{Heinze} {et~al.}(2013){Heinze}, {Metchev}, {Apai}, {Flateau},
  {Kurtev}, {Marley}, {Radigan}, {Burgasser}, {Artigau}, \&
  {Plavchan}}]{Heinze2013}
{Heinze}, A.~N., {Metchev}, S., {Apai}, D., {et~al.} 2013, \apj, 767, 173,
  \dodoi{10.1088/0004-637X/767/2/173}

\bibitem[{{Helling} {et~al.}(2008){Helling}, {Ackerman}, {Allard}, {Dehn},
  {Hauschildt}, {Homeier}, {Lodders}, {Marley}, {Rietmeijer}, {Tsuji}, \&
  {Woitke}}]{Helling2008}
{Helling}, C., {Ackerman}, A., {Allard}, F., {et~al.} 2008, \mnras, 391, 1854,
  \dodoi{10.1111/j.1365-2966.2008.13991.x}

\bibitem[{{Higson} {et~al.}(2019){Higson}, {Handley}, {Hobson}, \&
  {Lasenby}}]{higson19}
{Higson}, E., {Handley}, W., {Hobson}, M., \& {Lasenby}, A. 2019, Statistics
  and Computing, 29, 891, \dodoi{10.1007/s11222-018-9844-0}

\bibitem[{{Hinz} {et~al.}(2010){Hinz}, {Rodigas}, {Kenworthy}, {Sivanandam},
  {Heinze}, {Mamajek}, \& {Meyer}}]{Hinz2010}
{Hinz}, P.~M., {Rodigas}, T.~J., {Kenworthy}, M.~A., {et~al.} 2010, \apj, 716,
  417, \dodoi{10.1088/0004-637X/716/1/417}

\bibitem[{{Hoch} {et~al.}(2022){Hoch}, {Konopacky}, {Barman}, {Theissen},
  {Brock}, {Perrin}, {Ruffio}, {Macintosh}, \& {Marois}}]{Hoch2022}
{Hoch}, K. K.~W., {Konopacky}, Q.~M., {Barman}, T.~S., {et~al.} 2022, \aj, 164,
  155, \dodoi{10.3847/1538-3881/ac84d4}

\bibitem[{Hunter(2007)}]{Matplotlib}
Hunter, J.~D. 2007, Computing in Science \& Engineering, 9, 90,
  \dodoi{10.1109/MCSE.2007.55}

\bibitem[{{Ilin} {et~al.}(2021){Ilin}, {Poppenhaeger}, {Schmidt},
  {J{\"a}rvinen}, {Newton}, {Alvarado-G{\'o}mez}, {Pineda}, {Davenport},
  {Oshagh}, \& {Ilyin}}]{Ilin2021}
{Ilin}, E., {Poppenhaeger}, K., {Schmidt}, S.~J., {et~al.} 2021, \mnras, 507,
  1723, \dodoi{10.1093/mnras/stab2159}

\bibitem[{{Jaffe} {et~al.}(2016){Jaffe}, {Barnes}, {Brooks}, {Lee}, {Mace},
  {Pak}, {Park}, \& {Park}}]{Jaffe2016}
{Jaffe}, D.~T., {Barnes}, S., {Brooks}, C., {et~al.} 2016, in Society of
  Photo-Optical Instrumentation Engineers (SPIE) Conference Series, Vol. 9908,
  Ground-based and Airborne Instrumentation for Astronomy VI, ed. C.~J.
  {Evans}, L.~{Simard}, \& H.~{Takami}, 990821, \dodoi{10.1117/12.2232994}

\bibitem[{{Janson} {et~al.}(2010){Janson}, {Bergfors}, {Goto}, {Brandner}, \&
  {Lafreni{\`e}re}}]{Janson2010}
{Janson}, M., {Bergfors}, C., {Goto}, M., {Brandner}, W., \& {Lafreni{\`e}re},
  D. 2010, \apjl, 710, L35, \dodoi{10.1088/2041-8205/710/1/L35}

\bibitem[{{Johns}(2006)}]{Johns2006}
{Johns}, M. 2006, in Society of Photo-Optical Instrumentation Engineers (SPIE)
  Conference Series, Vol. 6267, Society of Photo-Optical Instrumentation
  Engineers (SPIE) Conference Series, ed. L.~M. {Stepp}, 626729,
  \dodoi{10.1117/12.670839}

\bibitem[{{Jones} {et~al.}(2013){Jones}, {Noll}, {Kausch}, {Szyszka}, \&
  {Kimeswenger}}]{Jones2013}
{Jones}, A., {Noll}, S., {Kausch}, W., {Szyszka}, C., \& {Kimeswenger}, S.
  2013, \aap, 560, A91, \dodoi{10.1051/0004-6361/201322433}

\bibitem[{{Jones} \& {Tsuji}(1997)}]{Jones&Tsuji1997}
{Jones}, H. R.~A., \& {Tsuji}, T. 1997, \apjl, 480, L39, \dodoi{10.1086/310619}

\bibitem[{{Karalidi} {et~al.}(2016){Karalidi}, {Apai}, {Marley}, \&
  {Buenzli}}]{karalidi16}
{Karalidi}, T., {Apai}, D., {Marley}, M.~S., \& {Buenzli}, E. 2016, \apj, 825,
  90, \dodoi{10.3847/0004-637X/825/2/90}

\bibitem[{{Kasting} {et~al.}(1993){Kasting}, {Whitmire}, \&
  {Reynolds}}]{kasting93}
{Kasting}, J.~F., {Whitmire}, D.~P., \& {Reynolds}, R.~T. 1993, \icarus, 101,
  108, \dodoi{10.1006/icar.1993.1010}

\bibitem[{{Kervella} {et~al.}(2019){Kervella}, {Arenou}, {Mignard}, \&
  {Th{\'e}venin}}]{Kervella2019}
{Kervella}, P., {Arenou}, F., {Mignard}, F., \& {Th{\'e}venin}, F. 2019, \aap,
  623, A72, \dodoi{10.1051/0004-6361/201834371}

\bibitem[{{Khokhlova}(1985)}]{Khokhlova1985}
{Khokhlova}, V.~L. 1985, \apspr, 4, 99

\bibitem[{{Khokhlova} {et~al.}(1986){Khokhlova}, {Rice}, \&
  {Wehlau}}]{Khokhlova1986}
{Khokhlova}, V.~L., {Rice}, J.~B., \& {Wehlau}, W.~H. 1986, \apj, 307, 768,
  \dodoi{10.1086/164462}

\bibitem[{{Kochukhov}(2016)}]{Kochukhov2016}
{Kochukhov}, O. 2016, in Lecture Notes in Physics, Berlin Springer Verlag, ed.
  J.-P. {Rozelot} \& C.~{Neiner}, Vol. 914, 177,
  \dodoi{10.1007/978-3-319-24151-7_9}

\bibitem[{{Kochukhov} {et~al.}(2010){Kochukhov}, {Makaganiuk}, \&
  {Piskunov}}]{Kochukhov2010}
{Kochukhov}, O., {Makaganiuk}, V., \& {Piskunov}, N. 2010, \aap, 524, A5,
  \dodoi{10.1051/0004-6361/201015429}

\bibitem[{{Kopparapu} {et~al.}(2013){Kopparapu}, {Ramirez}, {Kasting}, {Eymet},
  {Robinson}, {Mahadevan}, {Terrien}, {Domagal-Goldman}, {Meadows}, \&
  {Deshpande}}]{kopparapu13}
{Kopparapu}, R.~K., {Ramirez}, R., {Kasting}, J.~F., {et~al.} 2013, \apj, 765,
  131, \dodoi{10.1088/0004-637X/765/2/131}

\bibitem[{{Lagrange} {et~al.}(2012){Lagrange}, {De Bondt}, {Meunier},
  {Sterzik}, {Beust}, \& {Galland}}]{Lagrange2012}
{Lagrange}, A.~M., {De Bondt}, K., {Meunier}, N., {et~al.} 2012, \aap, 542,
  A18, \dodoi{10.1051/0004-6361/201117985}

\bibitem[{{Lagrange} {et~al.}(2009){Lagrange}, {Gratadour}, {Chauvin}, {Fusco},
  {Ehrenreich}, {Mouillet}, {Rousset}, {Rouan}, {Allard}, {Gendron}, {Charton},
  {Mugnier}, {Rabou}, {Montri}, \& {Lacombe}}]{Lagrange2009}
{Lagrange}, A.~M., {Gratadour}, D., {Chauvin}, G., {et~al.} 2009, \aap, 493,
  L21, \dodoi{10.1051/0004-6361:200811325}

\bibitem[{{Lagrange} {et~al.}(2010){Lagrange}, {Bonnefoy}, {Chauvin}, {Apai},
  {Ehrenreich}, {Boccaletti}, {Gratadour}, {Rouan}, {Mouillet}, {Lacour}, \&
  {Kasper}}]{Lagrange2010}
{Lagrange}, A.~M., {Bonnefoy}, M., {Chauvin}, G., {et~al.} 2010, Science, 329,
  57, \dodoi{10.1126/science.1187187}

\bibitem[{{Lagrange} {et~al.}(2020){Lagrange}, {Rubini}, {Nowak}, {Lacour},
  {Grandjean}, {Boccaletti}, {Langlois}, {Delorme}, {Gratton}, {Wang},
  {Flasseur}, {Galicher}, {Kral}, {Meunier}, {Beust}, {Babusiaux}, {Le
  Coroller}, {Thebault}, {Kervella}, {Zurlo}, {Maire}, {Wahhaj}, {Amorim},
  {Asensio-Torres}, {Benisty}, {Berger}, {Bonnefoy}, {Brandner}, {Cantalloube},
  {Charnay}, {Chauvin}, {Choquet}, {Cl{\'e}net}, {Christiaens}, {Coud{\'e} Du
  Foresto}, {de Zeeuw}, {Desidera}, {Duvert}, {Eckart}, {Eisenhauer},
  {Galland}, {Gao}, {Garcia}, {Garcia Lopez}, {Gendron}, {Genzel}, {Gillessen},
  {Girard}, {Hagelberg}, {Haubois}, {Henning}, {Heissel}, {Hippler},
  {Horrobin}, {Janson}, {Kammerer}, {Kenworthy}, {Keppler}, {Kreidberg},
  {Lapeyr{\`e}re}, {Le Bouquin}, {L{\'e}na}, {M{\'e}rand}, {Messina},
  {Molli{\`e}re}, {Monnier}, {Ott}, {Otten}, {Paumard}, {Paladini}, {Perraut},
  {Perrin}, {Pueyo}, {Pfuhl}, {Rodet}, {Rodriguez-Coira}, {Rousset}, {Samland},
  {Shangguan}, {Schmidt}, {Straub}, {Straubmeier}, {Stolker}, {Vigan},
  {Vincent}, {Widmann}, {Woillez}, \& {GRAVITY Collaboration}}]{Lagrange2020}
{Lagrange}, A.~M., {Rubini}, P., {Nowak}, M., {et~al.} 2020, \aap, 642, A18,
  \dodoi{10.1051/0004-6361/202038823}

\bibitem[{{Lane} {et~al.}(2007){Lane}, {Hallinan}, {Zavala}, {Butler}, {Boyle},
  {Bourke}, {Antonova}, {Doyle}, {Vrba}, \& {Golden}}]{Lane2007}
{Lane}, C., {Hallinan}, G., {Zavala}, R.~T., {et~al.} 2007, \apjl, 668, L163,
  \dodoi{10.1086/523041}

\bibitem[{{Lavie} {et~al.}(2017){Lavie}, {Mendon{\c{c}}a}, {Mordasini},
  {Malik}, {Bonnefoy}, {Demory}, {Oreshenko}, {Grimm}, {Ehrenreich}, \&
  {Heng}}]{Lavie2017}
{Lavie}, B., {Mendon{\c{c}}a}, J.~M., {Mordasini}, C., {et~al.} 2017, \aj, 154,
  91, \dodoi{10.3847/1538-3881/aa7ed8}

\bibitem[{{Lawrence} {et~al.}(2012){Lawrence}, {Warren}, {Almaini}, {Edge},
  {Hambly}, {Jameson}, {Lucas}, {Casali}, {Adamson}, {Dye}, {Emerson},
  {Foucaud}, {Hewett}, {Hirst}, {Hodgkin}, {Irwin}, {Lodieu}, {McMahon},
  {Simpson}, {Smail}, {Mortlock}, \& {Folger}}]{Lawrence2012}
{Lawrence}, A., {Warren}, S.~J., {Almaini}, O., {et~al.} 2012, VizieR Online
  Data Catalog, II/314

\bibitem[{{Liu} {et~al.}(2016){Liu}, {Dupuy}, \& {Allers}}]{Liu2016}
{Liu}, M.~C., {Dupuy}, T.~J., \& {Allers}, K.~N. 2016, \apj, 833, 96,
  \dodoi{10.3847/1538-4357/833/1/96}

\bibitem[{{Lord}(1992)}]{Lord1992}
{Lord}, S.~D. 1992, {A new software tool for computing Earth's atmospheric
  transmission of near- and far-infrared radiation}, NASA Technical Memorandum
  103957

\bibitem[{{Luger} {et~al.}(2021){Luger}, {Bedell}, {Foreman-Mackey},
  {Crossfield}, {Zhao}, \& {Hogg}}]{luger21a}
{Luger}, R., {Bedell}, M., {Foreman-Mackey}, D., {et~al.} 2021, arXiv e-prints,
  arXiv:2110.06271.
\newblock \doarXiv{2110.06271}

\bibitem[{{Luger} {et~al.}(2017){Luger}, {Sestovic}, {Kruse}, {Grimm},
  {Demory}, {Agol}, {Bolmont}, {Fabrycky}, {Fernandes}, {Van Grootel},
  {Burgasser}, {Gillon}, {Ingalls}, {Jehin}, {Raymond}, {Selsis}, {Triaud},
  {Barclay}, {Barentsen}, {Howell}, {Delrez}, {de Wit}, {Foreman-Mackey},
  {Holdsworth}, {Leconte}, {Lederer}, {Turbet}, {Almleaky}, {Benkhaldoun},
  {Magain}, {Morris}, {Heng}, \& {Queloz}}]{Luger2017}
{Luger}, R., {Sestovic}, M., {Kruse}, E., {et~al.} 2017, Nature Astronomy, 1,
  0129, \dodoi{10.1038/s41550-017-0129}

\bibitem[{{Macintosh} {et~al.}(2015){Macintosh}, {Graham}, {Barman}, {De Rosa},
  {Konopacky}, {Marley}, {Marois}, {Nielsen}, {Pueyo}, {Rajan}, {Rameau},
  {Saumon}, {Wang}, {Patience}, {Ammons}, {Arriaga}, {Artigau}, {Beckwith},
  {Brewster}, {Bruzzone}, {Bulger}, {Burningham}, {Burrows}, {Chen}, {Chiang},
  {Chilcote}, {Dawson}, {Dong}, {Doyon}, {Draper}, {Duch{\^e}ne}, {Esposito},
  {Fabrycky}, {Fitzgerald}, {Follette}, {Fortney}, {Gerard}, {Goodsell},
  {Greenbaum}, {Hibon}, {Hinkley}, {Cotten}, {Hung}, {Ingraham},
  {Johnson-Groh}, {Kalas}, {Lafreniere}, {Larkin}, {Lee}, {Line}, {Long},
  {Maire}, {Marchis}, {Matthews}, {Max}, {Metchev}, {Millar-Blanchaer},
  {Mittal}, {Morley}, {Morzinski}, {Murray-Clay}, {Oppenheimer}, {Palmer},
  {Patel}, {Perrin}, {Poyneer}, {Rafikov}, {Rantakyr{\"o}}, {Rice}, {Rojo},
  {Rudy}, {Ruffio}, {Ruiz}, {Sadakuni}, {Saddlemyer}, {Salama}, {Savransky},
  {Schneider}, {Sivaramakrishnan}, {Song}, {Soummer}, {Thomas}, {Vasisht},
  {Wallace}, {Ward-Duong}, {Wiktorowicz}, {Wolff}, \&
  {Zuckerman}}]{Macintosh2015}
{Macintosh}, B., {Graham}, J.~R., {Barman}, T., {et~al.} 2015, Science, 350,
  64, \dodoi{10.1126/science.aac5891}

\bibitem[{{Madhusudhan} {et~al.}(2011){Madhusudhan}, {Burrows}, \&
  {Currie}}]{Madhusudhan2011}
{Madhusudhan}, N., {Burrows}, A., \& {Currie}, T. 2011, \apj, 737, 34,
  \dodoi{10.1088/0004-637X/737/1/34}

\bibitem[{{Maihara} {et~al.}(1993){Maihara}, {Iwamuro}, {Yamashita}, {Hall},
  {Cowie}, {Tokunaga}, \& {Pickles}}]{Maihara1993}
{Maihara}, T., {Iwamuro}, F., {Yamashita}, T., {et~al.} 1993, \pasp, 105, 940,
  \dodoi{10.1086/133259}

\bibitem[{{Marconi} {et~al.}(2022){Marconi}, {Abreu}, {Adibekyan}, {Alberti},
  {Albrecht}, {Alcaniz}, {Aliverti}, {Allende Prieto}, {Alvarado G{\'o}mez},
  {Amado}, {Amate}, {Andersen}, {Artigau}, {Baker}, {Baldini}, {Balestra},
  {Barnes}, {Baron}, {Barros}, {Bauer}, {Beaulieu}, {Bellido-Tirado},
  {Benneke}, {Bensby}, {Bergin}, {Biazzo}, {Bik}, {Birkby}, {Blind}, {Boisse},
  {Bolmont}, {Bonaglia}, {Bonfils}, {Borsa}, {Brandeker}, {Brandner}, {Broeg},
  {Brogi}, {Brousseau}, {Brucalassi}, {Brynnel}, {Buchhave}, {Buscher},
  {Cabral}, {Calderone}, {Calvo-Ortega}, {Canto Martins}, {Cantalloube},
  {Carbonaro}, {Chauvin}, {Chazelas}, {Cheffot}, {Cheng}, {Chiavassa},
  {Christensen}, {Cirami}, {Cook}, {Cooke}, {Coretti}, {Covino}, {Cowan},
  {Cresci}, {Cristiani}, {Cunha Parro}, {Cupani}, {D'Odorico}, {de Castro
  Le{\~a}o}, {De Cia}, {De Medeiros}, {Debras}, {Debus}, {Demangeon},
  {Dessauges-Zavadsky}, {Di Marcantonio}, {Dionies}, {Doyon}, {Dunn},
  {Ehrenreich}, {Faria}, {Feruglio}, {Fisher}, {Fontana}, {Fumagalli}, {Fusco},
  {Fynbo}, {Gabella}, {Gaessler}, {Gallo}, {Gao}, {Genolet}, {Genoni},
  {Giacobbe}, {Giro}, {Gon{\c{c}}alves}, {Gonzalez}, {Gonz{\'a}lez
  Hern{\'a}ndez}, {Gracia T{\'e}mich}, {Haehnelt}, {Haniff}, {Hatzes},
  {Helled}, {Hoeijmakers}, {Huke}, {J{\"a}rvinen}, {J{\"a}rvinen}, {Kaminski},
  {Korn}, {Kouach}, {Kowzan}, {Kreidberg}, {Landoni}, {Lanotte}, {Lavail},
  {Li}, {Liske}, {Lovis}, {Lucatello}, {Lunney}, {MacIntosh}, {Madhusudhan},
  {Magrini}, {Maiolino}, {Malo}, {Man}, {Marquart}, {Marques}, {Martins},
  {Martins}, {Maslowski}, {Mason}, {Mason}, {McCracken}, {Mergo}, {Micela},
  {Mitchell}, {Molli{\`e}re}, {Monteiro}, {Montgomery}, {Mordasini}, {Morin},
  {Mucciarelli}, {Murphy}, {N'Diaye}, {Neichel}, {Niedzielski}, {Niemczura},
  {Nortmann}, {Noterdaeme}, {Nunes}, {Oggioni}, {Oliva}, {{\"O}nel}, {Origlia},
  {{\"O}stlin}, {Palle}, {Papaderos}, {Pariani}, {Pe{\~n}ate Castro}, {Pepe},
  {Perreault Levasseur}, {Petit}, {Pino}, {Piqueras}, {Pollo}, {Poppenhaeger},
  {Quirrenbach}, {Rauscher}, {Rebolo}, {Redaelli}, {Reffert}, {Reid},
  {Reiners}, {Richter}, {Riva}, {Rivoire}, {Rodr{\'\i}guez-L{\'o}pez},
  {Roederer}, {Romano}, {Rousseau}, {Rowe}, {Salvadori}, {Santos}, {Santos
  Diaz}, {Sanz-Forcada}, {Sarajlic}, {Sauvage}, {Sch{\"a}fer}, {Schiavon},
  {Schmidt}, {Selmi}, {Sivanandam}, {Sordet}, {Sordo}, {Sortino}, {Sosnowska},
  {Sousa}, {Stempels}, {Strassmeier}, {Su{\'a}rez Mascare{\~n}o}, {Sulich},
  {Sun}, {Tanvir}, {Tenegi-Sangin{\'e}s}, {Thibault}, {Thompson}, {Tozzi},
  {Turbet}, {Vall{\'e}e}, {Varas}, {Venn}, {V{\'e}ran}, {Verma}, {Viel},
  {Wade}, {Waring}, {Weber}, {Weder}, {Wehbe}, {Weingrill}, {Woche}, {Xompero},
  {Zackrisson}, {Zanutta}, {Zapatero Osorio}, {Zechmeister}, \&
  {Zimara}}]{Marconi2022}
{Marconi}, A., {Abreu}, M., {Adibekyan}, V., {et~al.} 2022, in Society of
  Photo-Optical Instrumentation Engineers (SPIE) Conference Series, Vol. 12184,
  Ground-based and Airborne Instrumentation for Astronomy IX, ed. C.~J.
  {Evans}, J.~J. {Bryant}, \& K.~{Motohara}, 1218424,
  \dodoi{10.1117/12.2628689}

\bibitem[{{Marley} {et~al.}(2012){Marley}, {Saumon}, {Cushing}, {Ackerman},
  {Fortney}, \& {Freedman}}]{Marley2012}
{Marley}, M.~S., {Saumon}, D., {Cushing}, M., {et~al.} 2012, \apj, 754, 135,
  \dodoi{10.1088/0004-637X/754/2/135}

\bibitem[{{Marley} {et~al.}(2010){Marley}, {Saumon}, \&
  {Goldblatt}}]{Marley2010}
{Marley}, M.~S., {Saumon}, D., \& {Goldblatt}, C. 2010, \apjl, 723, L117,
  \dodoi{10.1088/2041-8205/723/1/L117}

\bibitem[{{Marley} {et~al.}(2002){Marley}, {Seager}, {Saumon}, {Lodders},
  {Ackerman}, {Freedman}, \& {Fan}}]{Marley2002}
{Marley}, M.~S., {Seager}, S., {Saumon}, D., {et~al.} 2002, \apj, 568, 335,
  \dodoi{10.1086/338800}

\bibitem[{{Marois} {et~al.}(2008){Marois}, {Macintosh}, {Barman}, {Zuckerman},
  {Song}, {Patience}, {Lafreni{\`e}re}, \& {Doyon}}]{Marois2008}
{Marois}, C., {Macintosh}, B., {Barman}, T., {et~al.} 2008, Science, 322, 1348,
  \dodoi{10.1126/science.1166585}

\bibitem[{{Marois} {et~al.}(2010){Marois}, {Zuckerman}, {Konopacky},
  {Macintosh}, \& {Barman}}]{Marois2010}
{Marois}, C., {Zuckerman}, B., {Konopacky}, Q.~M., {Macintosh}, B., \&
  {Barman}, T. 2010, \nat, 468, 1080, \dodoi{10.1038/nature09684}

\bibitem[{{Martin} {et~al.}(2023){Martin}, {Sethi}, {Armitage}, {Gilbert},
  {Rodr{\'\i}guez Mart{\'\i}nez}, \& {Gilbert}}]{Martin2023}
{Martin}, D.~V., {Sethi}, R., {Armitage}, T., {et~al.} 2023, arXiv e-prints,
  arXiv:2301.10858.
\newblock \doarXiv{2301.10858}

\bibitem[{{Mawet} {et~al.}(2019){Mawet}, {Fitzgerald}, {Konopacky}, {Beichman},
  {Jovanovic}, {Dekany}, {Hover}, {Chisholm}, {Ciardi}, {Artigau}, {Banyal},
  {Beatty}, {Benneke}, {Blake}, {Burgasser}, {Canalizo}, {Chen}, {Do},
  {Doppmann}, {Doyon}, {Dressing}, {Fang}, {Greene}, {Hillenbrand}, {Howard},
  {Kane}, {Kataria}, {Kempton}, {Knutson}, {Kotani}, {Lafreni{\`e}re}, {Liu},
  {Nishiyama}, {Pandey}, {Plavchan}, {Prato}, {Rajaguru}, {Robertson}, {Salyk},
  {Sato}, {Schlawin}, {Sengupta}, {Sivarani}, {Skidmore}, {Tamura}, {Terada},
  {Vasisht}, {Wang}, \& {Zhang}}]{Mawet2019}
{Mawet}, D., {Fitzgerald}, M., {Konopacky}, Q., {et~al.} 2019, in Bulletin of
  the American Astronomical Society, Vol.~51, 134.
\newblock \doarXiv{1908.03623}

\bibitem[{{McCarthy} {et~al.}(2016){McCarthy}, {Fanson}, {Bernstein}, {Ashby},
  {Bigelow}, {Boyadjian}, {Bouchez}, {Chauvin}, {Donoso}, {Filgueira},
  {Goodrich}, {Groark}, {Jacoby}, \& {Pearce}}]{McCarthy2016}
{McCarthy}, P.~J., {Fanson}, J., {Bernstein}, R., {et~al.} 2016, in Society of
  Photo-Optical Instrumentation Engineers (SPIE) Conference Series, Vol. 9906,
  Ground-based and Airborne Telescopes VI, ed. H.~J. {Hall}, R.~{Gilmozzi}, \&
  H.~K. {Marshall}, 990612, \dodoi{10.1117/12.2234505}

\bibitem[{McKinney {et~al.}(2010)}]{Pandas}
McKinney, W., {et~al.} 2010, in Proceedings of the 9th Python in Science
  Conference, Vol. 445, Austin, TX, 51--56

\bibitem[{{Miles} {et~al.}(2022){Miles}, {Biller}, {Patapis}, {Worthen},
  {Rickman}, {Hoch}, {Skemer}, {Perrin}, {Chen}, {Mukherjee}, {Morley},
  {Moran}, {Bonnefoy}, {Petrus}, {Carter}, {Choquet}, {Hinkley}, {Ward-Duong},
  {Leisenring}, {Millar-Blanchaer}, {Pueyo}, {Ray}, {Stapelfeldt}, {Stone},
  {Wang}, {Absil}, {Balmer}, {Boccaletti}, {Bonavita}, {Booth}, {Bowler},
  {Chauvin}, {Christiaens}, {Currie}, {Danielski}, {Fortney}, {Girard},
  {Greenbaum}, {Henning}, {Hines}, {Janson}, {Kalas}, {Kammerer}, {Kenworthy},
  {Kervella}, {Lagage}, {Lew}, {Liu}, {Macintosh}, {Marino}, {Marley},
  {Marois}, {Matthews}, {Matthews}, {Mawet}, {McElwain}, {Metchev}, {Meyer},
  {Molliere}, {Pantin}, {Rebollido}, {Ren}, {Vasist}, {Wyatt}, {Zhou},
  {Briesemeister}, {Bryan}, {Calissendorff}, {Catalloube}, {Cugno}, {De Furio},
  {Dupuy}, {Factor}, {Faherty}, {Fitzgerald}, {Franson}, {Gonzales}, {Hood},
  {Howe}, {Kraus}, {Kuzuhara}, {Lawson}, {Lazzoni}, {Liu}, {Llop-Sayson},
  {Lloyd}, {Martinez}, {Mazoyer}, {Quanz}, {Adams Redai}, {Samland},
  {Schlieder}, {Tamura}, {Tan}, {Uyama}, {Vigan}, {Vos}, {Wagner}, {Wolff},
  {Ygouf}, {Zhang}, \& {Zhang}}]{Miles2022}
{Miles}, B.~E., {Biller}, B.~A., {Patapis}, P., {et~al.} 2022, arXiv e-prints,
  arXiv:2209.00620.
\newblock \doarXiv{2209.00620}

\bibitem[{{Millar-Blanchaer} {et~al.}(2020){Millar-Blanchaer}, {Girard},
  {Karalidi}, {Marley}, {van Holstein}, {Sengupta}, {Mawet}, {Kataria}, {Snik},
  {de Boer}, {Jensen-Clem}, {Vigan}, \& {Hinkley}}]{Millar-Blanchaer2020}
{Millar-Blanchaer}, M.~A., {Girard}, J.~H., {Karalidi}, T., {et~al.} 2020,
  \apj, 894, 42, \dodoi{10.3847/1538-4357/ab6ef2}

\bibitem[{{Molli{\`e}re} {et~al.}(2020){Molli{\`e}re}, {Stolker}, {Lacour},
  {Otten}, {Shangguan}, {Charnay}, {Molyarova}, {Nowak}, {Henning}, {Marleau},
  {Semenov}, {van Dishoeck}, {Eisenhauer}, {Garcia}, {Garcia Lopez}, {Girard},
  {Greenbaum}, {Hinkley}, {Kervella}, {Kreidberg}, {Maire}, {Nasedkin},
  {Pueyo}, {Snellen}, {Vigan}, {Wang}, {de Zeeuw}, \& {Zurlo}}]{Molliere2020}
{Molli{\`e}re}, P., {Stolker}, T., {Lacour}, S., {et~al.} 2020, \aap, 640,
  A131, \dodoi{10.1051/0004-6361/202038325}

\bibitem[{{Morley} {et~al.}(2014){Morley}, {Marley}, {Fortney}, {Lupu},
  {Saumon}, {Greene}, \& {Lodders}}]{Morley2014}
{Morley}, C.~V., {Marley}, M.~S., {Fortney}, J.~J., {et~al.} 2014, \apj, 787,
  78, \dodoi{10.1088/0004-637X/787/1/78}

\bibitem[{{Morris} {et~al.}(2018){Morris}, {Agol}, {Davenport}, \&
  {Hawley}}]{Morris2018}
{Morris}, B.~M., {Agol}, E., {Davenport}, J. R.~A., \& {Hawley}, S.~L. 2018,
  \apj, 857, 39, \dodoi{10.3847/1538-4357/aab6a5}

\bibitem[{{Morzinski} {et~al.}(2015){Morzinski}, {Males}, {Skemer}, {Close},
  {Hinz}, {Rodigas}, {Puglisi}, {Esposito}, {Riccardi}, {Pinna}, {Xompero},
  {Briguglio}, {Bailey}, {Follette}, {Kopon}, {Weinberger}, \&
  {Wu}}]{Morzinski2015}
{Morzinski}, K.~M., {Males}, J.~R., {Skemer}, A.~J., {et~al.} 2015, \apj, 815,
  108, \dodoi{10.1088/0004-637X/815/2/108}

\bibitem[{{Mulders} {et~al.}(2015){Mulders}, {Pascucci}, \&
  {Apai}}]{Mulders2015a}
{Mulders}, G.~D., {Pascucci}, I., \& {Apai}, D. 2015, \apj, 798, 112,
  \dodoi{10.1088/0004-637X/798/2/112}

\bibitem[{{National Academies of Sciences} \& Medicine(2021)}]{Decadal2021}
{National Academies of Sciences}, \& Medicine. 2021, {Pathways to Discovery in
  Astronomy and Astrophysics for the 2020s}, \dodoi{10.17226/26141}

\bibitem[{{Noll} {et~al.}(2000){Noll}, {Geballe}, {Leggett}, \&
  {Marley}}]{Noll2000}
{Noll}, K.~S., {Geballe}, T.~R., {Leggett}, S.~K., \& {Marley}, M.~S. 2000,
  \apjl, 541, L75, \dodoi{10.1086/312906}

\bibitem[{{Noll} {et~al.}(2012){Noll}, {Kausch}, {Barden}, {Jones}, {Szyszka},
  \& {Kimeswenger}}]{Noll2012}
{Noll}, S., {Kausch}, W., {Barden}, M., {et~al.} 2012, in EGU General Assembly
  Conference Abstracts, EGU General Assembly Conference Abstracts, 9813

\bibitem[{{O'Malley-James} \& {Kaltenegger}(2017)}]{O'Malley-James2017}
{O'Malley-James}, J.~T., \& {Kaltenegger}, L. 2017, \mnras, 469, L26,
  \dodoi{10.1093/mnrasl/slx047}

\bibitem[{{Oppenheimer} {et~al.}(1995){Oppenheimer}, {Kulkarni}, {Matthews}, \&
  {Nakajima}}]{Oppenheimer1995}
{Oppenheimer}, B.~R., {Kulkarni}, S.~R., {Matthews}, K., \& {Nakajima}, T.
  1995, Science, 270, 1478, \dodoi{10.1126/science.270.5241.1478}

\bibitem[{{Packham} {et~al.}(2014){Packham}, {Honda}, {Okamoto}, {Richter},
  {Chun}, {Kataza}, {Onaka}, {Sakon}, \& {MICHI Team}}]{Packham2014}
{Packham}, C., {Honda}, M., {Okamoto}, Y.~K., {et~al.} 2014, in Thirty Meter
  Telescope Science Forum, ed. M.~{Dickinson} \& H.~{Inami}, 20

\bibitem[{{Pai Asnodkar} {et~al.}(2022){Pai Asnodkar}, {Wang}, {Gaudi},
  {Cauley}, {Eastman}, {Ilyin}, {Strassmeier}, \& {Beatty}}]{PaiAsnodkar2021}
{Pai Asnodkar}, A., {Wang}, J., {Gaudi}, B.~S., {et~al.} 2022, \aj, 163, 40,
  \dodoi{10.3847/1538-3881/ac32c7}

\bibitem[{{Paudel} {et~al.}(2020){Paudel}, {Gizis}, {Mullan}, {Schmidt},
  {Burgasser}, \& {Williams}}]{Paudel2020}
{Paudel}, R.~R., {Gizis}, J.~E., {Mullan}, D.~J., {et~al.} 2020, \mnras, 494,
  5751, \dodoi{10.1093/mnras/staa1137}

\bibitem[{{Paudel} {et~al.}(2018){Paudel}, {Gizis}, {Mullan}, {Schmidt},
  {Burgasser}, {Williams}, \& {Berger}}]{Paudel2018}
---. 2018, \apj, 858, 55, \dodoi{10.3847/1538-4357/aab8fe}

\bibitem[{{Petit} {et~al.}(2008){Petit}, {Dintrans}, {Solanki}, {Donati},
  {Auri{\`e}re}, {Ligni{\`e}res}, {Morin}, {Paletou}, {Ramirez Velez},
  {Catala}, \& {Fares}}]{Petit2008}
{Petit}, P., {Dintrans}, B., {Solanki}, S.~K., {et~al.} 2008, \mnras, 388, 80,
  \dodoi{10.1111/j.1365-2966.2008.13411.x}

\bibitem[{{Petrus} {et~al.}(2022){Petrus}, {Chauvin}, {Bonnefoy}, {Tremblin},
  {Charnay}, {Delorme}, {Marleau}, {Bayo}, {Manjavacas}, {Lagrange},
  {Molli{\`e}re}, {Palma-Bifani}, \& {Biller James-S. Jenkins}}]{Petrus2022}
{Petrus}, S., {Chauvin}, G., {Bonnefoy}, M., {et~al.} 2022, arXiv e-prints,
  arXiv:2207.06622.
\newblock \doarXiv{2207.06622}

\bibitem[{{Piskunov} {et~al.}(1990){Piskunov}, {Tuominen}, \&
  {Vilhu}}]{Piskunov1990}
{Piskunov}, N.~E., {Tuominen}, I., \& {Vilhu}, O. 1990, \aap, 230, 363

\bibitem[{{Plummer} \& {Wang}(2022)}]{Plummer2022}
{Plummer}, M.~K., \& {Wang}, J. 2022, \apj, 933, 163,
  \dodoi{10.3847/1538-4357/ac75b9}

\bibitem[{{Radigan}(2014)}]{Radigan2014b}
{Radigan}, J. 2014, \apj, 797, 120, \dodoi{10.1088/0004-637X/797/2/120}

\bibitem[{{Radigan} {et~al.}(2012){Radigan}, {Jayawardhana}, {Lafreni{\`e}re},
  {Artigau}, {Marley}, \& {Saumon}}]{Radigan2012}
{Radigan}, J., {Jayawardhana}, R., {Lafreni{\`e}re}, D., {et~al.} 2012, \apj,
  750, 105, \dodoi{10.1088/0004-637X/750/2/105}

\bibitem[{{Radigan} {et~al.}(2014){Radigan}, {Lafreni{\`e}re}, {Jayawardhana},
  \& {Artigau}}]{Radigan2014a}
{Radigan}, J., {Lafreni{\`e}re}, D., {Jayawardhana}, R., \& {Artigau}, E. 2014,
  \apj, 793, 75, \dodoi{10.1088/0004-637X/793/2/75}

\bibitem[{{Reiners} \& {Basri}(2008)}]{Reiners&Basri2008}
{Reiners}, A., \& {Basri}, G. 2008, \apj, 684, 1390, \dodoi{10.1086/590073}

\bibitem[{{Reiners} \& {Basri}(2010)}]{Reiners&Basri2010}
---. 2010, \apj, 710, 924, \dodoi{10.1088/0004-637X/710/2/924}

\bibitem[{{Roettenbacher} \& {Kane}(2017)}]{Roettenbacher2017}
{Roettenbacher}, R.~M., \& {Kane}, S.~R. 2017, \apj, 851, 77,
  \dodoi{10.3847/1538-4357/aa991e}

\bibitem[{{Ruffio} {et~al.}(2021){Ruffio}, {Konopacky}, {Barman}, {Macintosh},
  {Hoch}, {De Rosa}, {Wang}, {Czekala}, \& {Marois}}]{Ruffio2021}
{Ruffio}, J.-B., {Konopacky}, Q.~M., {Barman}, T., {et~al.} 2021, \aj, 162,
  290, \dodoi{10.3847/1538-3881/ac273a}

\bibitem[{{Sabotta} {et~al.}(2021){Sabotta}, {Schlecker}, {Chaturvedi},
  {Guenther}, {Mu{\~n}oz Rodr{\'\i}guez}, {Mu{\~n}oz S{\'a}nchez}, {Caballero},
  {Shan}, {Reffert}, {Ribas}, {Reiners}, {Hatzes}, {Amado}, {Klahr}, {Morales},
  {Quirrenbach}, {Henning}, {Dreizler}, {Pall{\'e}}, {Perger}, {Azzaro},
  {Jeffers}, {Kaminski}, {K{\"u}rster}, {Lafarga}, {Montes}, {Passegger}, \&
  {Zechmeister}}]{Sabotta2021}
{Sabotta}, S., {Schlecker}, M., {Chaturvedi}, P., {et~al.} 2021, \aap, 653,
  A114, \dodoi{10.1051/0004-6361/202140968}

\bibitem[{{Saumon} \& {Marley}(2008)}]{Saumon2008}
{Saumon}, D., \& {Marley}, M.~S. 2008, \apj, 689, 1327, \dodoi{10.1086/592734}

\bibitem[{{Scalo} {et~al.}(2007){Scalo}, {Kaltenegger}, {Segura}, {Fridlund},
  {Ribas}, {Kulikov}, {Grenfell}, {Rauer}, {Odert}, {Leitzinger}, {Selsis},
  {Khodachenko}, {Eiroa}, {Kasting}, \& {Lammer}}]{scalo07}
{Scalo}, J., {Kaltenegger}, L., {Segura}, A.~G., {et~al.} 2007, Astrobiology,
  7, 85, \dodoi{10.1089/ast.2006.0125}

\bibitem[{{Schweitzer} {et~al.}(1996){Schweitzer}, {Hauschildt}, {Allard}, \&
  {Basri}}]{Schweitzer1996}
{Schweitzer}, A., {Hauschildt}, P.~H., {Allard}, F., \& {Basri}, G. 1996,
  \mnras, 283, 821, \dodoi{10.1093/mnras/283.3.821}

\bibitem[{{Showman} {et~al.}(2019){Showman}, {Tan}, \& {Zhang}}]{showman19}
{Showman}, A.~P., {Tan}, X., \& {Zhang}, X. 2019, \apj, 883, 4,
  \dodoi{10.3847/1538-4357/ab384a}

\bibitem[{{Shuster}(1993)}]{Shuster1993}
{Shuster}, M.~D. 1993, IEEE Transactions on Aerospace Electronic Systems, 29,
  263, \dodoi{10.1109/7.249140}

\bibitem[{{Skemer} {et~al.}(2012){Skemer}, {Hinz}, {Esposito}, {Burrows},
  {Leisenring}, {Skrutskie}, {Desidera}, {Mesa}, {Arcidiacono}, {Mannucci},
  {Rodigas}, {Close}, {McCarthy}, {Kulesa}, {Agapito}, {Apai}, {Argomedo},
  {Bailey}, {Boutsia}, {Briguglio}, {Brusa}, {Busoni}, {Claudi}, {Eisner},
  {Fini}, {Follette}, {Garnavich}, {Gratton}, {Guerra}, {Hill}, {Hoffmann},
  {Jones}, {Krejny}, {Males}, {Masciadri}, {Meyer}, {Miller}, {Morzinski},
  {Nelson}, {Pinna}, {Puglisi}, {Quanz}, {Quiros-Pacheco}, {Riccardi},
  {Stefanini}, {Vaitheeswaran}, {Wilson}, \& {Xompero}}]{Skemer2012}
{Skemer}, A.~J., {Hinz}, P.~M., {Esposito}, S., {et~al.} 2012, \apj, 753, 14,
  \dodoi{10.1088/0004-637X/753/1/14}

\bibitem[{{Skemer} {et~al.}(2014){Skemer}, {Marley}, {Hinz}, {Morzinski},
  {Skrutskie}, {Leisenring}, {Close}, {Saumon}, {Bailey}, {Briguglio},
  {Defrere}, {Esposito}, {Follette}, {Hill}, {Males}, {Puglisi}, {Rodigas}, \&
  {Xompero}}]{Skemer2014}
{Skemer}, A.~J., {Marley}, M.~S., {Hinz}, P.~M., {et~al.} 2014, \apj, 792, 17,
  \dodoi{10.1088/0004-637X/792/1/17}

\bibitem[{{Skidmore} {et~al.}(2015){Skidmore}, {TMT International Science
  Development Teams}, \& {Science Advisory Committee}}]{Skidmore2015}
{Skidmore}, W., {TMT International Science Development Teams}, \& {Science
  Advisory Committee}, T. 2015, Research in Astronomy and Astrophysics, 15,
  1945, \dodoi{10.1088/1674-4527/15/12/001}

\bibitem[{{Skilling}(2004)}]{Skilling2004}
{Skilling}, J. 2004, in American Institute of Physics Conference Series, Vol.
  735, Bayesian Inference and Maximum Entropy Methods in Science and
  Engineering: 24th International Workshop on Bayesian Inference and Maximum
  Entropy Methods in Science and Engineering, ed. R.~{Fischer}, R.~{Preuss}, \&
  U.~V. {Toussaint}, 395--405, \dodoi{10.1063/1.1835238}

\bibitem[{{Skilling}(2006)}]{skilling06}
{Skilling}, J. 2006, Bayesian Analysis, 1, 833, \dodoi{10.1214/06-BA127}

\bibitem[{{Snellen} {et~al.}(2014){Snellen}, {Brandl}, {de Kok}, {Brogi},
  {Birkby}, \& {Schwarz}}]{Snellen2014}
{Snellen}, I. A.~G., {Brandl}, B.~R., {de Kok}, R.~J., {et~al.} 2014, \nat,
  509, 63, \dodoi{10.1038/nature13253}

\bibitem[{{Snellen} \& {Brown}(2018)}]{2018Snellen&Brown}
{Snellen}, I.~A.~G., \& {Brown}, A.~G.~A. 2018, Nature Astronomy, 2, 883,
  \dodoi{10.1038/s41550-018-0561-6}

\bibitem[{{Speagle}(2020)}]{speagle20}
{Speagle}, J.~S. 2020, \mnras, 493, 3132, \dodoi{10.1093/mnras/staa278}

\bibitem[{{Su} {et~al.}(2009){Su}, {Rieke}, {Stapelfeldt}, {Malhotra},
  {Bryden}, {Smith}, {Misselt}, {Moro-Martin}, \& {Williams}}]{Su2009}
{Su}, K.~Y.~L., {Rieke}, G.~H., {Stapelfeldt}, K.~R., {et~al.} 2009, \apj, 705,
  314, \dodoi{10.1088/0004-637X/705/1/314}

\bibitem[{{Szentgyorgyi} {et~al.}(2018){Szentgyorgyi}, {Baldwin}, {Barnes},
  {Bean}, {Ben-Ami}, {Brennan}, {Budynkiewicz}, {Catropa}, {Chun}, {Conroy},
  {Contos}, {Crane}, {Durusky}, {Epps}, {Evans}, {Evans}, {Fishman}, {Frebel},
  {Gauron}, {Guzman}, {Hare}, {Jang}, {Jang}, {Jordan}, {Kim}, {Kim}, {Kim},
  {Lee}, {Lopez-Morales}, {Mendes de Oliveira}, {McCracken}, {McMuldroch},
  {Miller}, {Mueller}, {Oh}, {Onyuksel}, {Park}, {Park}, {Park}, {Paxson},
  {Phillips}, {Plummer}, {Podgorski}, {Rubin}, {Seifahrt}, {Stark}, {Steiner},
  {Uomoto}, {Walsworth}, \& {Yu}}]{Szentgyorgyi2018}
{Szentgyorgyi}, A., {Baldwin}, D., {Barnes}, S., {et~al.} 2018, in Society of
  Photo-Optical Instrumentation Engineers (SPIE) Conference Series, Vol. 10702,
  Ground-based and Airborne Instrumentation for Astronomy VII, ed. C.~J.
  {Evans}, L.~{Simard}, \& H.~{Takami}, 107021R, \dodoi{10.1117/12.2313539}

\bibitem[{{Thomas-Osip} {et~al.}(2008){Thomas-Osip}, {Prieto}, {Johns}, \&
  {Phillips}}]{Thomas-Osip2008}
{Thomas-Osip}, J.~E., {Prieto}, G., {Johns}, M., \& {Phillips}, M.~M. 2008, in
  Society of Photo-Optical Instrumentation Engineers (SPIE) Conference Series,
  Vol. 7012, Ground-based and Airborne Telescopes II, ed. L.~M. {Stepp} \&
  R.~{Gilmozzi}, 70121U, \dodoi{10.1117/12.789863}

\bibitem[{{Tsuji} {et~al.}(1996{\natexlab{a}}){Tsuji}, {Ohnaka}, \&
  {Aoki}}]{Tsuji1996a}
{Tsuji}, T., {Ohnaka}, K., \& {Aoki}, W. 1996{\natexlab{a}}, \aap, 305, L1

\bibitem[{{Tsuji} {et~al.}(1996{\natexlab{b}}){Tsuji}, {Ohnaka}, {Aoki}, \&
  {Nakajima}}]{Tsuji1996b}
{Tsuji}, T., {Ohnaka}, K., {Aoki}, W., \& {Nakajima}, T. 1996{\natexlab{b}},
  \aap, 308, L29

\bibitem[{{Tuomi} {et~al.}(2019){Tuomi}, {Jones}, {Butler}, {Arriagada},
  {Vogt}, {Burt}, {Laughlin}, {Holden}, {Shectman}, {Crane}, {Thompson},
  {Keiser}, {Jenkins}, {Berdi{\~n}as}, {Diaz}, {Kiraga}, \&
  {Barnes}}]{Tuomi2019}
{Tuomi}, M., {Jones}, H.~R.~A., {Butler}, R.~P., {et~al.} 2019, arXiv e-prints,
  arXiv:1906.04644.
\newblock \doarXiv{1906.04644}

\bibitem[{{van Leeuwen}(2007)}]{vanLeeuwen2007}
{van Leeuwen}, F. 2007, \aap, 474, 653, \dodoi{10.1051/0004-6361:20078357}

\bibitem[{{Vandal} {et~al.}(2020){Vandal}, {Rameau}, \& {Doyon}}]{Vandal2020}
{Vandal}, T., {Rameau}, J., \& {Doyon}, R. 2020, \aj, 160, 243,
  \dodoi{10.3847/1538-3881/abba30}

\bibitem[{{Vida} {et~al.}(2017){Vida}, {K{\H{o}}v{\'a}ri}, {P{\'a}l},
  {Ol{\'a}h}, \& {Kriskovics}}]{Vida2017}
{Vida}, K., {K{\H{o}}v{\'a}ri}, Z., {P{\'a}l}, A., {Ol{\'a}h}, K., \&
  {Kriskovics}, L. 2017, \apj, 841, 124, \dodoi{10.3847/1538-4357/aa6f05}

\bibitem[{Virtanen {et~al.}(2020)Virtanen, Gommers, Oliphant, Haberland, Reddy,
  Cournapeau, Burovski, Peterson, Weckesser, Bright, {van der Walt}, Brett,
  Wilson, Millman, Mayorov, Nelson, Jones, Kern, Larson, Carey, Polat, Feng,
  Moore, {VanderPlas}, Laxalde, Perktold, Cimrman, Henriksen, Quintero, Harris,
  Archibald, Ribeiro, Pedregosa, {van Mulbregt}, \& {SciPy 1.0
  Contributors}}]{scipy2020}
Virtanen, P., Gommers, R., Oliphant, T.~E., {et~al.} 2020, Nature Methods, 17,
  261, \dodoi{10.1038/s41592-019-0686-2}

\bibitem[{{Vogt} {et~al.}(1987){Vogt}, {Penrod}, \& {Hatzes}}]{vogt87}
{Vogt}, S.~S., {Penrod}, G.~D., \& {Hatzes}, A.~P. 1987, \apj, 321, 496,
  \dodoi{10.1086/165647}

\bibitem[{{Vos} {et~al.}(2017){Vos}, {Allers}, \& {Biller}}]{Vos2017}
{Vos}, J.~M., {Allers}, K.~N., \& {Biller}, B.~A. 2017, \apj, 842, 78,
  \dodoi{10.3847/1538-4357/aa73cf}

\bibitem[{{Vos} {et~al.}(2022){Vos}, {Burningham}, {Faherty}, {Alejandro},
  {Gonzales}, {Calamari}, {Bardalez Gagliuffi}, {Visscher}, {Tan}, {Morley},
  {Marley}, {Gemma}, {Whiteford}, {Gaarn}, \& {Park}}]{Vos2022b}
{Vos}, J.~M., {Burningham}, B., {Faherty}, J.~K., {et~al.} 2022, arXiv
  e-prints, arXiv:2212.07399.
\newblock \doarXiv{2212.07399}

\bibitem[{{Wang} {et~al.}(2018{\natexlab{a}}){Wang}, {David}, {Hillenbrand},
  {Mawet}, {Albrecht}, \& {Liu}}]{Wang18}
{Wang}, J., {David}, T.~J., {Hillenbrand}, L.~A., {et~al.} 2018{\natexlab{a}},
  \apj, 865, 141, \dodoi{10.3847/1538-4357/aadee8}

\bibitem[{{Wang} {et~al.}(2017){Wang}, {Prato}, \& {Mawet}}]{Wang17}
{Wang}, J., {Prato}, L., \& {Mawet}, D. 2017, \apj, 838, 35,
  \dodoi{10.3847/1538-4357/aa6345}

\bibitem[{{Wang} {et~al.}(2020){Wang}, {Wang}, {Ma}, {Chilcote}, {Ertel},
  {Guyon}, {Ilyin}, {Jovanovic}, {Kalas}, {Lozi}, {Macintosh}, {Strassmeier},
  \& {Stone}}]{Wang2020}
{Wang}, J., {Wang}, J.~J., {Ma}, B., {et~al.} 2020, \aj, 160, 150,
  \dodoi{10.3847/1538-3881/ababa7}

\bibitem[{{Wang} {et~al.}(2022{\natexlab{a}}){Wang}, {Kolecki}, {Ruffio},
  {Wang}, {Mawet}, {Baker}, {Bartos}, {Blake}, {Bond}, {Calvin}, {Cetre},
  {Delorme}, {Doppmann}, {Echeverri}, {Finnerty}, {Fitzgerald}, {Jovanovic},
  {Liu}, {Lopez}, {Morris}, {Pai Asnodkar}, {Pezzato}, {Ragland}, {Roy},
  {Ruane}, {Sappey}, {Schofield}, {Skemer}, {Venenciano}, {Kent Wallace},
  {Wallack}, {Wizinowich}, \& {Xuan}}]{Wang2022}
{Wang}, J., {Kolecki}, J.~R., {Ruffio}, J.-B., {et~al.} 2022{\natexlab{a}},
  \aj, 163, 189, \dodoi{10.3847/1538-3881/ac56e2}

\bibitem[{{Wang} {et~al.}(2018{\natexlab{b}}){Wang}, {Graham}, {Dawson},
  {Fabrycky}, {De Rosa}, {Pueyo}, {Konopacky}, {Macintosh}, {Marois}, {Chiang},
  {Ammons}, {Arriaga}, {Bailey}, {Barman}, {Bulger}, {Chilcote}, {Cotten},
  {Doyon}, {Duch{\^e}ne}, {Esposito}, {Fitzgerald}, {Follette}, {Gerard},
  {Goodsell}, {Greenbaum}, {Hibon}, {Hung}, {Ingraham}, {Kalas}, {Larkin},
  {Maire}, {Marchis}, {Marley}, {Metchev}, {Millar-Blanchaer}, {Nielsen},
  {Oppenheimer}, {Palmer}, {Patience}, {Perrin}, {Poyneer}, {Rajan}, {Rameau},
  {Rantakyr{\"o}}, {Ruffio}, {Savransky}, {Schneider}, {Sivaramakrishnan},
  {Song}, {Soummer}, {Thomas}, {Wallace}, {Ward-Duong}, {Wiktorowicz}, \&
  {Wolff}}]{JJWang2018}
{Wang}, J.~J., {Graham}, J.~R., {Dawson}, R., {et~al.} 2018{\natexlab{b}}, \aj,
  156, 192, \dodoi{10.3847/1538-3881/aae15010.48550/arXiv.1809.04107}

\bibitem[{{Wang} {et~al.}(2021){Wang}, {Ruffio}, {Morris}, {Delorme},
  {Jovanovic}, {Pezzato}, {Echeverri}, {Finnerty}, {Hood}, {Zanazzi}, {Bryan},
  {Bond}, {Cetre}, {Martin}, {Mawet}, {Skemer}, {Baker}, {Xuan}, {Wallace},
  {Wang}, {Bartos}, {Blake}, {Boden}, {Buzard}, {Calvin}, {Chun}, {Doppmann},
  {Dupuy}, {Duch{\^e}ne}, {Feng}, {Fitzgerald}, {Fortney}, {Freedman},
  {Knutson}, {Konopacky}, {Lilley}, {Liu}, {Lopez}, {Lupu}, {Marley},
  {Meshkat}, {Miles}, {Millar-Blanchaer}, {Ragland}, {Roy}, {Ruane}, {Sappey},
  {Schofield}, {Weiss}, {Wetherell}, {Wizinowich}, \& {Ygouf}}]{JJWang2021}
{Wang}, J.~J., {Ruffio}, J.-B., {Morris}, E., {et~al.} 2021, \aj, 162, 148,
  \dodoi{10.3847/1538-3881/ac1349}

\bibitem[{{Wang} {et~al.}(2022{\natexlab{b}}){Wang}, {Gao}, {Chilcote}, {Lozi},
  {Guyon}, {Marois}, {De Rosa}, {Sahoo}, {Groff}, {Vievard}, {Jovanovic},
  {Greenbaum}, \& {Macintosh}}]{JJWang2022}
{Wang}, J.~J., {Gao}, P., {Chilcote}, J., {et~al.} 2022{\natexlab{b}}, \aj,
  164, 143, \dodoi{10.3847/1538-3881/ac8984}

\bibitem[{{Wheatley} {et~al.}(2017){Wheatley}, {Louden}, {Bourrier},
  {Ehrenreich}, \& {Gillon}}]{Wheatley2017}
{Wheatley}, P.~J., {Louden}, T., {Bourrier}, V., {Ehrenreich}, D., \& {Gillon},
  M. 2017, \mnras, 465, L74, \dodoi{10.1093/mnrasl/slw192}

\bibitem[{{Wilson} {et~al.}(2021){Wilson}, {Froning}, {Duvvuri}, {France},
  {Youngblood}, {Schneider}, {Berta-Thompson}, {Brown}, {Buccino}, {Hawley},
  {Irwin}, {Kaltenegger}, {Kowalski}, {Linsky}, {Parke Loyd}, {Miguel},
  {Pineda}, {Redfield}, {Roberge}, {Rugheimer}, {Tian}, \&
  {Vieytes}}]{Wilson2021}
{Wilson}, D.~J., {Froning}, C.~S., {Duvvuri}, G.~M., {et~al.} 2021, \apj, 911,
  18, \dodoi{10.3847/1538-4357/abe771}

\bibitem[{{Witte} {et~al.}(2011){Witte}, {Helling}, {Barman}, {Heidrich}, \&
  {Hauschildt}}]{Witte2011}
{Witte}, S., {Helling}, C., {Barman}, T., {Heidrich}, N., \& {Hauschildt},
  P.~H. 2011, \aap, 529, A44, \dodoi{10.1051/0004-6361/201014105}

\bibitem[{{Wright} {et~al.}(2010){Wright}, {Eisenhardt}, {Mainzer}, {Ressler},
  {Cutri}, {Jarrett}, {Kirkpatrick}, {Padgett}, {McMillan}, {Skrutskie},
  {Stanford}, {Cohen}, {Walker}, {Mather}, {Leisawitz}, {Gautier}, {McLean},
  {Benford}, {Lonsdale}, {Blain}, {Mendez}, {Irace}, {Duval}, {Liu}, {Royer},
  {Heinrichsen}, {Howard}, {Shannon}, {Kendall}, {Walsh}, {Larsen}, {Cardon},
  {Schick}, {Schwalm}, {Abid}, {Fabinsky}, {Naes}, \& {Tsai}}]{WISE2010}
{Wright}, E.~L., {Eisenhardt}, P. R.~M., {Mainzer}, A.~K., {et~al.} 2010, \aj,
  140, 1868, \dodoi{10.1088/0004-6256/140/6/1868}

\bibitem[{{Yang} {et~al.}(2016){Yang}, {Apai}, {Marley}, {Karalidi}, {Flateau},
  {Showman}, {Metchev}, {Buenzli}, {Radigan}, {Artigau}, {Lowrance}, \&
  {Burgasser}}]{Yang2016}
{Yang}, H., {Apai}, D., {Marley}, M.~S., {et~al.} 2016, \apj, 826, 8,
  \dodoi{10.3847/0004-637X/826/1/8}

\bibitem[{{Zendejas} {et~al.}(2010){Zendejas}, {Segura}, \&
  {Raga}}]{zendejas10}
{Zendejas}, J., {Segura}, A., \& {Raga}, A.~C. 2010, \icarus, 210, 539,
  \dodoi{10.1016/j.icarus.2010.07.013}

\bibitem[{{Zhang}(2020)}]{Zhang2020}
{Zhang}, X. 2020, Research in Astronomy and Astrophysics, 20, 099,
  \dodoi{10.1088/1674-4527/20/7/99}

\bibitem[{{Zhang} \& {Showman}(2014)}]{zhang14}
{Zhang}, X., \& {Showman}, A.~P. 2014, \apjl, 788, L6,
  \dodoi{10.1088/2041-8205/788/1/L6}

\bibitem[{{Zhou} {et~al.}(2016){Zhou}, {Apai}, {Schneider}, {Marley}, \&
  {Showman}}]{Zhou2016}
{Zhou}, Y., {Apai}, D., {Schneider}, G.~H., {Marley}, M.~S., \& {Showman},
  A.~P. 2016, \apj, 818, 176, \dodoi{10.3847/0004-637X/818/2/176}

\bibitem[{{Zhou} {et~al.}(2022){Zhou}, {Bowler}, {Apai}, {Kataria}, {Morley},
  {Bryan}, {Skemer}, \& {Benneke}}]{Zhou2022}
{Zhou}, Y., {Bowler}, B.~P., {Apai}, D., {et~al.} 2022, arXiv e-prints,
  arXiv:2210.02464.
\newblock \doarXiv{2210.02464}

\bibitem[{{Zhou} {et~al.}(2020{\natexlab{a}}){Zhou}, {Bowler}, {Morley},
  {Apai}, {Kataria}, {Bryan}, \& {Benneke}}]{Zhou2020b}
{Zhou}, Y., {Bowler}, B.~P., {Morley}, C.~V., {et~al.} 2020{\natexlab{a}}, \aj,
  160, 77, \dodoi{10.3847/1538-3881/ab9e04}

\bibitem[{{Zhou} {et~al.}(2020{\natexlab{b}}){Zhou}, {Apai}, {Bedin}, {Lew},
  {Schneider}, {Burgasser}, {Manjavacas}, {Karalidi}, {Metchev},
  {Miles-P{\'a}ez}, {Cowan}, {Lowrance}, \& {Radigan}}]{Zhou2020}
{Zhou}, Y., {Apai}, D., {Bedin}, L.~R., {et~al.} 2020{\natexlab{b}}, \aj, 159,
  140, \dodoi{10.3847/1538-3881/ab6f65}

\bibitem[{{Zink} {et~al.}(2020){Zink}, {Hardegree-Ullman}, {Christiansen},
  {Petigura}, {Dressing}, {Schlieder}, {Ciardi}, \& {Crossfield}}]{Zink2020}
{Zink}, J.~K., {Hardegree-Ullman}, K.~K., {Christiansen}, J.~L., {et~al.} 2020,
  \aj, 160, 94, \dodoi{10.3847/1538-3881/aba123}

\bibitem[{{Zuckerman} {et~al.}(2006){Zuckerman}, {Bessell}, {Song}, \&
  {Kim}}]{Zuckerman2006}
{Zuckerman}, B., {Bessell}, M.~S., {Song}, I., \& {Kim}, S. 2006, \apjl, 649,
  L115, \dodoi{10.1086/508060}

\bibitem[{{Zuckerman} {et~al.}(2011){Zuckerman}, {Rhee}, {Song}, \&
  {Bessell}}]{Zuckerman2011}
{Zuckerman}, B., {Rhee}, J.~H., {Song}, I., \& {Bessell}, M.~S. 2011, \apj,
  732, 61, \dodoi{10.1088/0004-637X/732/2/61}

\bibitem[{{Zurlo} {et~al.}(2016){Zurlo}, {Vigan}, {Galicher}, {Maire}, {Mesa},
  {Gratton}, {Chauvin}, {Kasper}, {Moutou}, {Bonnefoy}, {Desidera}, {Abe},
  {Apai}, {Baruffolo}, {Baudoz}, {Baudrand}, {Beuzit}, {Blancard},
  {Boccaletti}, {Cantalloube}, {Carle}, {Cascone}, {Charton}, {Claudi},
  {Costille}, {de Caprio}, {Dohlen}, {Dominik}, {Fantinel}, {Feautrier},
  {Feldt}, {Fusco}, {Gigan}, {Girard}, {Gisler}, {Gluck}, {Gry}, {Henning},
  {Hugot}, {Janson}, {Jaquet}, {Lagrange}, {Langlois}, {Llored}, {Madec},
  {Magnard}, {Martinez}, {Maurel}, {Mawet}, {Meyer}, {Milli},
  {Moeller-Nilsson}, {Mouillet}, {Orign{\'e}}, {Pavlov}, {Petit}, {Puget},
  {Quanz}, {Rabou}, {Ramos}, {Rousset}, {Roux}, {Salasnich}, {Salter},
  {Sauvage}, {Schmid}, {Soenke}, {Stadler}, {Suarez}, {Turatto}, {Udry},
  {Vakili}, {Wahhaj}, {Wildi}, \& {Antichi}}]{Zurlo2016}
{Zurlo}, A., {Vigan}, A., {Galicher}, R., {et~al.} 2016, \aap, 587, A57,
  \dodoi{10.1051/0004-6361/201526835}

\end{thebibliography}
\bibliographystyle{aasjournal}

\end{CJK*}
\end{document}